\documentclass{jfm}

\usepackage{graphicx}
\usepackage{amsmath}
\usepackage{amssymb}
\usepackage[]{natbib}
\usepackage{hyperref}
\hypersetup{
    colorlinks = false,
    urlcolor   = blue,
    citecolor = black,
    citebordercolor  = {0 1 0},
}
\usepackage{comment}
\usepackage{bm}
\usepackage{float}

\newcommand{\RomanNumeralCaps}[1]
\linenumbers

\title{
       A swimming bacterium in a two-fluid model of a polymer solution
      }

\author{Sabarish V. Narayanan\aff{1}, Donald L. Koch\aff{1}
        \and Sarah Hormozi\aff{1}
        \corresp{\email{hormozi@cornell.edu}}
       }

\affiliation{
            \aff{1}Robert Frederick Smith School of Chemical and Biomolecular Engineering, Cornell University, Ithaca, NY 14853, USA.
            }

\begin{document}

\maketitle

\begin{abstract}
We analyse the motion of a flagellated bacterium in a two-fluid medium using slender body theory. The two-fluid model is useful for describing a body moving through a complex fluid with a microstructure whose length scale is comparable to the characteristic scale of the body. This is true for bacterial motion in biological fluids (entangled polymer solutions), where the entanglement results in a porous microstructure with typical pore diameters comparable to or larger than the flagellar bundle diameter but smaller than the diameter of the bacterial head. Thus the polymer and solvent satisfy different boundary conditions on the flagellar bundle and move with different velocities close to it. This gives rise to a screening length $L_B$ within which the fluids exchange momentum and the relative velocity between the two fluids decays. In this work, both the solvent and polymer of the two-fluid medium are modeled as Newtonian fluids with different viscosities $\mu_s$ and $\mu_p$ (viscosity ratio $\lambda$ = $\mu_p/\mu_s$), thereby capturing the effects solely introduced by the microstructure of the complex fluid. From our calculations, we observe an increased drag anisotropy for a rigid, slender flagellar bundle moving through this two-fluid medium, resulting in an enhanced swimming velocity of the organism. The results are sensitive to the interaction between the bundle and the polymer and we discuss two physical scenarios corresponding to two types of interaction. Our model provides an explanation for the experimentally observed enhancement of swimming velocity of bacteria in entangled polymer solutions and motivates further experimental investigations.
\end{abstract}

\begin{keywords}
%Authors should not enter keywords on the manuscript, as these must be chosen by the author during the online submission process and will then be added during the typesetting process (see \href{https://www.cambridge.org/core/journals/journal-of-fluid-mechanics/information/list-of-keywords}{Keyword PDF} for the full list).  Other classifications will be added at the same time.
\end{keywords}

\vspace{-0.5in}

\section{Introduction}\label{1}
Pathogenic bacteria are a persistent threat to human health incurring a heavy cost on the healthcare system\,\citep{1}. The motility of bacteria is an essential mechanism with which pathogens reach the membranes of susceptible cells or form harmful biofilms on tissues and implants\,\citep{2,3,4}. The majority of the cells and tissues prone to pathogenic infections in the human body are lined with a multi-scale complex biological fluid. For instance, the mucosal surfaces in the body including the epithelial cells in the respiratory tract, the human intestines, the urinary tract, the eyes etc. are lined with a slimy hydro-gel known as mucus\,\citep{5}. These are complex fluids, meaning that they possess a microstructure and often exhibit non-Newtonian rheology, which is a function of length scale. Therefore, a fundamental understanding of how the rheology and microstructure of biological fluids affect the motion of a swimming bacterium has applications ranging from designing therapeutic techniques by changing the properties of the biological fluids\,\citep{6} to designing synthetic swimmers for targeted drug delivery\,\citep{7,8,9}, and developing gene regulatory programs for bacteria to name a few. In this work, we develop a two-fluid model, where the complex fluid is modeled as a coupled, interpenetrating medium of two Newtonian fluids, and we analyse the motion of a flagellated bacterium (like {\it Eschericia Coli}) in it. This Newtonian model captures the effect of the microstructure present in these complex biological fluids by allowing for a relative motion between the solvent and polymer. This relative motion results from the fact that, in a fluid with microstructure having a length scale comparable to that of the flagellar bundle of the bacterium, the bundle interacts differently with the solvent and the polymer  exerting different forces on the two fluids. The model can be directly extended to a complex fluid with non-Newtonian rheology, using which, the combined effects of microstructure and non-Newtonian rheology of the complex biological fluid on the swimming bacterium can be analysed by numerical simulations. 

The motion of swimming microorganisms in Newtonian fluids has been a well-studied problem for decades \citep{10,10a,12,13}. Microorganisms, owing to their small size, essentially swim in a low Reynolds number ($Re$) environment, where viscous effects dominate inertial effects. In this Stokesian regime, the fluid flow is quasi-steady and linear making the flow time-reversal invariant. Therefore, the usual swimming strategies at the macroscopic scale, like periodic paddling motion, are ineffective and will result in no net motion \citep{Ludwig30}. To overcome this, microorganisms have evolved several successful propulsion strategies. There are many variations of these strategies among swimming microorganisms and several types of organisms exist that swim in low Reynolds number environments using different means \citep{12,13}. Specifically, there are two broad families of microorganisms namely prokaryotes (e.g. bacteria) and eukaryotes (e.g. sperm cells), and in this work, we restrict our attention to the motion of flagellated bacteria (prokaryotes). \citet{10a} showed that swimming bacteria, like \textit{Escherichia Coli} and \textit{Salmonella Typhimurium}, break the Stokesian symmetry by means of a rotating appendage - the flagellar bundle, made up of multiple individual flagellar filaments. The flagellum is a slender filament attached to the head of the bacterium by a hook and rotated by a molecular motor\,\citep{11}. {\it E.Coli} have prolate spheroidal heads of typical length $2-3 \mu m$ and width $1-2 \mu m$. The flagellar filament of {\it E.Coli} has a diameter of $\approx 20nm$ and traces out a helix with contour length $\approx 10 \mu m$. In the absence of external forces and moments, the helix is typically left-handed with a pitch $\approx 2.5 \mu m$ and a helical diameter $\approx 0.5 \mu m$\,\citep{Bergfig}. There are typically $\sim 5-8$ flagellar filaments per cell of {\it E.Coli}. When all the motors rotate in the same direction, all the filaments wrap into a helical bundle  of diameter $\approx 60 - 80 nm$ and rotate in unison. This generates thrust due to an anisotropy in the drag experienced by the flagellar bundle\,\citep{13}, propelling the bacterium forward - the motion being called a {\it run}. When one or more of the motors reverse, the corresponding filaments leave the bundle and undergo ‘polymorphic’ transformations which change the swimming direction of the cell - this process being called {\it tumbling}. Thus bacteria exhibit run and tumble motion in a fluid medium. This motion of bacteria in Newtonian fluids has been successfully modeled by resistive force theory (RFT)\,\citep{Gray,Chwang} and slender body theory (SBT)\,\citep{Batchelor,Cox,Johnson}, which treat the helical flagellar bundle as a slender fiber moving through a viscous fluid. Recently, these theories for slender objects have been experimentally verified by \citet{Swinney}, by comparing experimentally measured values of thrust, drag and torque on a slender helical fiber, that was rotated and translated in a viscous fluid, with those values predicted by the theories.

While the preceding discussion addressed bacterial motion in Newtonian fluids, bacterial motility in complex fluids is still an open question in many ways. There are several interesting characteristics exhibited by swimming bacteria in complex fluids\,\citep{SpagnolieARFM}. While the class of complex fluids is enormous, most of the attention so far has been centered on one type of complex fluid, namely, polymer solutions. The major motivation for this is that several biological fluids, which these organisms typically encounter, are polymer solutions\,\citep{13}. In these fluids, for instance, bacteria are known to swim in straighter trajectories\,\citep{Patteson15}, exhibit less frequent tumbling\,\citep{Breuer}, and form a flagellar bundle more rapidly\,\citep{Qu18}. 

A fundamental understanding of the aforementioned phenomena necessitates a thorough understanding of the swimming motion of bacteria in polymer solutions. Polymer solutions are non-Newtonian fluids possessing three primary characteristics namely: (i) viscoelasticity, (ii) shear-dependent viscosity, and (iii) a microstructure, and several studies have tried to understand the relative importance of these factors on the swimming motion of bacteria. Earlier theoretical studies on simple geometries like waving sheets\,\citep{Lauga} and waving filaments\,\citep{Fu1,Fu2} in viscoelastic polymer solutions with shear-independent viscosities (Boger fluids) as the swimming media, showed that the propulsive velocities of the sheets and fibers are smaller in viscoelastic fluids than in Newtonian solvents since the polymer solutions always have a larger viscosity than the Newtonian solvents (even with shear-thinning). However, later theoretical studies \,\citep{Teran,Spagnolie,Guy,Lauga2} on undulating sheets and helices, and experimental studies\,\citep{Liu,Zenit}  on artificial swimmers, extended the results of the earlier ones to show that swimming enhancement can result in a viscoelastic (Boger) media due to several factors like large amplitude oscillations\citep{Spagnolie,Liu}, stress-singularities at filament/sheet ends\citep{Teran}, dynamic balance of stresses\citep{Lauga2}, flexibility\citep{Zenit} and (elastic) stress-asymmetry\citep{Guy}. These results suggested that in viscoelastic media, the motion of microswimmers is highly dependent on the geometry of the swimmer, the generated waveform and the relaxation time of the medium, owing to its non-linearity. 

Experiments with actual bacteria, such as {\it E.Coli}, in shear-thinning, viscoelastic polymer solutions reported specifically that they swim at higher speeds than in Newtonian liquids having the same shear viscosity\,\citep{Berg,Patteson15,Qu18}. These studies proposed that shear-thinning of the polymer near the flagellar bundle, owing to its fast rotation, contributes most to the observed swimming enhancement of bacteria for polymers with small relaxation times (low $De$; $De$ being the Deborah number defined as the ratio of the polymer relaxation time to the flow time scale), while viscoelastic effects like normal stress differences and elastic stresses contribute significantly to enhancement with high relaxation time polymers (high $De$)\,\citep{Breuer}. This has motivated theoretical models\,\citep{Lauga3}, which use a two-layer approximation of the polymer solutions at low $De$ - with a layer of lower viscosity near the flagellar bundle and a layer of larger viscosity on the scale of the cell, that explain the observed experimental results. The experiments and the theoretical model mentioned above correspond to viscoelastic polymer solutions, with small polymer concentrations ($c < c^*$; $c$ is polymer concentration and $c^*$ is overlap concentration) as the swimming media, but biological fluids are usually concentrated polymer solutions ($c \geq c^*$)).

There are not many experimental or theoretical studies that address the fluid mechanics of bacterial motion in concentrated polymer solutions. \citet{Berg} first showed that bacteria can swim with higher velocities in concentrated polymer solutions, compared to polymer solutions with short chained polymers having same viscosity, but they wrongly attributed this enhancement to the presence of bacteria sized pores in the polymer network, which do not exist. \citet{Kudo} used this idea to develop a theory, which used different viscosities (resistances) for translation and rotation of both the head and flagellar bundle in a fluid medium modeling the entangled polymer solution. More recently, an experimental study by \citet{Poon} also showed that an enhancement in swimming speed is observed in a concentrated polymer solution ($c \sim c^*$), and the explanation offered was a combination of shear thinning and depletion of polymers from the vicinity of flagellar bundle, essentially making the flagellar bundle swim through a fluid of small viscosity. However, this explanation is not consistent with the fact, that for the polymer solution used in the experiment\,\citep{Poon}, rheological measurements do not show significant shear thinning at the shear rates assumed near the flagellar bundle, with the shear rate directly proportional to the bundle rotation rate. Moreover, the authors do not provide any explanation for there to be a depletion of polymers near the flagellar bundle. A computational work by \citet{Yeomans} on a bacterium swimming through a concentrated polymer solution showed an enhancement due to depletion of polymers near the flagellar bundle. The authors used coarse-grained molecular dynamics (MD) simulations, where the polymer was modeled as a chain of spherical (monomer) beads which were large, resulting in few degrees of freedom for the chain, and the observed depletion near the bundle may therefore be an overestimate that does not correspond to the actual scenario.

Recently the work of \citet{Cheng} has shown that, in the dilute and semi-dilute regimes, $c \lesssim c^*$, the colloidal nature of the polymer solutions quantitatively explain the observed swimming enhancement in the aforementioned experiments, where the enhancement scales with the radius of gyration of the polymer chains, and polymer dynamics may not be essential for capturing the phenomena. This suggests that the length scale of the polymer chains (microstructure) may be more relevant in this regime. Notably, the results of \citet{Cheng} with colloidal suspensions also quantitatively explain other features, namely, straighter trajectories of bacteria, reduced tumbling frequency etc., which were observed in the earlier experiments with dilute polymer solutions. Their findings seem to imply that the length scale of the polymer chains (microstructure), could be the most relevant parameter affecting the swimming velocity directly in  dilute and semi-dilute polymer solutions, while viscoelasticity leads to other consequences like straight trajectories, reduced tumbling etc., which affect velocity indirectly.

In concentrated polymer solutions, the question of relative importance of these characteristics on swimming motion is yet to be answered satisfactorily. Concentrated polymer solutions also exhibit shear-dependent viscosity and viscoelastic stresses, and crucially, they are entangled and possess a porous microstructure, where in some cases, the pore sizes may be comparable to the thickness of the flagellar bundle, but not the head of the bacterium. For instance, it is known that mucus has a microstructure that resembles a mesh, where the mucin filaments form a complicated network of entangled polymer fibers with pores ($\sim 100\,nm \text{ (in humans)}-\,400\,nm \text{ (in horses)}$ \citet{Lehr}), which are larger than the flagellar bundle diameter ($\sim 60-80 nm$ \citet{Bergfig}) and are filled with the Newtonian solvent \citep{Lai,Cone}. Notably, unlike dilute solutions, their viscoelastic response and shear-dependent viscosity cannot be explained by analogy to colloidal suspensions, and therefore the model of \citet{Cheng} cannot be applied. Also, the theoretical models mentioned earlier\,\citep{Kudo,Lauga3,Poon,Yeomans} assume bacteria sized pores, significant shear thinning or depletion near the flagellar bundle, and these assumptions are inappropriate for an entangled polymer solution like mucus. In such a medium, rather than a physical depletion of polymers or shear thinning, one has the flagellar bundle interacting differently with the solvent and polymer, exerting different forces on them, owing to the porous microstructure with pores having nearly the same size as the bundle diameter. This results in a relative motion between the solvent and polymer near the bundle, owing to the solvent and polymer being forced differently by it. Some earlier studies of waving sheets in entangled networks\,\citep{PowersTF,Wada,Du} have used this idea, with the polymer being treated as a purely elastic medium. They have not however used the model with helical geometries, like the flagellar bundle of a bacterium, which requires a slender body treatment and the polymer network in biological fluids like mucus are not perfectly elastic.

In this work, we propose a two-fluid model to accurately capture the effect of microstructure in an entangled polymer solution like mucus, wherein the pore scale is comparable to or larger than the diameter of the flagellar bundle. In our model, the polymer and solvent are both treated as Newtonian fluids with different viscosities $\mu_p$ and $\mu_s$; $\lambda = \mu_p/\mu_s$ being the viscosity ratio. This Newtonian approximation for the polymer is valid if the polymer has small non-Newtonian effects, with $De \ll 1$, an assumption that is fairly representative of the polymer solutions used in the experiments of\,\citet{Poon,Breuer}. In such a scenario, the flagellar bundle directly forces the solvent present in the pores, which then transmits the stresses to the polymer. These two fluids therefore move relative to each other leading to a Darcy drag term in the governing equations and hence a screening length $L_B$, within which the relative velocity of the two fluids decays. The resulting equations for the relative velocity of the solvent and polymer are similar to the Brinkman equations for flow through a porous media\,\citep{Brinkman}. Some earlier studies have developed resistive force theory (RFT) for a slender body\,\citep{Howells98,Leshansky} and a bacterium swimming in a Brinkman medium\,\citep{Pak}. In these studies, the pores result from a sparse random distribution of rigid bodies, whereas in this study, the porous structure results from polymers, which are also subject to motion. We first analyse the motion of a slender helical fiber in such a medium using slender body theory (SBT), and then use RFT to analyse the motion of a bacterium with a helical flagellar bundle in this medium. Our analysis indicates that bacterial motion is sensitive to the nature of the interaction between the flagellar bundle and the polymer, and predicts an increased drag anisotropy. This in turn leads to an enhancement in the swimming speed, compared to the case where the polymer solution is treated as a continuum mixture - a Newtonian medium with viscosity $\mu_s(1 + \lambda)$. We model two physical scenarios, corresponding to two possible polymer-flagellar bundle interactions: (i) a case where the polymer slips past the bundle and (ii) a case where the polymer is not subject to any direct continuum forcing by the bundle (no interaction). 

\section{The two-fluid model of the polymer solution} \label{2}
In this section we develop a two-fluid model of a polymer solution and analyse the motion of a sphere through it in order to explain it's features. Two-fluid models for polymer solutions were first introduced by \citet{Doi}.  While typical mixture models of polymer solutions assume that the polymers and solvent move with a common velocity, Doi's two-fluid model allowed for relative motion between the polymer and the solvent, owing to the fact that under certain conditions inhomogeneous fluid flow can create polymer concentration gradients and lead to diffusion of polymers relative to the solvent flow. The model describes a polymer solution composed of a Newtonian solvent phase with viscosity $\mu_s$ and a polymer phase with the two phases coexisting as interpenetrating continua. In general, the model permits a non-uniform concentration for the polymer, while treating the polymer as a non-Newtonian medium. In this work, the polymer is modeled as a Newtonian fluid with uniform concentration (as this fairly represents the conditions found in earlier experiments\,\citep{Poon,Breuer}), having a different viscosity $\mu_p$, where the viscosity is equivalent to the polymer's contribution to the zero-shear viscosity of the polymer solution. The polymer and the solvent are allowed to move relative to each other and the inertial effects in both fluids are considered to be negligible, with the Reynolds numbers ($Re$) based on both $\mu_s$ and $\mu_p$ being small; $Re \ll 1$.  Here the non-equilibrium condition between the polymer and solvent is created by the different forces they experience at the boundary of the rotating flagellar bundle, because of the microstructure \lq seen\rq\, by the flagellar bundle.  Unlike in \citet{Doi}, we consider the polymer to have a constant concentration and an incompressible mass conservation equation.  This condition can be approximated in the model of \citet{Doi} if the osmotic susceptibility of the polymer is small, so that the osmotic pressure (termed $p_p$ here) takes on whatever values are needed to impose the incompressibility of the polymer phase. The governing equations of our two-fluid model are  given by:
\begin{align}
  &\bm{\nabla}\cdot\bm{u}_s = 0,\quad \bm{\nabla}\cdot\bm{u}_p = 0 = 0 \label{Eq.1}\\
  &\mu_s \nabla^2 \bm{u}_s - \nabla p_s - \xi (\bm{u}_s - \bm{u}_p) = 0 \label{Eq.2} \\
  &\mu_p \nabla^2 \bm{u}_p - \nabla p_p + \xi (\bm{u}_s - \bm{u}_p) = 0 \label{Eq.3}
\end{align}
where, $\bm{u}_s, \bm{u}_p, p_s,$ and $p_p$ correspond to the solvent and polymer phase velocities and pressures respectively and $\xi$ is the Darcy resistance coefficient defined as $\xi = \mu_s/L_B^2$, where $L_B$ is the screening length of the two-fluid medium. As noted earlier, the form of the equations is similar to Brinkman's equations in a porous medium\,\citep{Brinkman}, except that here the polymers forming the porous network are capable of flowing. Thus, $L_B$ can be considered to be the length scale of hydrodynamic coupling in the polymer solution, which is $O(\phi^{-1/2} \log \phi^{1/2})$, $\phi$ being the polymer volume fraction, if the polymers are assumed to be fibers of finite length randomly oriented in space \citep{Howells98}. The above equations can be written in dimensionless form as:
\begin{align}
    &\bm{\nabla}\cdot\bm{u}_s = 0,\quad \bm{\nabla}\cdot\bm{u}_p = 0 \label{Eq.4}\\
    &\nabla^2 \bm{u}_s - \nabla p_s - \frac{1}{L_B^2}(\bm{u}_s - \bm{u}_p) = 0 \label{Eq.5}\\
    &\lambda \nabla^2 \bm{u}_p - \lambda \nabla p_p + \frac{1}{L_B^2}(\bm{u}_s - \bm{u}_p) = 0 \label{Eq.6}
\end{align}
where we have non-dimensionalised the lengths with a characteristic length scale $l$, the velocity with characteristic velocity scale $U$ and the solvent and polymer pressures with $\mu_s U / l$ and $\mu_p U/l$. %We now look at the canonical case of a sphere translating and rotating in the two-fluid medium  to elucidate the physical features of motion in the two-fluid medium.

\subsection{A translating and rotating sphere in the two-fluid medium} \label{2.1}
\begin{figure}
\centering
\includegraphics[trim = {0cm 0cm 0cm 0.5cm}, clip, scale = 0.45]{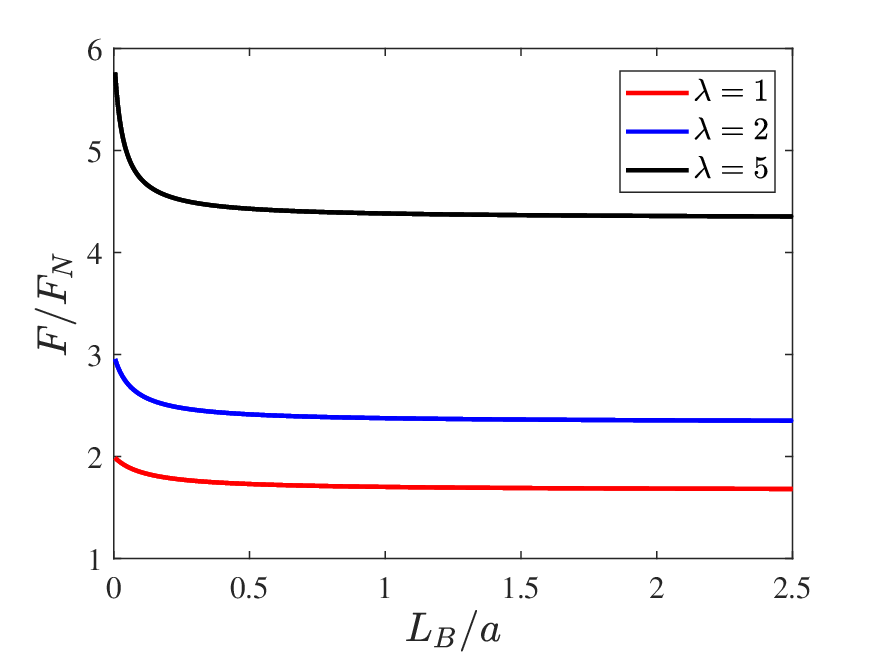}
\includegraphics[trim = {0cm 0cm 0cm 0.5cm}, clip, scale = 0.45]{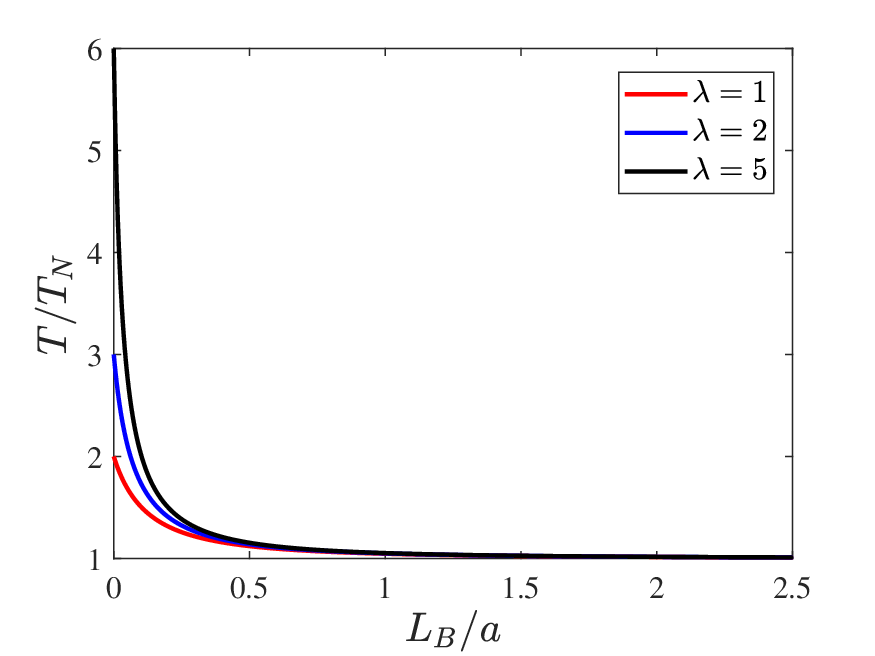}
\caption{Plot of the (a) drag force normalised by $F_N = 6 \pi U \mu_s a$ on a sphere of radius $a$ translating with velocity $U$ and (b) torque normalised by $T_N = 8 \pi \mu_s a^3 \omega$ on a sphere rotating with angular velocity $\omega$ in a two-fluid medium with a slipping polymer, as a function of $L_B/a$.}
\label{fig:1}
\end{figure}

Before embarking on the more challenging problems of studying the motion of a helix and then a bacterium in the two-fluid medium, we study simple flows that provide insight into the response of the medium.  Consider a sphere of radius \textquoteleft$a$' moving with a velocity $\bm{U}$ and rotating with an angular velocity $\bm{\omega}$ through the quiescent two-fluid medium. As mentioned in the introduction, our calculations consider two sets of boundary conditions corresponding to two physical scenarios - i) the polymer slipping past the solid body and ii) the polymer not interacting with the solid body. The first case is relevant when the body moves through an entangled polymer solution with pore sizes comparable to the characteristic length scale of the body (in this example, this is the sphere diameter $2a$) and the second case is relevant when the pore size is much larger than the characteristic length scale of the body, so that the polymer is not directly forced by the moving body, but is forced indirectly by the solvent which is affected by the motion. The solvent satisfies no-slip in both cases. %In such a scenario, the polymer, being a macro-molecule and owing to its entanglement, can slip past the solid boundary, if the solid boundary is non-adsorbing.
Thus, the boundary conditions for the first case are:
\begin{align}
  &\bm{u}_s, \bm{u}_p \rightarrow 0 \; \text{as} \; r \rightarrow \infty \label{Eq.7}\\
  &\bm{u}_s = \bm{U} + \bm{\omega} \times \bm{r} \; \text{at} \; r = a \label{Eq.8}\\
  &\bm{u}_p\cdot\bm{n} = \bm{U}\cdot\bm{n} \; \text{at} \; r = a \label{Eq.9}\\
  &(\bm{I} - \bm{nn})\cdot(\bm{\sigma}_p\cdot\bm{n}) = 0 \; \text{at} \; r = a \label{Eq.10}
\end{align}
which are respectively, the far-field conditions, no slip condition for the solvent, no penetration condition for the polymer and zero tangential polymer stress on the sphere surface (a completely slipping polymer). Here, $r = |\bm{r}|$ is the radial distance and $\bm{n} = \bm{r}/r$ is the unit normal. Therefore, the polymer will not resist tangential motion (shearing) but will resist normal motion (pressure). The solution procedure and the exact expressions for the velocities and pressures for both fluids are given in Appendix \ref{App.A} and it involves solving the above set of coupled PDEs by defining two new fields $\bm{u}_m = \bm{u}_s + \lambda \bm{u}_p$ and $\bm{u}_d = \bm{u}_p - \bm{u}_s$ (similarly for $p_m$, $p_d$ and other varables). These two fields define a mixture field satisfying Stokes equations and a difference field satisfying Brinkman equations, for which solutions are easily derived. Fig.\ref{fig:1} shows the normalised drag force on a translating sphere and the torque on a rotating sphere as functions of the screening length $L_B/a$ for different values of $\lambda$, where the normalisation is with respect to drag and torque in the solvent (of viscosity $\mu_s$). From the figure, we see that as the screening length approaches zero, the dimensional drag on the sphere approaches $6 \pi \mu_s |\bm{U}| (1 + \lambda) a$  for translation and the torque approaches $8 \pi \mu_s a^3 (1+\lambda) |\bm{\omega}|$ for the case of rotation.  These are the corresponding values for drag and torque, in a medium that is a mixture of the two fluids with viscosity $\mu_s(1 + \lambda)$. This suggests that for $L_B/a \rightarrow 0$, the medium behaves like a mixture implying that there is no relative motion between the two fluids, even if one of them is allowed to slip past the solid boundary. The other limit of $L_B/a \rightarrow \infty$ corresponds to decoupled solvent and polymer acting independently of each other, which results in a dimensional drag and torque of $6 \pi \mu_s |\bm{U}| (1 + \frac{2 \lambda}{3}) a$ and $8 \pi \mu_s a^3 (1+\frac{2\lambda}{3}) |\bm{\omega}|$ respectively. The factor $2/3$ arises because, in this limit, the sphere acts like a bubble moving through the polymer on account of the zero tangential stress condition on the polymer. This calculation shows that one can go from a mixture-like behavior to a completely decoupled behavior of the two fluids using the two-fluid model.

\begin{figure}
\centering
\includegraphics[trim = {0cm 0cm 0cm 0.5cm}, clip, scale = 0.45]{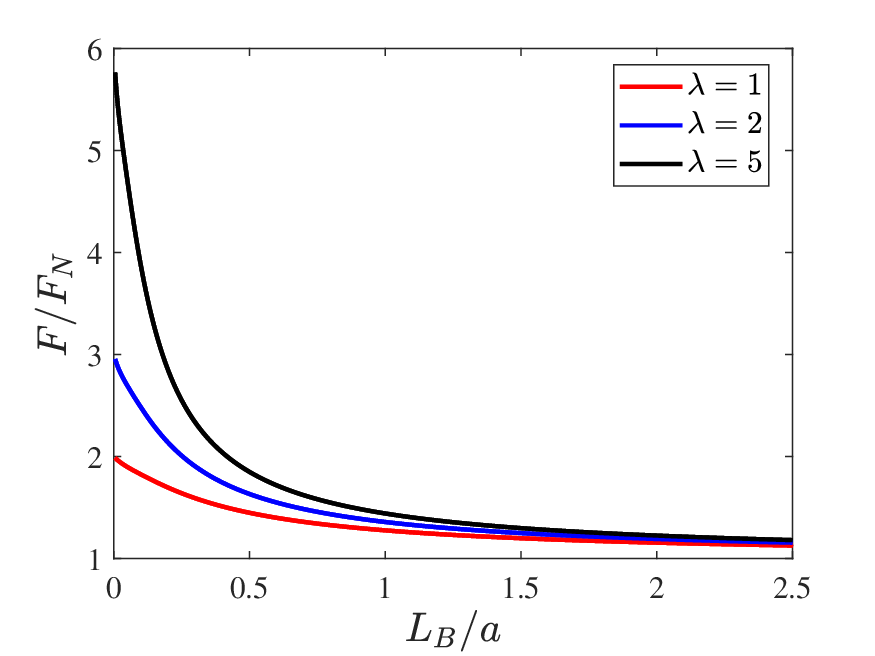}
\includegraphics[trim = {0cm 0cm 0cm 0.5cm}, clip, scale = 0.45]{torque}
\caption{Plot of the (a) drag force normalised by $F_N = 6 \pi U \mu_s a$ on a sphere of radius $a$ translating with velocity $U$ and (b) torque normalised by $T_N = 8 \pi \mu_s a^3 \omega$ on a sphere rotating with angular velocity $\omega$ in a two fluid medium with no polymer-sphere interaction, as a function of $L_B/a$.}
\label{fig:1n}
\end{figure}

A similar calculation can be done for the second case, where there is no polymer-sphere interaction with the boundary conditions now given by:
\begin{align}
  &\bm{u}_s, \bm{u}_p \rightarrow 0 \; \text{as} \; r \rightarrow \infty \label{Eq.7a}\\
  &\bm{u}_s = \bm{U} + \bm{\omega} \times \bm{r}\; \text{at} \; r = a \label{Eq.8a}\\
  &\bm{\sigma}_p\cdot\bm{n} = 0 \; \text{at} \; r = a \label{Eq.9a}
\end{align}
For this case, the plots of normalised drag and torque are given in Fig.\ref{fig:1n}, which are similar to the ones in Fig.\ref{fig:1} (the torque on the sphere being exactly the same). The primary difference between this scenario and the previous one arises in the drag force acting on the sphere in the limit of $L_B/a \rightarrow \infty$, for which the drag on the sphere is $6 \pi \mu_s |\bm{U}| a$. This is consistent with the fact that the polymer is not forced by the sphere and in the limit of large screening length, where the fluids act independently, only the solvent contributes to the drag. 

The takeaway from this sample calculation is that, using the two-fluid model, one can go from mixture-like behavior (a single fluid with viscosity $\mu_s + \mu_p$ - similar to the canonical treatment of polymer solutions) to completely decoupled behavior of the two-fluids by varying $L_B$. Thus, the screening length $L_B$ is equivalent to the characteristic length scale of the microstructure in the polymer solution. In the next section, we derive solutions for a slender helical fiber moving through the two-fluid medium, satisfying the same two sets of boundary conditions as the sphere, and analyse the effect of microstructure on it's motion.

\vspace{-0.1in}
\subsection{Fundamental solutions of the two-fluid equations} \label{2.1}
In order to derive the SBT in the two-fluid medium, we need the fundamental solutions for the two-fluid equations, which are derived  here. The dimensionless governing equations are now given by:
\begin{align}
 \nabla^2 \bm{u}_s - \nabla p_s - \frac{1}{L^2_B} (\bm{u}_s - \bm{u}_p) = \bm{F}_s \delta(\bm{r}) \label{Eq.B1}\\
 \lambda \nabla^2 \bm{u}_p - \lambda \nabla p_p + \frac{1}{L^2_B} (\bm{u}_s - \bm{u}_p) = \lambda \bm{F}_p \delta(\bm{r}) \label{Eq.B2}
\end{align}
with arbitrary forcing on both the solvent ($\bm{F}_s$) and the polymer ($\bm{F}_p$). The solution can be found by writing the above equation in terms of a mixture flow and difference flow, given by:
\begin{align}
 \nabla^2 \bm{u}_m - \nabla p_m = \bm{F}_m \delta(\bm{r}) \label{Eq.B3}\\
 \nabla^2 \bm{u}_d - \nabla p_p - \frac{1 + \lambda}{\lambda L^2_B} (\bm{u}_d) = \bm{F}_d \delta(\bm{r}) \label{Eq.B4}
\end{align}
where $\bm{u}_m = \bm{u}_s + \lambda \bm{u}_p$ and likewise for $p_m$ and $\bm{F}_m$ and $\bm{u}_d = \bm{u}_p - \bm{u}_s$ and similar definitions follow for $p_d$ and $\bm{F}_d$. Since the mixture flow and difference flow equations are the well known Stokes and Brinkman equations, one can find the Green's function for the two-fluid medium using the Green's functions of the Stokes ($\bm{G}_{St}$) and Brinkman media ($\bm{G}_{Br}$). Thus, this Green's function is a tensor $\bm{G}$ consisting of four elements namely $\bm{G}_{SS}$, $\bm{G}_{SP}$, $\bm{G}_{PS}$ and $\bm{G}_{PP}$, i.e.
\begin{align}
 \bm{G} =
 \begin{bmatrix}
    \bm{G}_{SS} & \bm{G}_{SP} \\
    \bm{G}_{PS} & \bm{G}_{PP}
 \end{bmatrix} \label{Eq.B5}
\end{align}
where $\bm{G}_{ij}$ gives the velocity of fluid $i$ due to a force acting on fluid $j$. In order to find these functions, one can write the equation for $\bm{u}_m$ and $\bm{u}_d$ in terms of these functions and equate it to the known Stokesian ($\bm{G}_{St}$) and Brinkman ($\bm{G}_{Br}$) Green's functions. This is given by:
\begin{align}
 \bm{u}_m &= \bm{F}_m \cdot \bm{G}_{St} = \bm{F}_s \cdot (\bm{G}_{SS} + \lambda \bm{G}_{SP}) + \lambda \bm{F}_p\cdot(\bm{G}_{PS} + \lambda \bm{G}_{PP}) \label{Eq.B6}\\
 \bm{u}_d &= \bm{F}_d \cdot \bm{G}_{Br} = \lambda \bm{F}_p \cdot (\bm{G}_{PP} -  \bm{G}_{PS}) - \bm{F}_s \cdot (\bm{G}_{SS} -  \bm{G}_{SP}). \label{Eq.B7}
\end{align}
Solving Eq.\ref{Eq.B6}-\ref{Eq.B7} for the four elements of $\bm{G}$, we get,
\begin{align}
 \bm{G}_{SS} &= \frac{1}{1+\lambda} (\bm{G}_{St} + \lambda \bm{G}_{Br}) \label{Eq.B8}\\
 \bm{G}_{SP} &= \frac{1}{1+\lambda} (\bm{G}_{St} - \bm{G}_{Br}) \label{Eq.B9}\\
 \bm{G}_{PS} &= \frac{1}{1+\lambda} (\bm{G}_{St} - \bm{G}_{Br}) \label{Eq.B10}\\
 \bm{G}_{PP} &= \frac{1}{\lambda(1+\lambda)} (\lambda \bm{G}_{St} + \bm{G}_{Br}). \label{Eq.B11}
\end{align}
Here, the Stokes and Brinkman Green's functions \citep{Howells} are given by:
\begin{align}
 \bm{G}_{St} &= \frac{1}{8 \pi} \left( \frac{\bm{I}}{r} + \frac{\bm{nn}}{r} \right) \label{Eq.B12}\\
 \bm{G}_{Br} &= (\bm{\nabla} \bm{\nabla} - \bm{I}\nabla^2) \left(\frac{2 \lambda  L^2_B \left(1- e^{-\frac{\sqrt{\frac{\lambda +1}{\lambda }} r}{L_B}}\right)}{(\lambda +1) r}\right) \label{Eq.B13}
\end{align}
where $\bm{n}=\bm{r}/r$.

\section{Slender body theory for the two-fluid medium} \label{3}
\begin{figure}
\centerline{\includegraphics[scale = 0.3]{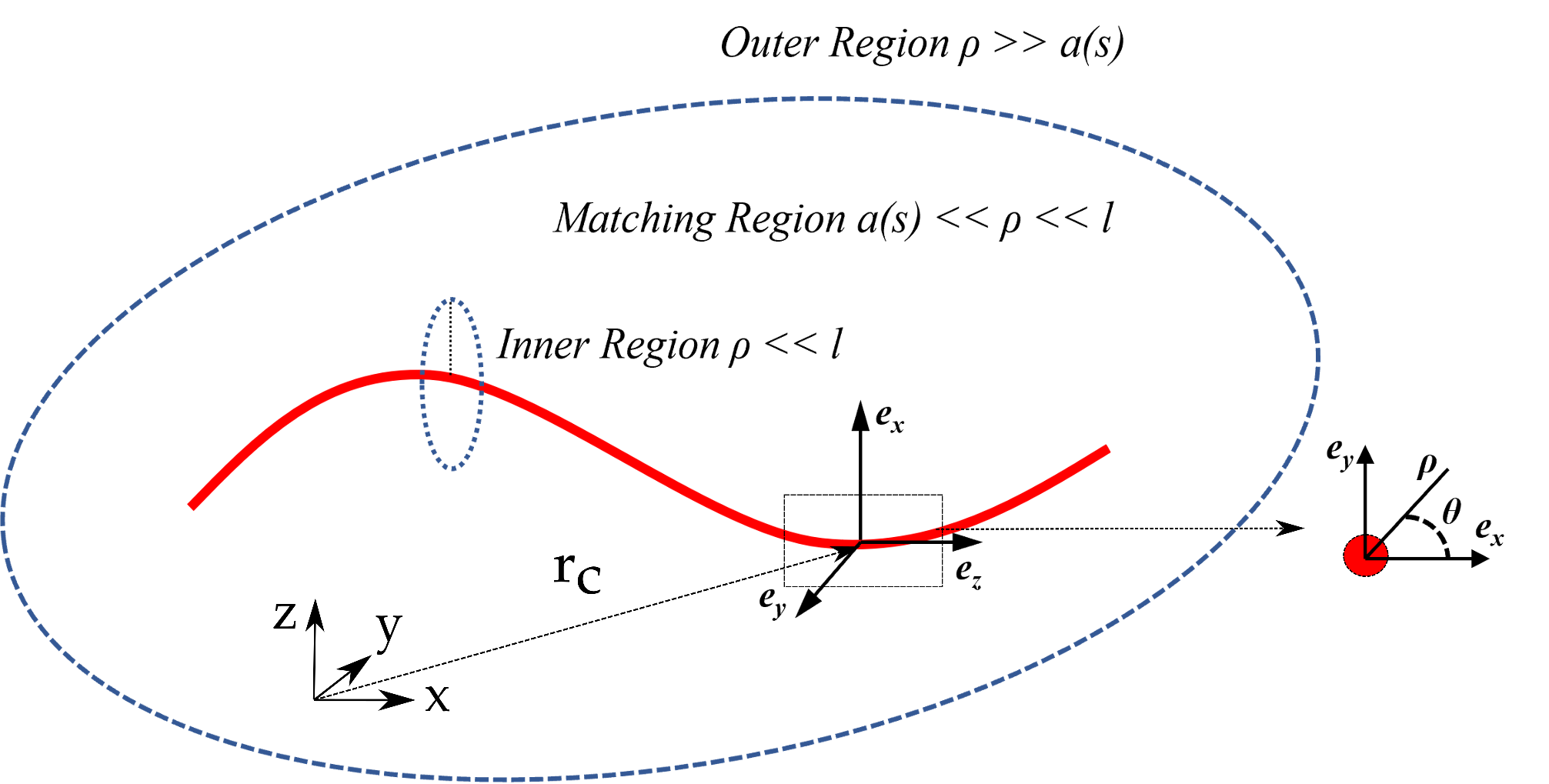}}
\caption{Local coordinate system in the particle reference frame for a general curved body; $\bm{e_z}$ is along the tangent to the filament axis, $\bm{e_x}$ is along the normal and $\bm{e_y}$ is pointed along the binormal to the centerline of the slender body ($\bm{r_c}$).}
\label{fig:2}
\end{figure}

Herein, the velocity disturbance created by a slender fiber with a circular cross-section, when placed in the two-fluid medium, is described using slender body theory (SBT). SBT allows for an approximate solution of the flow produced by bodies which are long and thin in the Stokesian regime\,\citep{Batchelor,Cox,Keller,Johnson,Borker}. The basic idea in SBT is to obtain the strength of a line of singularities placed along the centerline of the slender filament that approximates the field of interest around the filament far away from the cross-sectional surface, termed as the outer region, i.e. $ a \ll \rho$. Here $\rho$ is the radial distance from the centerline of the slender filament, and \textquoteleft $a$' is a measure of the cross-sectional size of the particle at a certain location along the centerline of the slender body as shown in Fig.\ref{fig:2}. The singularity for a Stokes flow problem is a point force. The strength of the singularities is found by matching the field approximated in the outer region, termed as the outer solution, to a field obtained from the inner region ($\rho \ll l$, where $l$ is the length of the slender filament). In the inner region, any curved slender body with $O(1)$ curvature appears locally as a straight infinite cylinder to a first approximation. The velocity field in the inner region is therefore obtained by assuming flow over an infinite cylinder, which is two-dimensional. Thus, the flow along and transverse to the cylinder is solved separately. Any coupling between these flows arises due to curvature and finite aspect ratio of the particle and leads to algebraic $O(\gamma^{-2})$ corrections ($\gamma = l/(2a)$ being the aspect ratio of the fiber) to the velocity disturbance \citep{Cox,Johnson} which are not considered here. Placing higher-order singularities along the centerline of the slender filament gives a better estimate of the field of interest. In Stokes flow, these singularities would include doublets, rotlets, sources, stresslets and quadrupoles \citep{Cox,Keller,Johnson}. These higher order singularities are also not considered in this work.

In this work, we consider a slender body with circular cross-section having a characteristic radius $a$ and length $l$, with aspect ratio $\gamma = l/(2a) \gg 1$. The body is assumed to have a curved centerline, with the curvature ($\kappa$) assumed to be much smaller than the slenderness parameter, i.e $\kappa \ll \gamma$. The radius of cross-section is allowed to vary along the longitudinal direction of the slender body, i.e. $a(s) = a\times \overline{a}(s)$, where $s$ is the arc length along the centerline, with $a$ being chosen as the cross-section radius at the mid-point of the curved centerline. The position vector is denoted by $\bm{r}$ and $\bm{r_c}$ denotes the position of the centerline of the slender body. A local coordinate system ($\bm{e_x}, \bm{e_y}, \bm{e_z}$) is chosen based on the tangent ($\bm{e_z}$ ), normal ($\bm{e_x}$) and binormal ($\bm{e_y}$) to the centerline of the slender body, as shown in Fig.\ref{fig:2} and is mathematically given by:
\begin{align}
    \bm{e_z} = \frac{\partial \bm{r_c}}{\partial s}, \; \bm{e_x} = \frac{1}{\kappa}\frac{\partial^2 \bm{r_c}}{\partial s^2}, \; \bm{e_y} = \bm{e_z}\times \bm{e_x} \label{Eq.11}
\end{align}
where $\kappa$ is the local curvature of the body centerline ($\kappa = \left| \frac{\partial^2 \bm{r_c}}{\partial s^2}\right|$). The velocity on the particle surface ($\bm{r} = \bm{r_s}$) is given by
\begin{align}
    \bm{u}(\bm{r} = \bm{r_s}) = \bm{U} + \bm{\omega} \times \bm{r_s} = \bm{U} + \bm{\omega} \times \bm{r_c} + \bm{\omega} \times (\bm{r_s} - \bm{r_c}) \label{Eq.12}
\end{align}
In canonical SBT for Stokes flow, the only relevant length scale in the inner region is $a$ and all other length scales are assumed to be in the outer region. For the case of a two-fluid medium, we have one other length scale, the screening length $L_B$, which can either be considered part of the inner or outer region, resulting in two different formulations of SBT for a slender fiber. However, these two formulations overlap when $L_B$ is of the same order as the length scale of the matching region. Additionally, one can have different versions of SBT corresponding to different polymer-fiber interactions, which affect the solutions in the inner region. In our study, we consider two types of polymer-fiber interactions: (i) polymer slipping over the fiber and (ii) polymer not interacting with the fiber. For the first case, we consider $L_B$ to be in the inner, outer and matching region and for the second case, $L_B$ is in the outer region, owing to the fact that the no-interaction boundary condition is only applicable if the microstructure length scale is larger than the characteristic length scale of the moving body (here, the fiber cross-sectional diameter $2a$).

\subsection{Slender body theory for polymer slip condition with $L_B$ in the inner region}
For a slipping polymer, when the screening length is in the inner region ($L_B/a \sim O(1)$), the inner solution corresponds to the disturbance field due to the motion of a circular cylinder in the two-fluid medium. In the outer region, the screening length satisfies the limit $L_B/l \ll 1$. From Section \ref{2.1}, we recall that this limit corresponds to the mixture-like behavior of the two-fluid medium, which is essentially a single fluid medium with viscosity $\mu_s(1+\lambda)$. Thus, the outer solution is the velocity disturbance due to the distribution of Stokeslets along the centerline ($\bm{r_c}$) of the fiber in a medium of viscosity $\mu_s(1+\lambda)$. The inner and outer solutions given below for this case are then matched to obtain a governing equation for the singularity strength. Note that in all the cases presented hereafter, the velocities in the inner and outer region are presented in dimensionless form. The inner solution is made dimensionless by choosing $a$, $U$, $\mu_sU/a$ and $\mu_pU/a$ as the length, velocity, and solvent and polymer stress scales. For the outer solution, we choose $l$, $U$, $\mu_sU/l$ and $\mu_pU/l$ as the length, velocity, solvent and polymer stress scales. Also, all the equations are derived for a translating fiber for simplicity, and the rotation of the fiber can be included by simply adding the surface velocity due to rotation to the translation velocity.

\subsubsection{Inner solution ($\rho \ll l$)}
The velocity field around a cylinder of radius $a$ in the two-fluid medium, can be derived by solving the governing equations Eq.\ref{Eq.1}-\ref{Eq.3} following a  procedure similar to that given for a sphere in Appendix \ref{App.A}, subject to the boundary conditions,
\begin{align}
  &\bm{u}_s = \bm{U} \; \text{at} \; \bm{r} = \bm{r_s} \label{Eq.13}\\
  &\bm{u}_p\cdot\bm{n} = \bm{U} \cdot \bm{n} \; \text{at} \; \bm{r} = \bm{r_s} \label{Eq.14}\\
  &(\bm{I} - \bm{nn})\cdot(\bm{\sigma}_p \cdot \bm{n}) = 0 \; \text{at} \; \bm{r} = \bm{r_s} \label{Eq.15}\\
  &\int (\bm{\sigma}_s + \bm{\sigma}_p) \cdot \bm{n} \, dA = \bm{f} \; \text{at} \; \bm{r} = \bm{r_s} \label{Eq.16}
\end{align}
where the last boundary condition is the (unknown) drag force per unit length acting on the cylinder surface, and is the same as the Stokeslet strength of the outer solution. Here $\bm{r} = s \bm{e_z} + \bm{\rho}$ written in terms of a polar coordinate system, where $\bm{\rho} = \bm{e_x} + \bm{e_y}$, with $|\bm{\rho}| = \rho$, which are defined in Eq.\ref{Eq.11}), is normal to the axis of the cylinder and $\bm{e_z}$ is along the axis of the cylinder with $\bm{n} = \frac{\bm{\rho}}{\rho}$ and $\bm{r_s} = a \bm{n}$. In the matching region, the velocity fields are subject to the limit $\rho \gg a$. Since the outer solution for this case is the velocity field in the mixture of two fluids, having viscosity $\mu_s(1+\lambda)$, the inner velocity field is also written for the mixture of solvent and polymer ($\bm{u}_s + \lambda \bm{u}_p$), so as to match it to the outer solution. Therefore, the outer limit ($\rho/a \gg 1$) of the inner mixture velocity field (dimensionless) for transverse and longitudinal motions of the cylinder is written as:
\begin{equation}
\begin{split}
  \bm{u}^{in} &= \bm{U}(1+\lambda) - \left[\frac{\bm{f} \cdot (\bm{I} + \bm{e_z e_z})}{4\pi} \log \left(\rho \right) - \frac{(\bm{f} \cdot \bm{n})\bm{n}}{4 \pi} + \frac{\bm{f}}{4 \pi} (\frac{1}{2} + g(\lambda,L_B)) \right. \\
  &\left. + \frac{(\bm{f} \cdot \bm{e_z})\bm{e_z}}{4 \pi} \left[h(\lambda,L_B) - g(\lambda,L_B) -\frac{1}{2} \right] + O(\frac{1}{\rho^2}) \right]
  \end{split}
  \label{Eq.17}
\end{equation}
Note, that the $\rho$ and $L_B$ that appear inside the logarithm and $g,h$ are dimensionless. The functions $g(\lambda,L_B)$ and $h(\lambda,L_B)$ are given by:
\begin{align}
g &= \frac{\lambda }{ \left(\frac{\frac{1}{L_B} \sqrt{\frac{1+\lambda }{\lambda}} K_1\left(\sqrt{\frac{1+\lambda}{\lambda} \frac{1}{L_B}} \right)}{K_0\left(\sqrt{\frac{1+\lambda}{\lambda} \frac{1}{L_B}} \right)}+2 \lambda +2\right)} \\
h &= \frac{2 \lambda  K_0\left(\sqrt{\frac{1+\lambda}{\lambda} \frac{1}{L_B}} \right)}{\frac{1}{L_B} \sqrt{\frac{\lambda +1}{\lambda}} K_1\left(\sqrt{\frac{1+\lambda}{\lambda} \frac{1}{L_B}} \right)}
\end{align}
where $K_0$, $K_1$ are modified Bessel functions. Note, that $g \rightarrow 0$ and $h \rightarrow 0$ for $L_B \rightarrow 0$ and Eq.\ref{Eq.17} reduces to the solution in a single fluid medium \citep{Keller}.

\subsubsection{Outer solution ($\rho \gg a$)}
The outer solution for this case is the velocity disturbance produced by a distribution of Stokeslets along the centerline of the fiber in a fluid with viscosity $\mu_s(1+\lambda)$, since $L_B$ is in the inner region. Thus, one has:
\begin{align}
 \bm{u}^{out}(\bm{r}) = \bm{U}_{\infty}(\bm{r}) + \frac{1}{8 \pi} \int_{\bm{r_c}} \bm{f}(\bm{r'}) \cdot \left[\frac{\bm{I}}{|\bm{r - r'}|} + \frac{ (\bm{r - r'})(\bm{r - r'}) }{|\bm{r - r'}|^3} \right] ds' \label{Eq.18}
\end{align}
where $\bm{r}$ is the point at which the velocity is evaluated, $\bm{r'}$ takes all values along the
centerline and $ds'$ is the elemental length along the centerline of the slender body. As $\bm{r'} \rightarrow \bm{r}$, the integral diverges as $\log \rho$. One can add and subtract an analytically integrable term that captures the diverging part of the integral as shown by \citet{Keller}. Using $|\bm{r - r'}| = \sqrt{(s-s')^2 + \rho^2}$ in terms of the local polar coordinate system, the resulting expression for the inner limit ($\rho \ll l$) of the outer solution is given by:
\begin{equation}
    \begin{split}
    \bm{u}(\bm{r}) \approx\, &\bm{U}_{\infty}(\bm{r}) + \frac{\bm{f} \cdot (\bm{I + e_z e_z})}{4 \pi} \left(\log \left(\frac{2(\sqrt{s(1-s)})}{\rho} \right) \right) - \frac{\bm{f} \cdot \bm{e_z e_z}}{4 \pi} + \frac{\bm{f} \cdot \bm{nn}}{4 \pi} + \\
    &\frac{1}{8\pi} \int \left[\left( \frac{\bm{I}}{|\bm{r_c}(s) - \bm{r_c}(s')|} + \frac{(\bm{r_c}(s) - \bm{r_c}(s'))(\bm{r_c}(s) - \bm{r_c}(s'))}{|\bm{r_c}(s) - \bm{r_c}(s')|^3} \right) \cdot \bm{f}(\bm{r_c}(s')) \right.\\
    &\left.- \left(\frac{(\bm{I} + \bm{e_z e_z})}{|s - s'|} \right) \cdot \bm{f}(\bm{r_c}(s)) \right] ds'
    \end{split} \label{Eq.19}
\end{equation}
where $\bm{n}$ is the radial unit vector in the $\bm{e_x}-\bm{e_y}$ plane. The integral on the right-hand side of Eq.\ref{Eq.19} is shown to have a finite limit by \citet{Keller}. The $\log \rho$ term in Eq.\ref{Eq.19}, matches the $\log \rho$ term from the inner solution in Eq.\ref{Eq.17} and is a portion of the velocity disturbance produced by an infinite cylinder with the same force per unit length at each point. Here again, the velocity field is dimensionless with lengths non-dimensionalised by $l$, the length of the fiber.

\subsubsection{Matching region ($a \ll \rho \ll l$)}
The velocity produced from the inner solution for $\rho \gg a$, should asymptotically match
the velocity field from the outer solution for $\rho \ll l$ as the velocity field cannot abruptly change in this matching region (i.e. $a \ll \rho \ll l$). Matching the velocity fields from the inner and outer solutions, using Eq.\ref{Eq.17},\ref{Eq.19}, leads to an integral equation for the force per unit length given by:
\begin{equation}
\begin{split}
    \bm{U} = &\frac{\bm{f}(\bm{r}) \cdot (\bm{I} + \bm{e_z}\bm{e_z})}{4 \pi (1 + \lambda)} \left[ \log 2\gamma + \log \left(\frac{2\sqrt{s(1-s)}}{\overline{a}(s)} \right) \right] + \frac{\bm{f}(\bm{r})}{4 \pi (1 + \lambda)} (\frac{1}{2} + g(\lambda,L_B)) +\\
    &\frac{(\bm{f} \cdot \bm{e_z})\bm{e_z}}{4 \pi (1 + \lambda)} \left[h(\lambda,L_B) - g(\lambda,L_B) -\frac{3}{2} \right] + \\
    &\frac{1}{8\pi (1 + \lambda)} \int \left[\left( \frac{\bm{I}}{|\bm{r_c}(s) - \bm{r_c}(s')|} + \frac{(\bm{r_c}(s) - \bm{r_c}(s'))(\bm{r_c}(s) - \bm{r_c}(s'))}{|\bm{r_c}(s) - \bm{r_c}(s')|^3} \right) \cdot \bm{f}(\bm{r_c}(s')) \right.\\
    &\left.- \left(\frac{(\bm{I} + \bm{e_z e_z})}{|s - s'|} \right) \cdot \bm{f}(\bm{r_c}(s)) \right] ds'
\end{split}\label{Eq.20}
\end{equation}
where $\overline{a}(s)$ denotes the changing cross-section of the fiber along it's centerline (note, $a(s) = a \times \overline{a}(s)$, $a$ being the radius at the mid-point of the centerline). Note that the term $\log (2 \gamma) = \log (l/a)$ arises because the inner and outer solutions are non-dimensionalised by different length scales $a$ and $l$ respectively. The error in the above integral equation is $O(\gamma^{-2})$ and so this gives the force per unit length with errors of $O(\gamma^{-2})$. Solution of this integral equation gives the unknown force strength in terms of the known surface velocity and it can be obtained numerically or by an asymptotic expansion in $\epsilon = 1/\log 2\gamma$.

\subsubsection{Resistive force theory (RFT)}
The leading order force per unit length from Eq. \ref{Eq.20} is given by:
\begin{equation}
\begin{split}
    \bm{U}(1+\lambda) = &\frac{\bm{f} \cdot (\bm{I} + \bm{e_z}\bm{e_z})}{4 \pi} \log 2\gamma
\end{split}\label{Eq.21}
\end{equation}
which suggests that a slender filament of any arbitrary cross-section experiences an $O(1/\log 2\gamma)$ viscous drag equal to the viscous drag per unit length experienced by a long cylinder due to its motion relative to the local fluid velocity, in a medium of viscosity $\mu_s(1+\lambda)$ (mixture). The higher-order terms in Eq.\ref{Eq.20} include the additional drag due to relative motion between the two fluids as well as a contribution that comes from the velocity disturbance created by the particle itself. This leading order relation constitutes what is termed the resistive force theory (RFT) \citep{Gray,Chwang,13} for the two-fluid medium for $L_B/a \sim O(1)$. Using, Eq.\ref{Eq.21} above, one can determine the drag per unit length experienced by a locally straight fiber of circular cross-section for motions parallel and perpendicular to the fiber axis with a unit velocity. For the two-fluid medium with the polymer slipping past the fiber, and with $L_B/a \sim O(1)$, these are given by:
\begin{align}
    f_{\perp} = \frac{4 \pi (1 + \lambda)}{\log 2\gamma} \label{Eq.22}\\
    f_{||} = \frac{2 \pi (1 + \lambda)}{\log 2\gamma} \label{Eq.23}
\end{align}
The ratio of the expressions above give the drag anisotropy:
\begin{align}
    \frac{f_{\perp}}{f_{||}} = 2 \label{Eq.24}
\end{align}
which is the same as the anisotropy obtained from RFT for a single fluid medium \citep{Gray,Chwang,13}. This is consistent with the fact that $L_B$ is in the inner region and of the same length scale as the diameter of the fiber. In the outer region, this corresponds to the limit $L_B/l \ll 1$. Therefore, to the leading order in $\epsilon = 1/\log 2\gamma$, the slender fiber essentially swims in a mixture with viscosity $\mu_s(1+\lambda)$, and hence results in the same anisotropy as the single-fluid case.

\subsection{SBT for polymer slip condition with $L_B$ in the outer region}
In this scenario, we consider the screening length as part of the outer region ($L_B/a \gg O(1)$), such that in the inner region, one has the fiber moving in two decoupled fluids, the solvent and polymer, and in the outer region, the two-fluid behavior persists. Accordingly, the inner solution is the disturbance field due to an infinite cylinder moving in solvent and polymer, satisfying independent boundary conditions and the outer solution is approximated by a distribution of the fundamental singularities of the two-fluid medium, along the fiber length. 

\subsubsection{Inner solution}
In the inner region, the solvent satisfies a no-slip condition while the polymer exerts zero tangential stress on the cylinder. These are given by:
\begin{align}
    &\bm{u}_s^{in} = \bm{U} \: \text{at } \bm{r} = \bm{r_s} \label{Eq.32} \\
    &(\bm{I} - \bm{nn})\cdot(\bm{\sigma}_p^{in} \cdot \bm{n}) = 0 \; \text{at } \bm{r} = \bm{r_s} \label{Eq.33} \\
    &\bm{u}_p^{in} \cdot \bm{n} = \bm{U} \cdot \bm{n}\: \text{at } \bm{r} = \bm{r_s} \label{Eq.33A}
\end{align}
Using these conditions and the fact that the solvent and polymer exert a drag per unit length of $\bm{f}_s$ and $\bm{f}_p$, the outer limit ($\rho/a \gg 1$) of the inner solution (dimensionless) is given by:
\begin{equation}
    \bm{u}_s^{in} = \bm{U} - \frac{\bm{f}_s\cdot(\bm{I} + \bm{e_z e_z})}{4 \pi} \log \rho + \frac{\bm{f}_s}{4 \pi}\cdot\left[ \bm{nn} - \frac{\bm{I} - \bm{e_ze_z}}{2}\right] + O(\frac{1}{\rho^2})\label{Eq.34}
\end{equation}
\begin{equation}
    \bm{u}_p^{in} = \bm{U} \cdot (\bm{I} - (1-c)\bm{e_ze_z}) - \frac{\bm{f}_p\cdot(\bm{I} - \bm{e_z e_z})}{4 \pi} \log \rho + \frac{\bm{f}_p}{4 \pi}\cdot \left[ \bm{nn} - (\bm{I} - \bm{e_ze_z}) \right] + O(\frac{1}{\rho^2}) \label{Eq.35}
\end{equation}
The difference in the polymer and solvent fields arise from the difference in the boundary conditions satisfied by the two fluids at the cylinder surface. Importantly, since the polymer is assumed to exert no tangential stress, it follows that $\bm{f}_p \cdot\bm{e_z} = 0$, which renders the surface velocity in the tangential direction arbitrary, denoted here by $\bm{u}_p^{in} \cdot \bm{e_z} = c\; \bm{U} \cdot \bm{e_z}$, where $c$ is an arbitrary constant. This velocity field will be matched to the (inner limit of the) outer solution as before in order to get a governing equation for the force strengths $\bm{f}_s$ and $\bm{f}_p$.

\subsubsection{Outer solution}
In the outer region, the fiber is approximated by a uniform distribution of the fundamental singularities of the two-fluid model - \lq two-fluidlets\rq\, (see Section \ref{2.1}), which are combinations of Stokeslets (associated with the mixture flow satisfying Stokes equations) and  \textquoteleft shielded' Stokeslets (associated with the difference flow satisfying Brinkman equations). Accordingly, the dimensionless velocity field in the outer region for the two fluids are given by:
\begin{equation}
    \bm{u}_s^{out} = \bm{U}_\infty + \frac{1}{8 \pi} \int_{\bm{r}_c} \left[\bm{f}_s \cdot \bm{G}_{SS} + \lambda \bm{f}_p \cdot \bm{G}_{PS}\right] ds' \label{Eq.36}
\end{equation}
\begin{equation}
    \bm{u}_p^{out} = \bm{U}_\infty + \frac{1}{8 \pi} \int_{\bm{r}_c} \left[\lambda \bm{f}_p \cdot \bm{G}_{PP} + \bm{f}_s \cdot \bm{G}_{SP}\right] ds' \label{Eq.37}
\end{equation}
Substituting for the Green's functions $\bm{G}_{SS}$, $\bm{G}_{PP}$, $\bm{G}_{PS}$ and $\bm{G}_{SP}$ from Section \ref{2.1}, we have:
\begin{equation}
\begin{split}
    \bm{u}_s^{out} = &\bm{U}_\infty + \frac{1}{8\pi} \int_{\bm{r}_c} \bm{f}_s(\bm{r'}) \cdot \bm{G}_{St}(\bm{r}-\bm{r'})\, ds' + \\
    &\frac{1}{8 \pi} \int_{\bm{r}_c} \frac{\lambda}{1 + \lambda} \left[\bm{f}_s(\bm{r'}) - \bm{f}_p(\bm{r'}) \right]\cdot\left[\bm{G}_{Br} - \bm{G}_{St} \right](\bm{r}-\bm{r'})\,ds'
    \end{split}\label{Eq.38}
\end{equation}
\begin{equation}
\begin{split}
    \bm{u}_p^{out} = &\bm{U}_\infty + \frac{1}{8\pi } \int_{\bm{r}_c} \bm{f}_p(\bm{r'})\cdot\bm{G}_{St}(\bm{r}-\bm{r'})\, ds' + \\
    &\frac{1}{8 \pi } \int_{\bm{r}_c} \frac{1}{1 + \lambda} \left[\bm{f}_p(\bm{r'}) - \bm{f}_s(\bm{r'}) \right]\cdot\left[\bm{G}_{Br} - \bm{G}_{St} \right](\bm{r}-\bm{r'})\, ds'
    \end{split}\label{Eq.39}
\end{equation}
where we have added and subtracted $\lambda(\bm{f}_s(\bm{r'}) - \bm{f}_p(\bm{r'})) \cdot \bm{G}_{St}$ from Eq.\ref{Eq.38} and $(\bm{f}_p(\bm{r'}) - \bm{f}_s(\bm{r'}) \cdot \bm{G}_{St}$ from Eq.\ref{Eq.39}, with $\bm{G}_{Br}(\bm{r''})$ and $\bm{G}_{St}(\bm{r''})$ given by:
\begin{equation}
    \bm{G}_{Br}(\bm{r''}) = (\bm{\nabla \nabla - I}\nabla^2) \left(\frac{2}{\alpha^2 r''^2} ( 1 - e^{-\alpha r''})\right) \label{Eq.40}
\end{equation}
\begin{equation}
    \bm{G}_{St}(\bm{r''}) = \frac{\bm{I}}{r''} - \frac{\bm{r''r''}}{r''^3} \label{Eq.41}
\end{equation}
where, $\bm{r''} = \bm{r} - \bm{r'}$ and $\alpha = \frac{1}{L_B} \sqrt{\frac{1 + \lambda}{\lambda}}$. The terms involving the difference $\bm{G}_{Br} - \bm{G}_{St}$ do not diverge for $\bm{r} \rightarrow \bm{r'}$. However the term with $\bm{G}_{St}$ does and needs to be treated the same way as before by adding and subtracting a singularity, that asymptotically cancels the divergence in these terms for $\bm{r} \rightarrow \bm{r'}$. The inner limits of the outer velocity fields ($\rho \ll l$) after this simplification are given by:
\begin{equation}
\begin{split}
 \bm{u}_s^{out} &=  \bm{U}_\infty + \frac{\bm{f}_s\cdot (\bm{I + e_z e_z})}{4 \pi} \left(\log \left(\frac{2(\sqrt{s(1-s)})}{\rho}\right) \right) - \frac{\bm{f}_s \cdot \bm{e_z e_z}}{4 \pi} + \frac{\bm{f}_s \cdot \bm{n}}{4 \pi} + \\
 &\frac{1}{8\pi (1 + \lambda)} \int \left[\left( \frac{\bm{I}}{|\bm{r_c}(s) - \bm{r_c}(s')|} + \frac{(\bm{r_c}(s) - \bm{r_c}(s'))(\bm{r_c}(s) - \bm{r_c}(s'))}{|\bm{r_c}(s) - \bm{r_c}(s')|^3} \right) \cdot\bm{f}_{s}(\bm{r_c}(s')) \right.\\
 &\left.- \left(\frac{(\bm{I} + \bm{e_z e_z})}{|s - s'|} \right) \cdot \bm{f}_{s}(\bm{r_c}(s)) \right] ds' + \\
 &\frac{1}{8 \pi} \int \frac{\lambda}{1 + \lambda} \left[\bm{f}_s(\bm{r_c}(s')) - \bm{f}_p(\bm{r_c}(s')) \right]\cdot \left[\bm{G}_{Br} - \bm{G}_{St} \right]\,ds'
 \end{split} \label{Eq.42}
\end{equation}
\begin{equation}
\begin{split}
 \bm{u}_p^{out} &=  \bm{U}_\infty + \frac{\bm{f}_p\cdot (\bm{I + e_z e_z})}{4 \pi} \left(\log \left(\frac{2(\sqrt{s(1-s)})}{\rho}\right) \right) - \frac{\bm{f}_p \cdot\bm{e_z e_z}}{4 \pi} + \frac{\bm{f}_p \cdot \bm{nn}}{4 \pi} + \\
 &\frac{1}{8\pi (1 + \lambda)} \int \left[\left( \frac{\bm{I}}{|\bm{r_c}(s) - \bm{r_c}(s')|} + \frac{(\bm{r_c}(s) - \bm{r_c}(s'))(\bm{r_c}(s) - \bm{r_c}(s'))}{|\bm{r_c}(s) - \bm{r_c}(s')|^3} \right) \cdot \bm{f}_p(\bm{r_c}(s')) \right.\\
 &\left.- \left(\frac{(\bm{I} + \bm{e_z e_z})}{|s - s'|} \right) \cdot \bm{f}_p(\bm{r_c}(s)) \right] ds' + \\
 &\frac{1}{8 \pi} \int \frac{1}{1 + \lambda} \left[\bm{f}_p(\bm{r_c}(s')) - \bm{f}_s(\bm{r_c}(s')) \right]\cdot \left[\bm{G}_{Br} - \bm{G}_{St} \right]\,ds'
 \end{split} \label{Eq.43}
\end{equation}

\subsubsection{Matching Region}
After matching the inner and outer solutions, we get:
\begin{equation}
 \begin{split}
 &\bm{U} = \frac{\bm{f}_s\cdot(\bm{I} + \bm{e_z e_z})}{4 \pi} \left( \log 2\gamma + \log \left(\frac{2\sqrt{s(1 - s)}}{\overline{a}(s)} \right)\right) + \frac{\bm{f}_s\cdot(\bm{I}-3\bm{e_z e_z})}{8 \pi} + \\
 &\frac{1}{8\pi} \int \left[\left( \frac{\bm{I}}{|\bm{r_c}(s) - \bm{r_c}(s')|} + \frac{(\bm{r_c}(s) - \bm{r_c}(s'))(\bm{r_c}(s) - \bm{r_c}(s'))}{|\bm{r_c}(s) - \bm{r_c}(s')|^3} \right) \cdot \bm{f}_s(\bm{r_c}(s')) \right.\\
 &\left.- \left(\frac{(\bm{I} + \bm{e_z e_z})}{|s - s'|} \right) \cdot \bm{f}_s(\bm{r_c}(s)) \right] ds' + \frac{1}{8 \pi} \int \frac{\lambda \left[\bm{f}_s(\bm{r_c}(s')) - \bm{f}_p(\bm{r_c}(s')) \right]}{1 + \lambda} \cdot \left[\bm{G}_{Br} - \bm{G}_{St} \right]\,ds'
 \end{split} \label{Eq.44}
\end{equation}
\begin{equation}
 \begin{split}
 &\bm{U} \cdot(\bm{I} - (1-c)\bm{e_z e_z}) = \frac{\bm{f}_p}{4 \pi } \left( \log 2\gamma + \log \left(\frac{2\sqrt{s(1 - s)}}{\overline{a}(s)} \right)\right) +  \frac{\bm{f}_p\cdot(\bm{I}-2\bm{e_z e_z})}{4 \pi}\\
 &+ \frac{1}{8\pi} \int \left[\left( \frac{\bm{I}}{|\bm{r_c}(s) - \bm{r_c}(s')|} + \frac{(\bm{r_c}(s) - \bm{r_c}(s'))(\bm{r_c}(s) - \bm{r_c}(s'))}{|\bm{r_c}(s) - \bm{r_c}(s')|^3} \right) \cdot\bm{f}_p(\bm{r_c}(s')) \right.\\
 &\left.- \left(\frac{(\bm{I} + \bm{e_z e_z})}{|s - s'|} \right) \cdot \bm{f}_p(\bm{r_c}(s)) \right] ds' + \frac{(\bm{f}_p \cdot\bm{e_z})\bm{e_z}}{4 \pi } \left( \log \frac{2l\sqrt{s(1 - s)}}{\rho} - \log \frac{\rho}{a} \right) +\\
 &\frac{1}{8 \pi } \int \frac{1}{1 + \lambda} \left[\bm{f}_p(\bm{r_c}(s')) - \bm{f}_s(\bm{r_c}(s')) \right]\cdot\left[\bm{G}_{Br} - \bm{G}_{St} \right]\,ds' \\
 \end{split} \label{Eq.45}
\end{equation}
where the equation for $\bm{f}_p$ is accompanied by an additional condition given by $\bm{f}_p \cdot \bm{e_z} = 0$. Here, the tensor $\bm{G}_{Br} - \bm{G}_{St}$ is given by:
\begin{equation}
 \begin{split}
 \bm{G}_{Br} - \bm{G}_{St} = &\bm{I} \left( \frac{2}{\alpha^2 r''^3} \left( e^{-\alpha r''}\left(1 + \alpha r''+ \alpha^2 r''^2\right) - 1\right) - \frac{1}{r''} \right) + \\
 &(\bm{r_c}(s) - \bm{r_c}(s')) \left(\frac{6}{\alpha^2 r''^5} \left( 1 - e^{-\alpha r''}\left(1 + \alpha r''+\frac{\alpha^2 r''^2}{3}\right) \right) - \frac{1}{r''^3} \right)
 \end{split} \label{Eq.46}
\end{equation}
\begin{equation}
 \bm{G}_{Br} - \bm{G}_{St} = \mathcal{F}_1(\alpha,r'') \, \bm{I}  + \mathcal{F}_2(\alpha,r'')\, (\bm{r_c}(s) - \bm{r_c}(s')) \label{Eq.46a}
\end{equation}
where $r'' = |(\bm{r_c}(s) - \bm{r_c}(s'))|$. Contracting Eq.\ref{Eq.45} with $\bm{e_z}$ to get the equation for the arbitrary constant $c$, we have:
\begin{equation}
\begin{split}
    c\;\bm{U}\cdot\bm{e_z} = &\frac{\bm{f}_p \cdot \bm{e_z}}{4 \pi } \left( \log 2\gamma + \log \left(\frac{2\sqrt{s(1 - s)}}{\overline{a}(s)} \right)\right) -  \frac{\bm{f}_p\cdot\bm{ e_z}}{4 \pi} + \frac{(\bm{f}_p\bm{.e_z})}{4 \pi } \left( \log \frac{2l\sqrt{s(1 - s)}}{\rho} - \log \frac{\rho}{a} \right) \\
    &+\frac{1}{8\pi} \int \left[\bm{f}_p(\bm{r_c}(s'))\cdot\left( \frac{\bm{I}}{|\bm{r_c}(s) - \bm{r_c}(s')|} + \frac{(\bm{r_c}(s) - \bm{r_c}(s'))(\bm{r_c}(s) - \bm{r_c}(s'))}{|\bm{r_c}(s) - \bm{r_c}(s')|^3} \right) \cdot \bm{e_z} \right. \\
    &\left. - \frac{2 \bm{f}_p(\bm{r_c}(s)) \cdot \bm{e_z}}{|s -s'|} \right] \, ds' + \frac{1}{8 \pi } \int \frac{(\bm{f}_p(\bm{r_c}(s')) - \bm{f}_s(\bm{r_c}(s')))}{1 + \lambda} \cdot \left[\bm{G}_{Br} - \bm{G}_{St} \right] \cdot\bm{e_z} \, ds'
\end{split} \label{Eq.47}
\end{equation}
Substituting Eq.\ref{Eq.47} in Eq.\ref{Eq.45}, we get for $\bm{f}_p$:
\begin{equation}
 \begin{split}
 &\bm{U}\cdot(\bm{I} - \bm{e_z e_z}) = \frac{\bm{f}_p \cdot (\bm{I} - \bm{e_z e_z}) }{4 \pi } \left( \log 2\gamma + \log \left(\frac{2\sqrt{s(1 - s)}}{\overline{a}(s)} \right)\right) +  \frac{\bm{f}_p\cdot(\bm{I}-\bm{e_z e_z})}{4 \pi}\\
 &+ \frac{1}{8\pi} \int \left[\left( \frac{\bm{I}}{|\bm{r_c}(s) - \bm{r_c}(s')|} + \frac{(\bm{r_c}(s) - \bm{r_c}(s'))(\bm{r_c}(s) - \bm{r_c}(s'))}{|\bm{r_c}(s) - \bm{r_c}(s')|^3} \right) \cdot \bm{f}_p(\bm{r_c}(s')) \right.\\
 &\left.- \left(\frac{(\bm{I} + \bm{e_z e_z})}{|s - s'|} \right) \cdot \bm{f}_p(\bm{r_c}(s)) \right] ds' +
 \frac{1}{8 \pi } \int \frac{\left[\bm{f}_p(\bm{r_c}(s'))- \bm{f}_s(\bm{r_c}(s')) \right]}{1 + \lambda} \cdot\left[\bm{G}_{Br} - \bm{G}_{St} \right]\,ds' \\
 &-\frac{1}{8\pi} \int \left\{ \left[ \bm{f}_p(\bm{r_c}(s'))\cdot\left( \frac{\bm{I}}{|\bm{r_c}(s) - \bm{r_c}(s')|} + \frac{(\bm{r_c}(s) - \bm{r_c}(s'))(\bm{r_c}(s) - \bm{r_c}(s'))}{|\bm{r_c}(s) - \bm{r_c}(s')|^3} \right) \cdot \bm{e_z}  \right. \right. \\
 &\left. \left. -\frac{2 \bm{f}_p(\bm{r_c}(s)) \cdot \bm{e_z}}{|s -s'|} \right] \right\} \bm{e_z}
 -\frac{1}{8 \pi } \int \left \{ \frac{(\bm{f}_p(\bm{r_c}(s')) - \bm{f}_s(\bm{r_c}(s')))}{1 + \lambda} \cdot \left[\bm{G}_{Br} - \bm{G}_{St} \right] \cdot\bm{e_z} \right \}\bm{e_z} \, ds'
 \end{split} \label{Eq.45N}
\end{equation}
with the condition $\bm{f}_p \cdot \bm{e}_z = 0$. The force strengths are obtained by simultaneously solving Eq.\ref{Eq.44} and Eq.\ref{Eq.45N}, with the definitions of $\bm{G}_{Br}$ and $\bm{G}_{St}$ given in Eq.\ref{Eq.46},\ref{Eq.46a}.

\subsubsection{Resistive force theory}
The leading order solution to the force strengths $\bm{f}_s$ and $\bm{f}_p$ for this scenario ($L_B/a \gg O(1)$) are given by:
\begin{equation}
    \bm{U} = \frac{\bm{f}_s \cdot(\bm{I} + \bm{e_z e_z})}{4 \pi} \log 2\gamma \label{Eq.48}
\end{equation}
\begin{equation}
    \bm{U}\cdot(\bm{I} - \bm{e_z e_z}) = \frac{\bm{f}_p \cdot (\bm{I} - \bm{e_z e_z})}{4 \pi} \log 2\gamma \label{Eq.49}
\end{equation}
The total force defined as $\bm{f} = \bm{f}_s + \lambda \bm{f}_p$, is therefore:
\begin{equation}
    \bm{f} = 4 \pi \bm{U} \cdot \left[ (\bm{I} - \frac{\bm{e_z e_z}}{2}) + \lambda (\bm{I} - \bm{e_z e_z}) \right] \frac{1}{\log 2\gamma} \label{Eq.50}
\end{equation}
to the leading order in $\epsilon = 1/\log (2 \gamma)$. The components of the force for translation parallel and perpendicular to the local filament axis (with unit velocity) are:
\begin{equation}
    f_{\perp} = \frac{4 \pi (1 + \lambda)}{\log 2\gamma} \label{Eq.51}
\end{equation}
\begin{equation}
    f_{||} = \frac{2 \pi}{\log 2\gamma} \label{Eq.52}
\end{equation}
and the anisotropy for this case is given by:
\begin{equation}
    \frac{f_{\perp}}{f_{||}} = 2(1 + \lambda) \label{Eq.53}
\end{equation}
which is a factor of $1+\lambda$ larger than the case with $L_B/a \sim O(1)$. Thus, in a scenario where the polymer slips past a fiber, with the screening length larger than the fiber diameter, the drag anisotropy increases, and is proportional to the viscosity ratio $\lambda$.

\subsection{Slender body theory for polymer slip condition with $L_B$ in the matching region}
\begin{figure}
\centerline{\includegraphics[scale = 0.65]{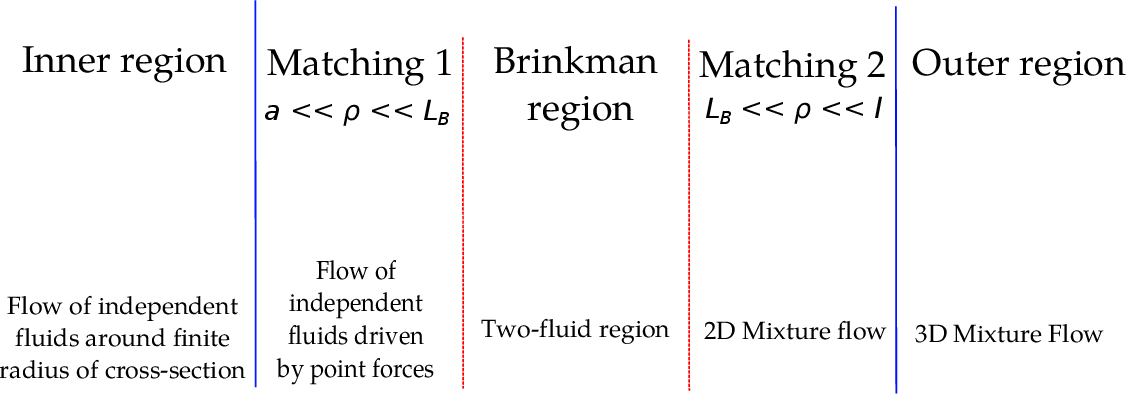}}
\caption{A schematic showing the inner, outer and Brinkman regions for the two-fluid model with $a \ll L_B \ll l$.}
\label{fig:2A}
\end{figure}

When $L_B$ is in the matching region ($a \ll L_B \ll l$), the outer solution remains a $3D$ mixture flow, where the fiber can be approximated as a smooth distribution of Stokeslets along the centerline. The inner solution corresponds to the flow disturbance in decoupled solvent and polymer fluids produced by the moving cylinder. However, for $a \ll L_B \ll l$, there exists a Brinkman region in between the two, where the flow remains two-dimensional, but has coupled two-fluid behavior. This is sketched in Fig.\ref{fig:2A}. In order to obtain a governing integral equation for the force strengths, $\bm{f}_s$ and $\bm{f}_p$, one needs to perform two matching procedures as opposed to just one employed in the previous cases. The first matching is done in matching region 1, where $a \ll \rho \ll L_B$ and the second matching is done in matching region 2, where $L_B \ll \rho \ll l$.

A detailed derivation for this case is given in Appendix \ref{App.C}, and below, we directly give the governing integral equation for the force strengths $\bm{f}_s$ and $\bm{f}_p$ for $a \ll L_B \ll l$.
\begin{align}
\begin{split}
    &\bm{U} =\; \bm{U}_{\infty} + \frac{(\bm{f}_s + \lambda \bm{f}_p) \cdot (\bm{I} + \bm{e_z e_z})}{4\pi (1+\lambda)} \left[ \log (2 \gamma) + \log \left(\frac{2\sqrt{s(1-s)}}{\overline{a}(s)} \right) \right]+ \\
    &\frac{1}{8\pi (1 + \lambda)} \int \left[\left( \frac{\bm{I}}{|\bm{r_c}(s) - \bm{r_c}(s')|} + \frac{(\bm{r_c}(s) - \bm{r_c}(s'))(\bm{r_c}(s) - \bm{r_c}(s'))}{|\bm{r_c}(s) - \bm{r_c}(s')|^3} \right)\cdot(\bm{f}_s + \lambda \bm{f}_p)(\bm{r_c}(s')) \right.\\
    &\left.- \left(\frac{(\bm{I} + \bm{e_z e_z})}{|s - s'|} \right)\cdot(\bm{f}_s + \lambda \bm{f}_p)(\bm{r_c}(s)) \right] ds' + \frac{(\bm{f}_s + \lambda \bm{f}_p)\cdot(\bm{I}-3\bm{e_z e_z})}{8 \pi (1+\lambda)}\\
    &+\frac{\lambda (\bm{f}_s - \bm{f}_p)\cdot(\bm{I} + \bm{e_z e_z})}{4 \pi (1+\lambda)} \left[(\log 2 - \Gamma) + \log\left(\frac{L_B}{a} \right)\right]
\end{split}\label{Eq.53A}
\end{align}
\begin{align}
\begin{split}
    &\bm{U}\cdot(\bm{I} - (1-c)\bm{e_z e_z}) =\; \bm{U}_{\infty} + \frac{(\bm{f}_s + \lambda \bm{f}_p) \cdot (\bm{I} + \bm{e_z e_z})}{4\pi (1+\lambda)} \left[ \log (2 \gamma) + \log \left(\frac{2\sqrt{s(1-s)}}{\overline{a}(s)} \right) \right]+ \\
    &\frac{1}{8\pi (1 + \lambda)} \int \left[\left( \frac{\bm{I}}{|\bm{r_c}(s) - \bm{r_c}(s')|} + \frac{(\bm{r_c}(s) - \bm{r_c}(s'))(\bm{r_c}(s) - \bm{r_c}(s'))}{|\bm{r_c}(s) - \bm{r_c}(s')|^3} \right)\cdot(\bm{f}_s + \lambda \bm{f}_p)(\bm{r_c}(s')) \right.\\
    &\left.- \left(\frac{(\bm{I} + \bm{e_z e_z})}{|s - s'|} \right)\cdot(\bm{f}_s + \lambda \bm{f}_p)(\bm{r_c}(s)) \right] ds' + \frac{ (\bm{f}_s + \lambda \bm{f}_p)\cdot(\bm{I}-\bm{e_z e_z})}{8 \pi (1+\lambda)} + \frac{(\bm{f}_s + \lambda\bm{f}_p)\cdot \bm{e_ze_z}}{4 \pi (1+\lambda)} \\
    &+\frac{\bm{f}_p \cdot (\bm{I} - \bm{e_z e_z}) }{8 \pi} +\left(\frac{\bm{f}_p}{4 \pi (1+\lambda)}-\frac{\bm{f}_s\cdot(\bm{I} + \bm{e_z e_z})}{4 \pi (1+\lambda)}\right) \left[(\log 2 - \Gamma) +  \log\left(\frac{L_B}{a} \right)\right]
\end{split}\label{Eq.53Bo}
\end{align}
where $\Gamma$ is the Euler-Mascheroni constant. Note, that the factor $\log(L_B/a)$ does not lead to a divergence as $L_B/a \rightarrow 0$, since it is multiplied by $\bm{f}_s - \bm{f}_p$, which tends to zero as $L_B/a \rightarrow 0$. Eq.\ref{Eq.53Bo} can again be contracted with $\bm{e_z}$ to obtain $c$ as:
\begin{align}
\begin{split}
    &c(\bm{U}\cdot\bm{e_z}) =\; \frac{(\bm{f}_s + \lambda \bm{f}_p) \cdot \bm{e_z}}{2\pi (1+\lambda)} \left[ \log (2 \gamma) + \log \left(\frac{2\sqrt{s(1-s)}}{\overline{a}(s)} \right) \right] +\frac{(\bm{f}_s + \lambda\bm{f}_p)\cdot\bm{e_z}}{4 \pi (1+\lambda)} \\
    &+ \frac{1}{8\pi (1 + \lambda)} \int \left[ (\bm{f}_s + \lambda \bm{f}_p)(\bm{r_c}(s')) \cdot \left( \frac{\bm{I}}{|\bm{r_c}(s) - \bm{r_c}(s')|} + \frac{(\bm{r_c}(s) - \bm{r_c}(s'))(\bm{r_c}(s) - \bm{r_c}(s'))}{|\bm{r_c}(s) - \bm{r_c}(s')|^3} \right)\cdot \bm{e_z} \right.\\
    &\left.- \left(\frac{(\bm{f}_s + \lambda \bm{f}_p)(\bm{r_c}(s)) \cdot \bm{e_z}}{|s - s'|} \right) \right] ds' + \left(\frac{\bm{f}_p \cdot \bm{e_z}}{4 \pi (1+\lambda)}-\frac{\bm{f}_s\cdot \bm{e_z}}{2 \pi (1+\lambda)}\right) \left[(\log 2 - \Gamma) +  \log\left(\frac{L_B}{a} \right)\right]
\end{split}\label{Eq.53Ba}
\end{align}
which can be substituted into Eq.\ref{Eq.53Bo} to obtain:
\begin{align}
\begin{split}
    &\bm{U}\cdot(\bm{I} - \bm{e_z e_z}) =\; \bm{U}_{\infty} + \frac{(\bm{f}_s + \lambda \bm{f}_p) \cdot (\bm{I} - \bm{e_z e_z})}{4\pi (1+\lambda)} \left[ \log (2 \gamma) + \log \left(\frac{\sqrt{s(1-s)}}{\overline{a}(s)} \right) \right]+ \\
    &\frac{1}{8\pi (1 + \lambda)} \int \left[\left( \frac{\bm{I}}{|\bm{r_c}(s) - \bm{r_c}(s')|} + \frac{(\bm{r_c}(s) - \bm{r_c}(s'))(\bm{r_c}(s) - \bm{r_c}(s'))}{|\bm{r_c}(s) - \bm{r_c}(s')|^3} \right)\cdot(\bm{f}_s + \lambda \bm{f}_p)(\bm{r_c}(s')) \right.\\
    &\left.- \left(\frac{(\bm{I} + \bm{e_z e_z})}{|s - s'|} \right)\cdot(\bm{f}_s + \lambda \bm{f}_p)(\bm{r_c}(s)) \right] ds' + \frac{ \bm{f}_p\cdot(\bm{I}-\bm{e_z e_z})}{8 \pi} + \\
    &\frac{(\bm{f}_s + \lambda\bm{f}_p) \cdot (\bm{I} - \bm{e_z e_z})}{8 \pi (1+\lambda)} +\frac{(\bm{f}_p - \bm{f}_s)\cdot(\bm{I} - \bm{e_z e_z})}{4 \pi (1+\lambda)} \left[(\log 2 - \Gamma) + \log\left(\frac{L_B}{a} \right)\right] -\\
    &\frac{1}{8\pi (1 + \lambda)} \int \left \{ \left[ (\bm{f}_s + \lambda \bm{f}_p)(\bm{r_c}(s')) \cdot \left( \frac{\bm{I}}{|\bm{r_c}(s) - \bm{r_c}(s')|} + \frac{(\bm{r_c}(s) - \bm{r_c}(s'))(\bm{r_c}(s) - \bm{r_c}(s'))}{|\bm{r_c}(s) - \bm{r_c}(s')|^3} \right)\cdot \bm{e_z} \right. \right.\\
    &\left. \left.- \left(\frac{(\bm{f}_s + \lambda \bm{f}_p)(\bm{r_c}(s) \cdot \bm{e_z}}{|s - s'|} \right) \right] \right \} \cdot \bm{e_z} \, ds'
\end{split}\label{Eq.53B}
\end{align}
with the condition $\bm{f}_p \cdot\bm{e_z} = 0$. The equations Eq.\ref{Eq.20}, Eq.\ref{Eq.44},\ref{Eq.45N} and Eq.\ref{Eq.53A}-\ref{Eq.53B} correspond to the three formulations of slender body theory when one has a polymer that slips past the fiber. Each version has its own domain of validity depending on the screening length $L_B$. The three versions, however, can be combined into a single equation by using the formula:
\begin{align}
    \text{SBT}_{\text{Uniformly Valid}} = \text{SBT}_{L_B \sim O(a)} + \text{SBT}_{L_B \gg O(a)} - \text{SBT}_{a \ll L_B \ll l} \label{Eq.54A}
\end{align}
which is uniformly valid for all $L_B$.

\subsubsection{Resistive Force Theory}
For the case with $a \ll L_B \ll l$, the leading order solutions to the force strengths from Eq.\ref{Eq.53A},\ref{Eq.53B} are given by:
\begin{align}
    &\bm{U} = \frac{(\bm{f}_s + \lambda \bm{f}_p) \cdot (\bm{I} + \bm{e_z e_z})}{4 \pi (1 + \lambda)} \log (2 \gamma) + \frac{\lambda(\bm{f}_s-\bm{f}_p) \cdot (\bm{I} + \bm{e_z e_z})}{4\pi(1+\lambda)} \log \left( \frac{L_B}{a} \right)\label{Eq.55A} \\
    &\bm{U} \cdot (\bm{I} - \bm{e_z e_z}) = \frac{(\bm{f}_s + \lambda \bm{f}_p) \cdot (\bm{I} - \bm{e_z e_z})}{4 \pi (1 + \lambda)} \log (2 \gamma) + \frac{(\bm{f}_p-\bm{f}_s) \cdot (\bm{I} - \bm{e_z e_z})}{4\pi(1+\lambda)} \log \left( \frac{L_B}{a} \right)\label{Eq.55B}
\end{align}
where the leading order equation contains the term with coefficient $\log(L_B/a)$ as $L_B \gg a$, when the screening length is in the matching region.
Simplifying Eq.\ref{Eq.55A}, we get:
\begin{align}
\bm{U} \cdot \left(\bm{I} - \frac{\bm{e_z e_z}}{2}\right) = \frac{(\bm{f}_s + \lambda \bm{f}_p)}{4 \pi (1 + \lambda)} \log (2 \gamma) + \frac{\lambda(\bm{f}_s-\bm{f}_p)}{4\pi(1+\lambda)} \log \left( \frac{L_B}{a} \right)\label{Eq.55C}.
\end{align}
Eq.\ref{Eq.55B} directly gives the perpendicular component of the force strengths and Eq.\ref{Eq.55C} can be used to obtain the parallel component of the force strength. Using $\bm{f} = \bm{f}_s + \lambda \bm{f}_p$ and $\bm{f}_p \cdot \bm{e_z} = 0$, the parallel and perpendicular components of the total force on the fiber (for unit velocity of motion) are:
\begin{align}
    f_{\perp} &= \frac{4 \pi (1 + \lambda)}{\log 2\gamma} \label{Eq.55D} \\
    f_{||} &= \frac{2 \pi (1+\lambda)}{\log 2\gamma +  \lambda \log \left(\frac{L_B}{a} \right)} \label{Eq.55E}
\end{align}
and the anisotropy for this case is given by:
\begin{equation}
    \frac{f_{\perp}}{f_{||}} = \frac{2(\log 2\gamma + \lambda \log \left(\frac{L_B}{a} \right))}{\log (2\gamma)} = 2 \left(1 + \lambda\frac{\log \left(\frac{L_B}{a} \right)}{\log (2 \gamma)} \right) \label{Eq.55F}
\end{equation}
which reduces to the drag anisotropy for the case with screening length in the inner region for $L_B = a$ and to the drag anisotropy for the case with screening length in the outer region for $L_B = 2l$. Thus, one can combine the leading order solutions to $f_{\perp}$ and $f_{||}$ for the case of slipping polymer with $L_B$ in the inner, matching and outer region into the following piecewise-continuous form given by:
\begin{equation}
    f_{\perp} = \frac{4 \pi (1+\lambda)}{\log (2 \gamma)}; \quad \text{for} L_B \in [0,\infty], \label{Eq.55G}
\end{equation}
and
\begin{equation}
    f_{||} = \begin{cases}
        \frac{2 \pi (1 + \lambda)}{\log (2 \gamma)}; \quad &\text{for} L_B \in [0,a] \\
        \frac{2 \pi (1 + \lambda)}{\log (2 \gamma) + \lambda \log \left(\frac{L_B}{a} \right)}; \quad &\text{for} L_B \in [a, 2l] \\
        \frac{2 \pi}{\log (2 \gamma)}; \quad &\text{for} L_B \in [2l,\infty)
    \end{cases}.\label{Eq.55H}
\end{equation}
which provides the leading order solution to force strengths for the fiber in a two-fluid medium with polymer slip valid for all $L_B$.

\subsection{SBT with no polymer-fiber interaction}
We now consider the second type of polymer-fiber interaction, where the polymer in the two-fluid medium does not exert any direct force on the fiber, i.e. the polymer satisfies,
\begin{equation}
    \bm{\sigma}_p^{in} \cdot \bm{n} = 0 \; \text{at } \bm{r}  = \bm{r_s}\label{Eq.54}
\end{equation}
while the solvent still satisfies no-slip and no-penetration condition on the fiber surface, given by:
\begin{equation}
    \bm{u}_s^{in} = \bm{U} \; \text{at } \bm{r} = \bm{r_s} \label{Eq.55}
\end{equation}
This essentially implies, that the polymer can now move with an arbitrary velocity even in the plane perpendicular to the filament axis when $a \le \rho \le L_B$. This model nevertheless captures the physical scenario when the fiber is much smaller than the pores of the underlying microstructure in the complex fluid, because the polymers do not experience any direct forcing from the fiber motion. For this case, the screening length $L_B$ is considered part of the outer region, since $L_B$ is equivalent to the length scale of the microstructure and for these boundary conditions to hold, $L_B \gg a$. 

\subsubsection{Inner solution}
The inner solution for this scenario only involves the solvent velocity field, satisfying no-slip and no-penetration conditions, which is given by:
\begin{equation}
    \bm{u}_s^{in} = \bm{U} - \frac{\bm{f}_s\cdot(\bm{I} + \bm{e_z e_z})}{4 \pi} \log \frac{\rho}{a} + \frac{\bm{f}_s}{4 \pi}\cdot\left[ \bm{nn} - \frac{\bm{I} - \bm{e_ze_z}}{2}\right] + O(\frac{1}{\rho^2})\label{Eq.56}
\end{equation}
for $\rho \gg a$. As already noted the polymer can have an arbitrary velocity in the inner region given by $\bm{u}_p^{in} = c \bm{U}$.

\subsubsection{Outer solution}
The outer solution is obtained by approximating the fiber as a uniform distribution of two-fluidlets with the constraint $\bm{f}_p = 0$. Thus, we have for the outer solution,
\begin{equation}
    \bm{u}_s^{out} = \bm{U}_\infty + \frac{1}{8 \pi} \int_{\bm{r}_c} \left[\bm{f}_s \cdot \bm{G}_{SS} \right] ds' \label{Eq.57}
\end{equation}
\begin{equation}
    \bm{u}_p^{out} = \bm{U}_\infty + \frac{1}{8 \pi} \int_{\bm{r}_c} \left[\bm{f}_s \cdot \bm{G}_{SP}\right] ds' \label{Eq.58}
\end{equation}
where we have applied the constraint $\bm{f}_p = 0$. The above equations can again be simplified using the expressions for the two-fluid Green's functions. Taking the inner limit of the resulting outer solution ($\rho \ll l$) yields:
\begin{equation}
 \begin{split}
 \bm{u}_s^{out} &=  \bm{U}_\infty + \frac{\bm{f}_s\cdot (\bm{I + e_z e_z})}{4 \pi} \left(\log \left(\frac{2(\sqrt{s(1-s)})}{\rho}\right) \right) - \frac{\bm{f}_s\cdot\bm{e_z e_z}}{4 \pi} + \frac{\bm{f}_s \cdot\bm{nn}}{4 \pi} + \\
 &\frac{1}{8\pi} \int \left[\left( \frac{\bm{I}}{|\bm{r_c}(s) - \bm{r_c}(s')|} + \frac{(\bm{r_c}(s) - \bm{r_c}(s'))(\bm{r_c}(s) - \bm{r_c}(s'))}{|\bm{r_c}(s) - \bm{r_c}(s')|^3} \right) \cdot \bm{f}_{s}(\bm{r_c}(s')) \right.\\
 &\left.- \left(\frac{(\bm{I} + \bm{e_z e_z})}{|s - s'|} \right) \cdot \bm{f}_{s}(\bm{r_c}(s)) \right] ds' + \frac{\lambda}{8 \pi (1+\lambda)} \int_{\bm{r}_c}  \left[\bm{f}_s(\bm{r_c}(s')) \right]\cdot\left[\bm{G}_{Br} - \bm{G}_{St} \right]\,ds'
 \end{split} \label{Eq.59}
\end{equation}
\begin{equation}
    \bm{u}_p^{out} = \bm{U}_\infty + \frac{1}{8 \pi (1 + \lambda)} \int_{\bm{r}_c} \bm{f}_s \cdot \left[ \bm{G}_{St} - \bm{G}_{Br} \right] ds'. \label{Eq.60}
\end{equation}

\subsubsection{Matching region}
Matching the inner and outer solution, one gets,
\begin{equation}
 \begin{split}
 \bm{U} = &\frac{\bm{f}_s\cdot(\bm{I} + \bm{e_z e_z})}{4 \pi} \left( \log 2\gamma + \log \left(\frac{2\sqrt{s(1 - s)}}{\overline{a}(s)} \right)\right) + \frac{\bm{f}_s\cdot(\bm{I}-3\bm{e_z e_z})}{8 \pi} + \\
 &\frac{1}{8\pi} \int \left[\left( \frac{\bm{I}}{|\bm{r_c}(s) - \bm{r_c}(s')|} + \frac{(\bm{r_c}(s) - \bm{r_c}(s'))(\bm{r_c}(s) - \bm{r_c}(s'))}{|\bm{r_c}(s) - \bm{r_c}(s')|^3} \right)\cdot\bm{f}_s(\bm{r_c}(s')) \right.\\
 &\left.- \left(\frac{(\bm{I} + \bm{e_z e_z})}{|s - s'|} \right) \cdot\bm{f}_s(\bm{r_c}(s)) \right] ds' +\\
 &\frac{\lambda}{8 \pi(1+\lambda)} \int_{\bm{r}_c}  \left[\bm{f}_s(\bm{r_c}(s')) \right]\cdot\left[\bm{G}_{Br} - \bm{G}_{St} \right]\,ds'
 \end{split} \label{Eq.61}
\end{equation}
and
\begin{equation}
    c \; \bm{U} = \frac{1}{8 \pi (1 + \lambda)} \int_{\bm{r}_c} \bm{f}_s \cdot\left[ \bm{G}_{St} - \bm{G}_{Br} \right] ds' \label{Eq.62}
\end{equation}
which gives the arbitrary constant $c$.

\subsubsection{Resistive Force Theory}
To the leading order, the force on the fiber is given by:
\begin{equation}
    \bm{f}_s = \bm{f} = \frac{4 \pi \bm{U} \cdot (\bm{I} - \frac{\bm{e_z e_z}}{2})}{\log 2\gamma} \label{Eq.63}
\end{equation}
which results in an anisotropy of
\begin{equation}
    \frac{f_{\perp}}{f_{||}} = 2 \label{Eq.64}
\end{equation}
which is the same as in a single fluid medium. Thus the leading order anisotropy in this case is smaller than in the case where polymer slips past the fiber (with $L_B \geq a$). 

\section{Results of two-fluid SBT for a helical fiber} \label{4}
The different versions of SBT mentioned above are solved numerically for a helical fiber moving in a two-fluid medium, by adopting a simple numerical procedure described in \citet{Swinney}. The numerical technique involves discretizing the helix into $N$ segments per pitch and using the trapezium rule of numerical integration for the integrals. This results in a linear system of equations for the singularity strength $\bm{f}_i$ on the $i^{th}$ segment, which is then solved to obtain the strengths in terms of the known surface velocity of the segment $\bm{U}_i$. While the numerical procedure is the same as in \citet{Swinney}, it's adoption for the two-fluid model requires small changes which are described in detail and validated with exemplary SBT results in Appendix \ref{App.B}.

\subsection{Outline of results}
\begin{figure}
\centering
\includegraphics[scale = 0.5]{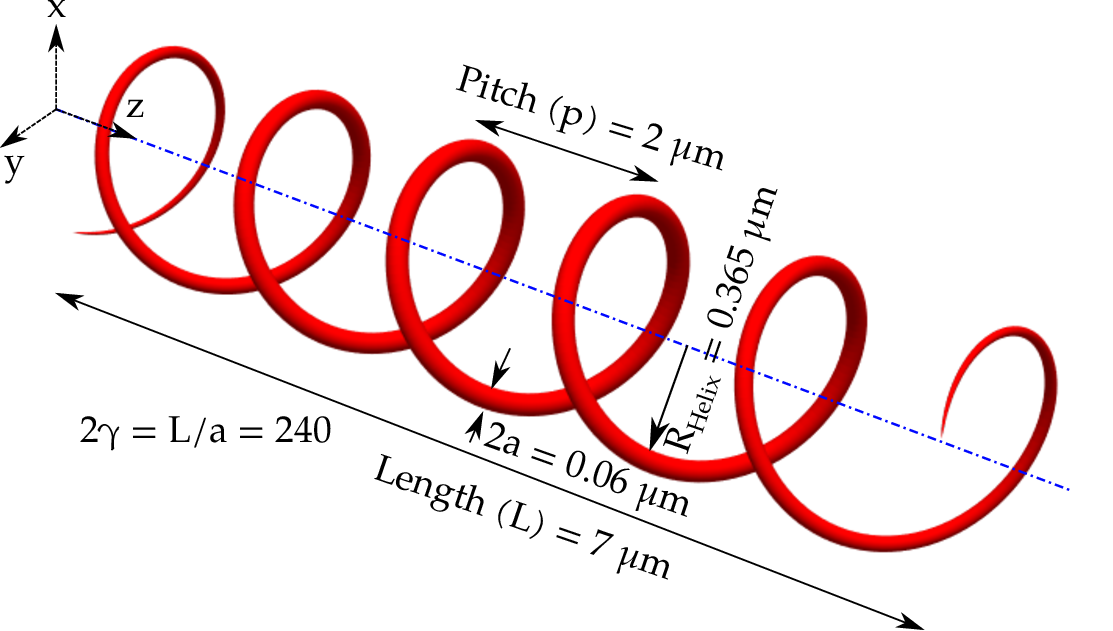}
\caption{Geometry of the helix used in the numerical calculation}
\label{fig:6N}
\end{figure}

In the discussion that follows, we first present the results for a slender helical fiber with prolate spheroidal cross-section ($a(s) = 2a\sqrt{s(1-s)}$; $a$ is the radius of the fiber at the mid point along it's centerline), that rotates and translates in a single Newtonian fluid with slip on it's surface. This calculation is done to elucidate the effect of slip on the fiber motion, since in the two-fluid model, one of the cases involves a slipping polymer medium. This is followed by a discussion of results for a slender helical fiber with the same cross-section in the two-fluid medium satisfying both polymer slip and no polymer-fiber interaction conditions. Here we discuss how the presence of microstructure affects the motion of the fiber. The results are presented in the form of thrust, torque and drag on a fiber that is rotating and translating, as a function of $L_B$ and viscosity ratio $\lambda$.  The thrust is the force along the axis of a helical fiber when it rotates on its axis, while we report the component of torque along the axis. The drag is the force opposing translation of the helical fiber along its axis. The dimensions of the helical fiber are chosen to be the dimensions of the helical flagellar bundle of {\it E.Coli} \citep{Berg} listed in Fig.\ref{fig:6N} and we vary $L_B$ and $\lambda$. All our calculations use these dimensions for the fiber unless otherwise mentioned, and have $N = 30$ segments per pitch (with $N = 110$ segments for the whole length). The assumed spheroidal cross-section of the fiber has been shown to be an accurate description of the flagellar bundle of the bacterium\,\citep{Das}, and also avoids ill-conditioned matrices that arise from discretizing a fiber of constant cross-sectional radius\,\citep{Mackaplow94}.

\subsection{A slender helical fiber in a fluid medium with slip}\label{App.D}
\begin{figure}
\centering
\includegraphics[trim = {0.3cm 0cm 1cm 0cm}, clip, scale = 0.36]{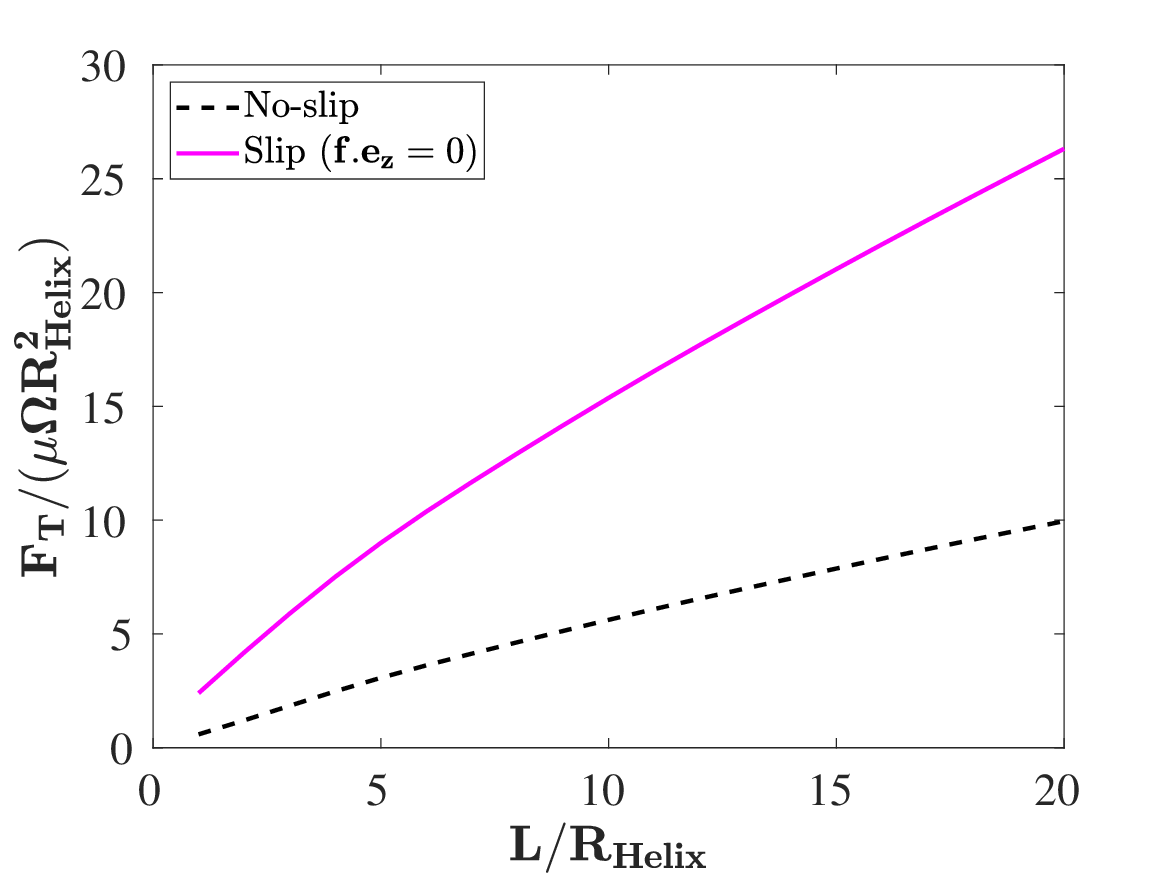}
\includegraphics[trim = {0.3cm 0cm 1cm 0cm}, clip, scale = 0.36]{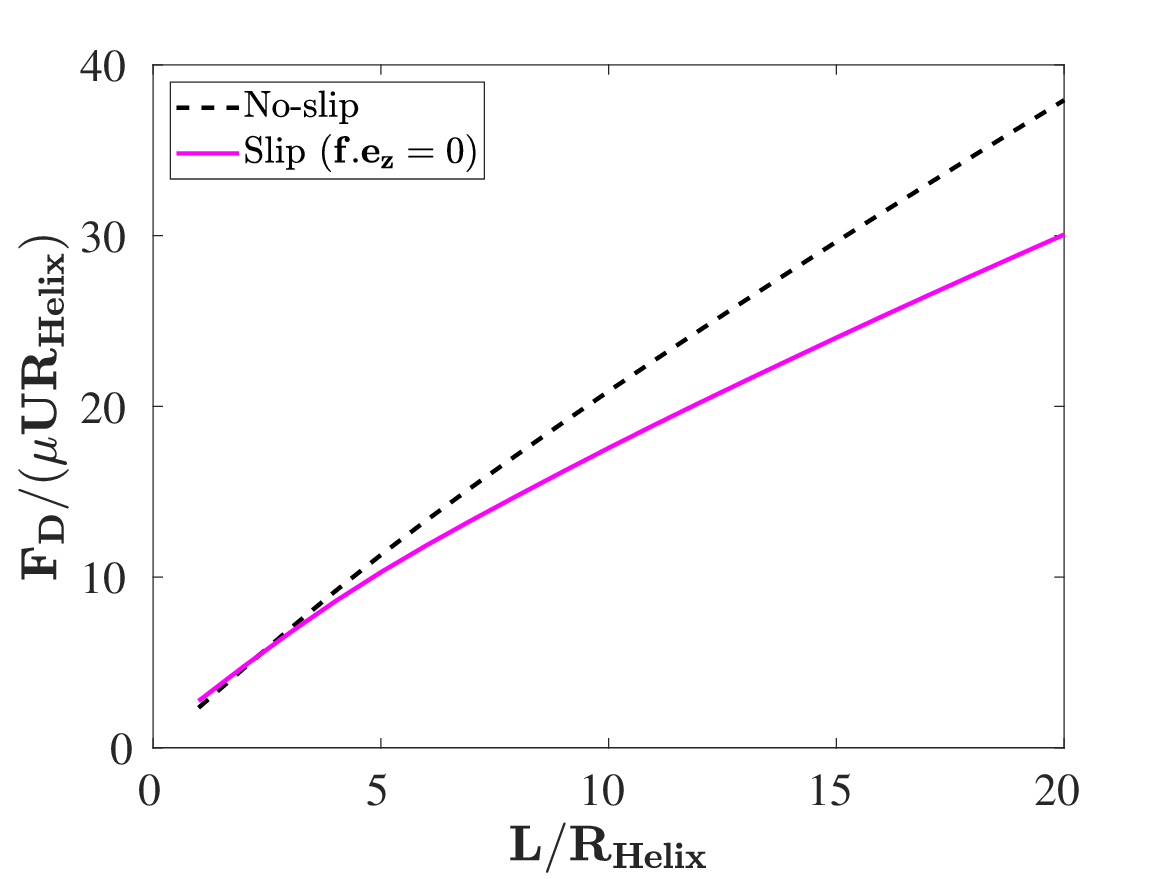}
\includegraphics[trim = {0.3cm 0cm 1cm 0cm}, clip, scale = 0.36]{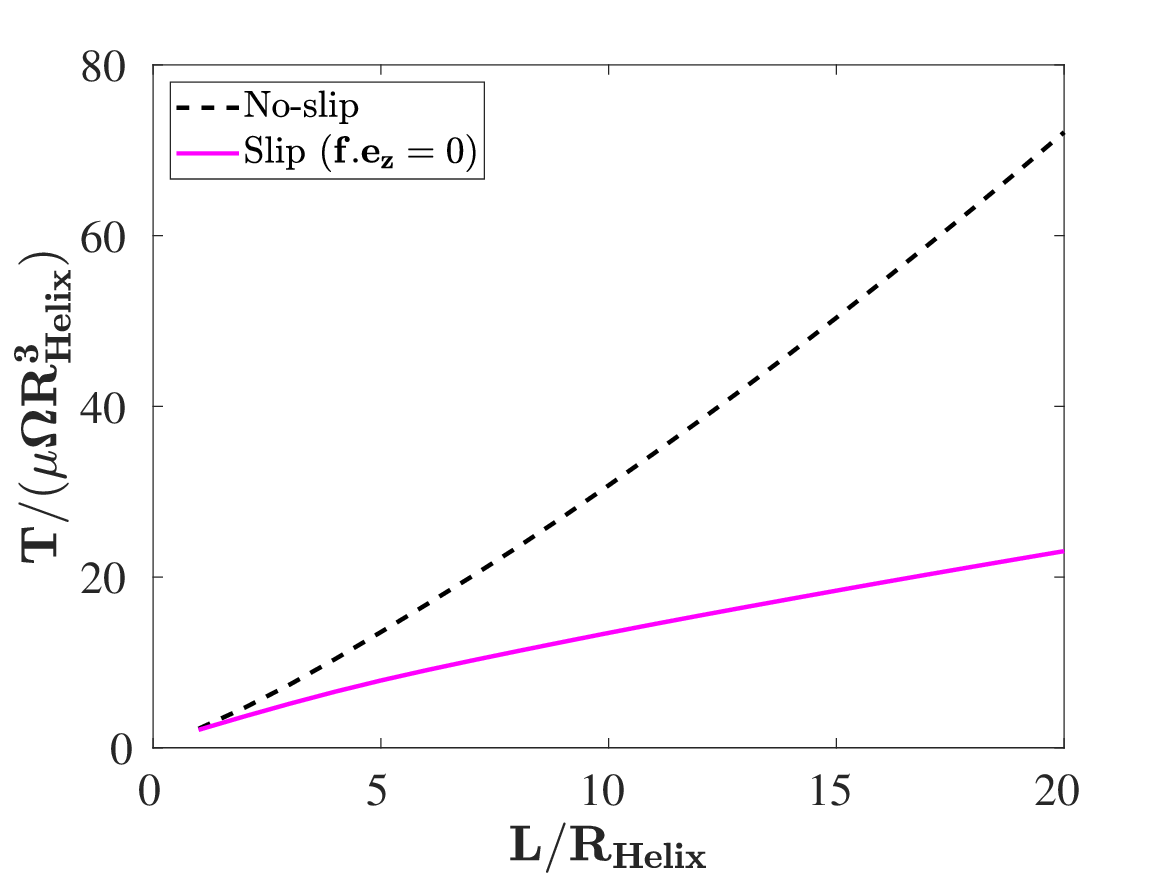}
\caption{Plots of normalised (a) thrust, (b) drag,  and (c) torque for a slender helical fiber with spheroidal cross-section translating with $U$ and rotating with $\Omega$ in a single fluid medium, as a function of $L/R_{Helix}$ from the numerical solutions of single-fluid SBT with no-slip and slip ($\bm{f.e_z} = 0$).}
\label{fig:7}
\end{figure}

When a slender fiber moves through a single-fluid medium with slip on its surface, the integral equation for the force strength ($\bm{f}$) along the fiber centerline using SBT is given by:
\begin{equation}
 \begin{split}
 &\bm{U} \cdot (\bm{I} - (1-c)\bm{e_z e_z}) = \frac{\bm{f}}{4 \pi } \left( \log 2\gamma + \log \left(\frac{2\sqrt{s(1 - s)}}{\overline{a}(s)} \right)\right) +  \frac{\bm{f}\cdot(\bm{I}-2\bm{e_z e_z})}{4 \pi}\\
 &+ \frac{1}{8\pi} \int_{\bm{r}_c} \left[\left( \frac{\bm{I}}{|\bm{r_c}(s) - \bm{r_c}(s')|} + \frac{(\bm{r_c}(s) - \bm{r_c}(s'))(\bm{r_c}(s) - \bm{r_c}(s'))}{|\bm{r_c}(s) - \bm{r_c}(s')|^3} \right)\cdot\bm{f}(\bm{r'}) \right.\\
 &\left.- \left(\frac{(\bm{I} + \bm{e_z e_z})}{|s - s'|} \right)\cdot\bm{f}(\bm{r}) \right] ds'
 \end{split} \label{Eq.76}
\end{equation}
This case is equivalent to a slender bubble moving through a fluid \citep{Hinch}. In Fig.\ref{fig:7}(a)-(c), we have plotted the drag, thrust and torque on a helical fiber moving with axial velocity $U$ and rotating with an angular velocity $\Omega$ calculated using Eq.\ref{Eq.76} and compared it with the results of the case when the fluid satisfies no-slip on the helix surface. In these calculations, the helix has the same dimensions as shown in Fig.\ref{fig:6N} except that the length was varied from $1 \mu m$ to the dimension in Fig.\ref{fig:6N} (while keeping $\gamma$ fixed at the value shown in Fig.\ref{fig:6N}). From the plot, we note, that while the drag and torque on the helix are smaller for the case with slip, slip leads to an increased thrust. This implies that the helix with slip can move at a higher velocity for a given rotation rate when the motion is force-free.
\begin{figure}
\centerline{\includegraphics[scale = 0.5]{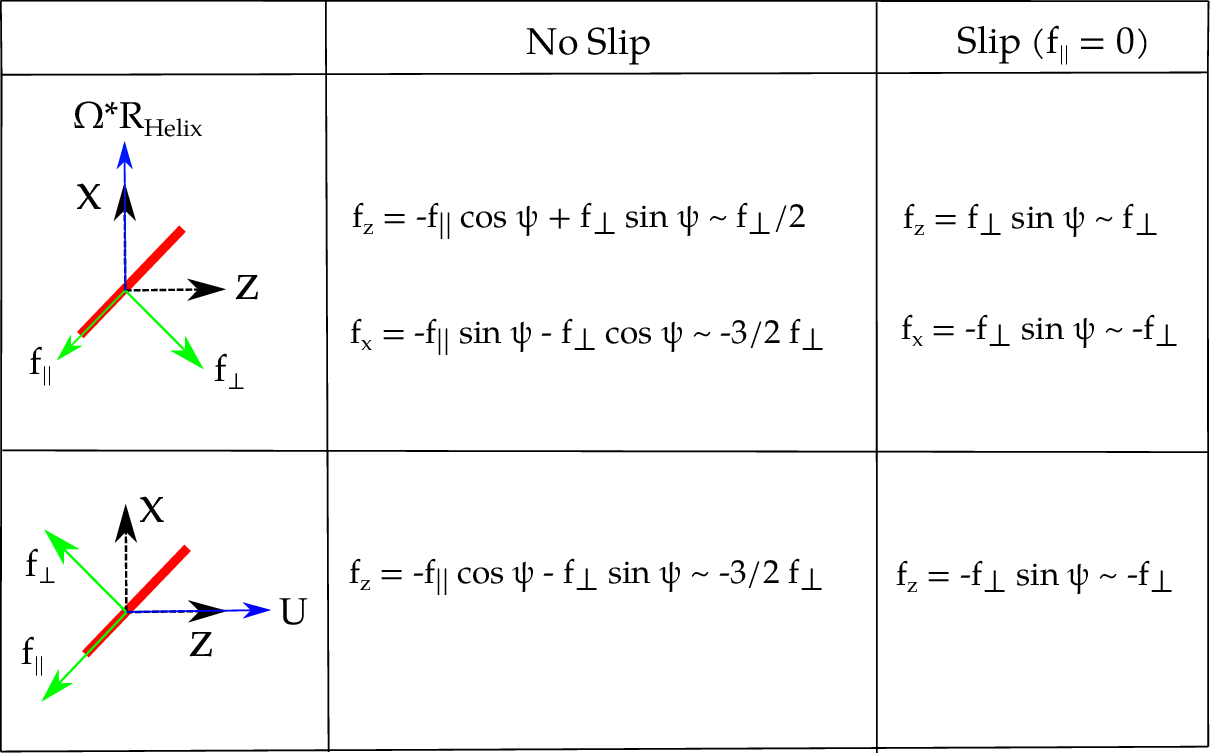}}
\caption{A schematic of a segment of helix showing the forces acting on the segment due to rotation and translation. The thrust and torque correspond to $f_{z}$ and $f_{x}$ in the former case and drag corresponds to $f_{z}$ in the latter. Note for simplicity we choose $\psi = \pi/4$ as the pitch angle of the helix, denoting the orientation of the segment with respect to the motion. }
\label{fig:8}
\end{figure}

This increased thrust can be understood in terms of the increased drag anisotropy for this case, as is shown in Fig.\ref{fig:8}. Here, a segment of the helix (with pitch angle $\psi = \pi/4$) that rotates with $\Omega$ (along $-z$) and translates with $U$ (along $z$) are shown. In the case of no-slip boundary condition, the segment is subject to resistance in both parallel and perpendicular direction ($f_{||}$ and $f_{\perp}$, where we assume $f_{\perp} / f_{||} = 2$), while for the case where the fluid can slip $f_{||} = 0$, with $f_{\perp}$ being the same as that for the no-slip case. This is because with perfect slip, the rigid slender body behaves like a bubble, and to leading order in $\epsilon = 1/\log (2 \gamma)$, the transverse force strength is the same as that for a body with no-slip\,\citep{Hinch}. This implies that the thrust and torque on the segment during rotation of the helix, where each segment locally moves with velocity $\Omega/R_{Helix}$ along $x$, are proportional to,
\begin{equation}
    \text{Thrust }(f_{z}) = \frac{1}{2} f_{\perp} \label{Eq.77}
\end{equation}
\begin{equation}
    \text{Torque }(f_{x}) = -\frac{3}{2} f_{\perp} \label{Eq.78}
\end{equation}
for no-slip condition and
\begin{equation}
    \text{Thrust }(f_{z}) = f_{\perp} \label{Eq.79}
\end{equation}
\begin{equation}
    \text{Torque }(f_{x}) = -f_{\perp} \label{Eq.80}
\end{equation}
for the slip condition. Here, the angular velocity vector for the helix is directed along $-z$ leading to thrust along $z$. Similarly, the drag on the segment (locally) translating with velocity $U$ along $z$ is,
\begin{equation}
    \text{Drag }(f_{z}) = -\frac{3}{2} f_{\perp} \label{Eq.81}
\end{equation}
for no slip and
\begin{equation}
    \text{Drag }(f_{z}) = -f_{\perp} \label{Eq.82}
\end{equation}
for the slip condition, where the negative sign indicates the force is opposite to the direction of motion ($z$). Taking the ratio between the slip and no-slip case, we get,
\begin{equation}
    \frac{\text{Thrust (slip)}}{\text{Thrust (no-slip)}} \approx 2 \label{Eq.83}
\end{equation}
\begin{equation}
    \frac{\text{Torque (slip)}}{\text{Torque (no-slip)}} \approx \frac{2}{3} \label{Eq.84}
\end{equation}
\begin{equation}
    \frac{\text{Drag (slip)}}{\text{Drag (no-slip)}} \approx \frac{2}{3} \label{Eq.85}
\end{equation}

Thus, we see that the slip condition results in a higher thrust and smaller drag and torque on the helix compared to the no-slip condition. Note that the actual values of these ratios from the numerical calculations (Fig.\ref{fig:7}) are different owing to the facts that the pitch angle for the helix in our numerical calculation is greater than $\pi/4$ and the ratio $f_{\perp}/f_{||} < 2$ and not exactly two, which makes the ratios for thrust and torque larger and drag smaller. The decreased value of $f_{\perp}/f_{||}$ arises primarily because the spheroidal shape of the flagellar bundle cross-section  allows the unit normal to the surface to have a component parallel to the local filament axis.  On the other hand, the effect of slip on $f_{\perp}$ (the increase compared to $f_{\perp}$ on a no-slip boundary, that occurs at $O(1/\log(2 \gamma)^2)$) is only modest numerically. 

\subsection{A slender helical fiber in a two-fluid medium}
In this section we calculate the drag, thrust and torque acting on a slender helical fiber translating and rotating in a two-fluid medium. First, we plot the results for the scenario where the polymer slips on the helix using the uniformly valid SBT (Eq.\ref{Eq.54A}) and then move on to the case where the polymer does not directly interact with the helix (Eq.\ref{Eq.61}).

\subsubsection{A slender helical fiber in a two-fluid medium with polymer slip}
\begin{figure}
\centering
\includegraphics[width = 0.495\textwidth]{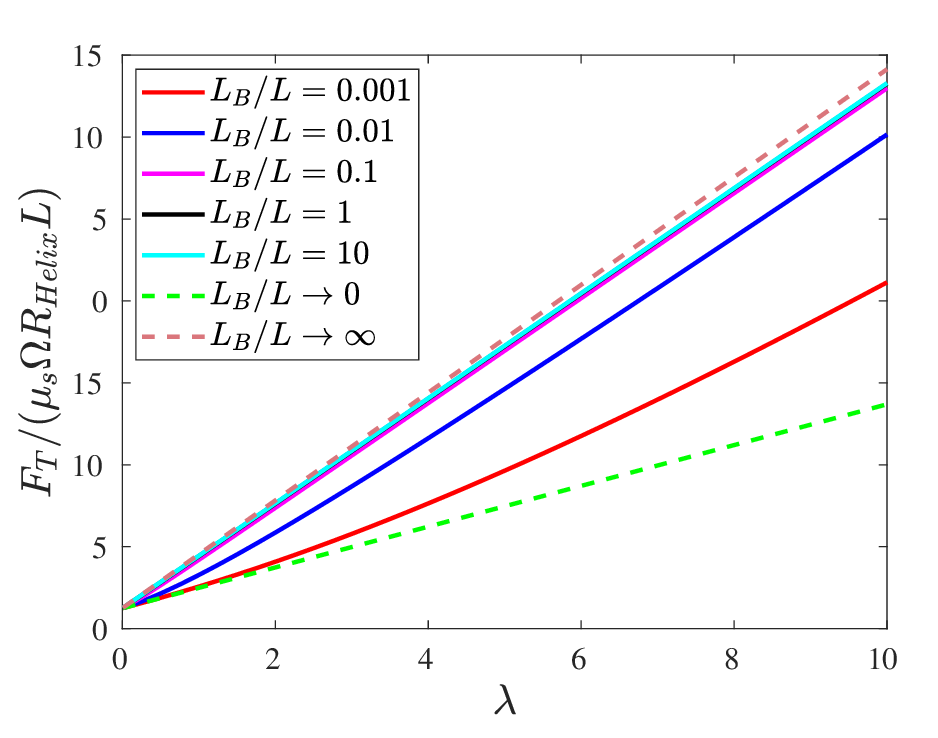}
\includegraphics[width = 0.495\textwidth]{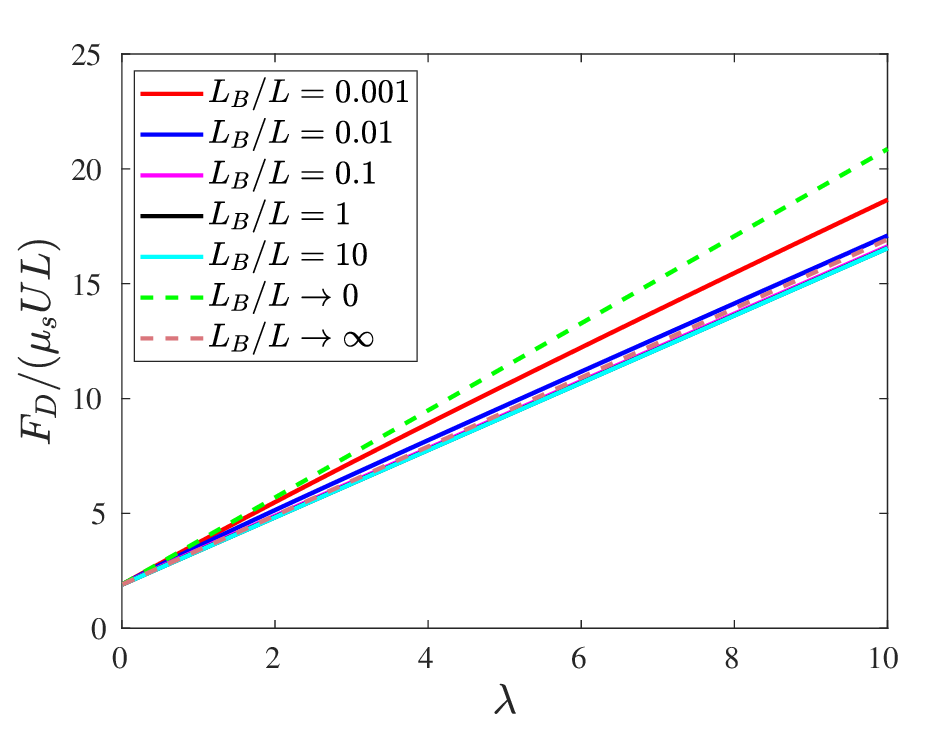}
\includegraphics[width = 0.495\textwidth]{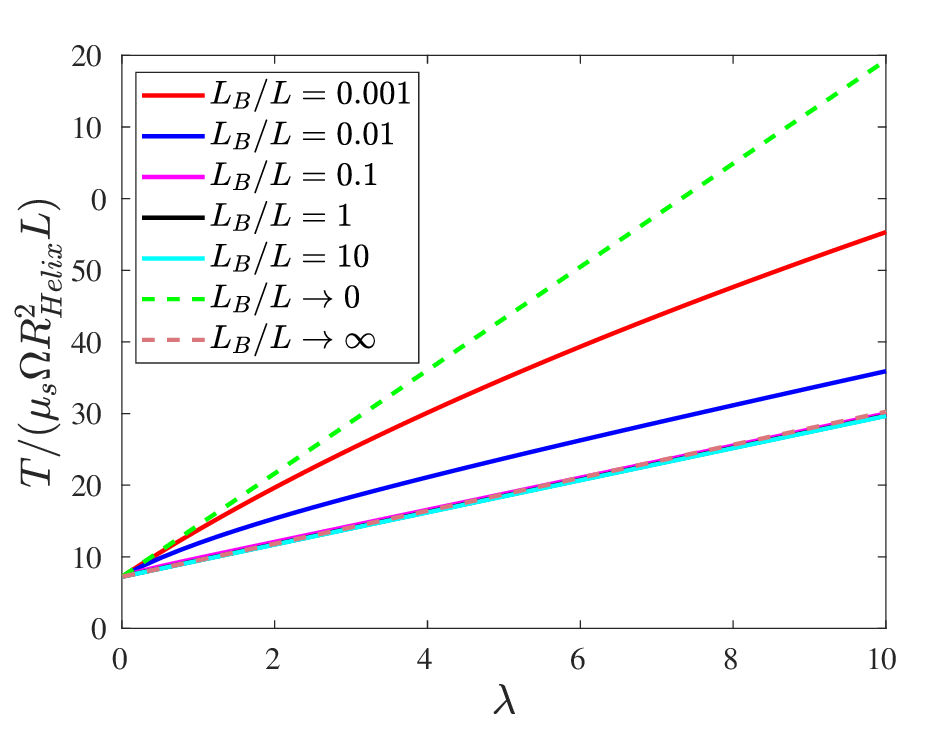}
\caption{Plots of normalised (a) thrust, (b) drag, and (c) torque for a slender helical fiber with spheroidal cross-section, as a function of $\lambda$, from the numerical solutions for axial translation ($U$) and rotation ($\Omega$) of the helix in a two-fluid medium with polymer slip, where the curves correspond to different $L_B/L$. Here, $R_{Helix}/L \approx 0.052$ and $a/L \approx 0.0043$.}
\label{fig:9}
\end{figure}
\begin{figure}
\centering
\includegraphics[width = 0.7\textwidth]{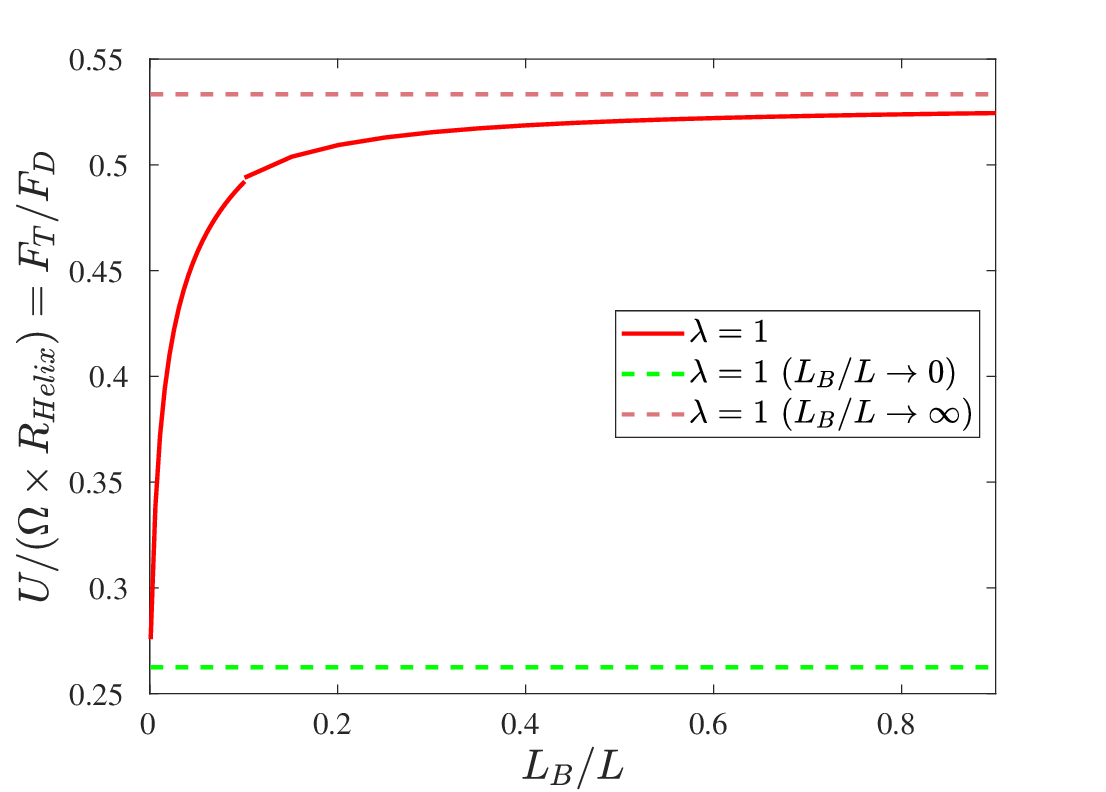}
\caption{Plot of the ratio of thrust and drag on a helix rotating and translating in a two-fluid medium with polymer slip ($\lambda = 1$). Green, dashed line is the ratio for the same helix rotating in a mixture ($L_B/L \rightarrow 0$) and brown, dashed line is the ratio in two decoupled fluids ($L_B/L \rightarrow \infty$). Here, as in Fig.\ref{fig:9} $R_{Helix}/L \approx 0.052$ and $a/L \approx 0.0043$.}
\label{fig:9N}
\end{figure}

Fig.\ref{fig:9} shows the thrust, drag and torque on a helical fiber in a two-fluid medium and the results are compared to cases where the helix moves in a mixture with viscosity $\mu_s(1 + \lambda)$, and in two independent fluids, with one of the fluids slipping past the helix. From the plot we see that the thrust on the helix increases and the drag decreases with increasing $L_B/L$, compared to the helix moving in a mixture, with this behavior being more pronounced at large $\lambda$. The behavior of drag and thrust in this case can be understood by considering a segment of the fiber as shown in Fig.\ref{fig:8} and repeating a similar exercise for the two fluid case, where now the polymer has slip, while the solvent satisfies no-slip. The drag anisotropy ($f_{\perp}/f_{||}$) for this case is now directly proportional to $\lambda$ (Eq.\ref{Eq.53}), while for the mixture, it still remains $\approx 2$. Proceeding with this exercise, one can show that the ratio $\text{Thrust}_{\text{Two-fluid}}/\text{Thrust}_{\text{Mixture}}$, will be a function of the viscosity ratio $\lambda$ and will always be greater than unity. Similarly, it can be shown that the ratios $\text{Drag}_{\text{Two-fluid}}/\text{Drag}_{\text{Mixture}}$ and $\text{Torque}_{\text{Two-fluid}}/\text{Torque}_{\text{Mixture}}$ will always be less than unity. This increased drag anisotropy in the two-fluid medium implies that the helix can move with an enhanced velocity if the motion is force-free. This is clearly seen in Fig.\ref{fig:9N}, where we plot the ratio of the dimensionless thrust and drag for a helix that translates and rotates in a two-fluid medium. Here the drag is normalised with $\mu_s U L$ and the thrust with $\mu_s \Omega R_{Helix} L$ as was done in Fig.\ref{fig:9}. This ratio is, therefore, equal to the ratio $U/(\Omega\, R_{Helix})$ of the force-free translation velocity of a helix to the rotation velocity that induces this motion.

\subsubsection{A slender helical fiber in a two-fluid medium with no polymer-fiber interaction}
\begin{figure}
\centering
\includegraphics[width = 0.495\textwidth]{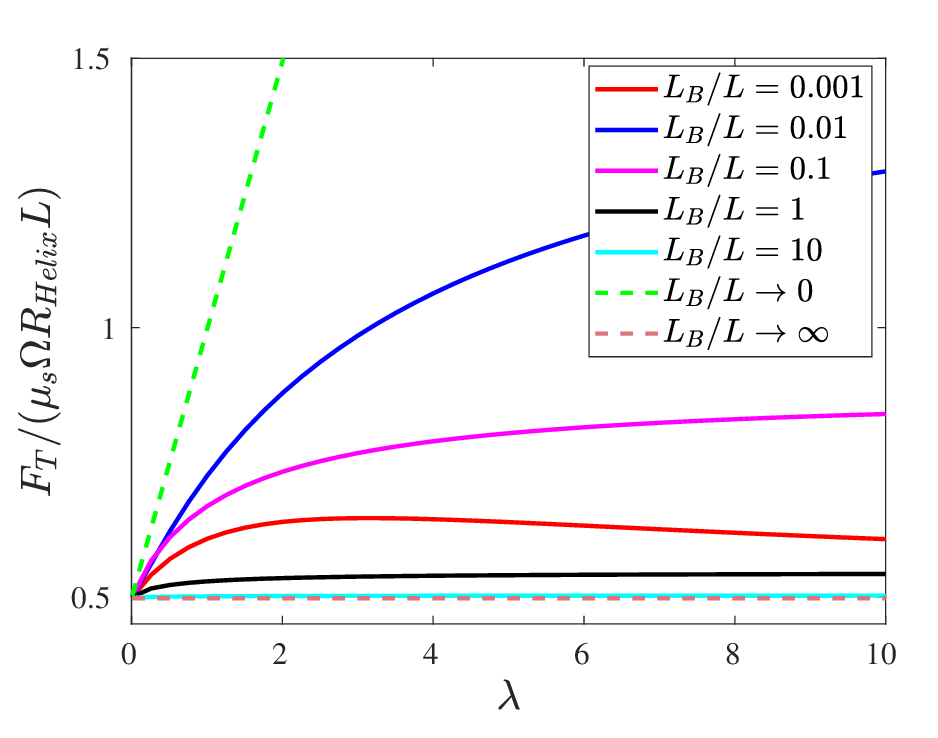}
\includegraphics[width = 0.495\textwidth]{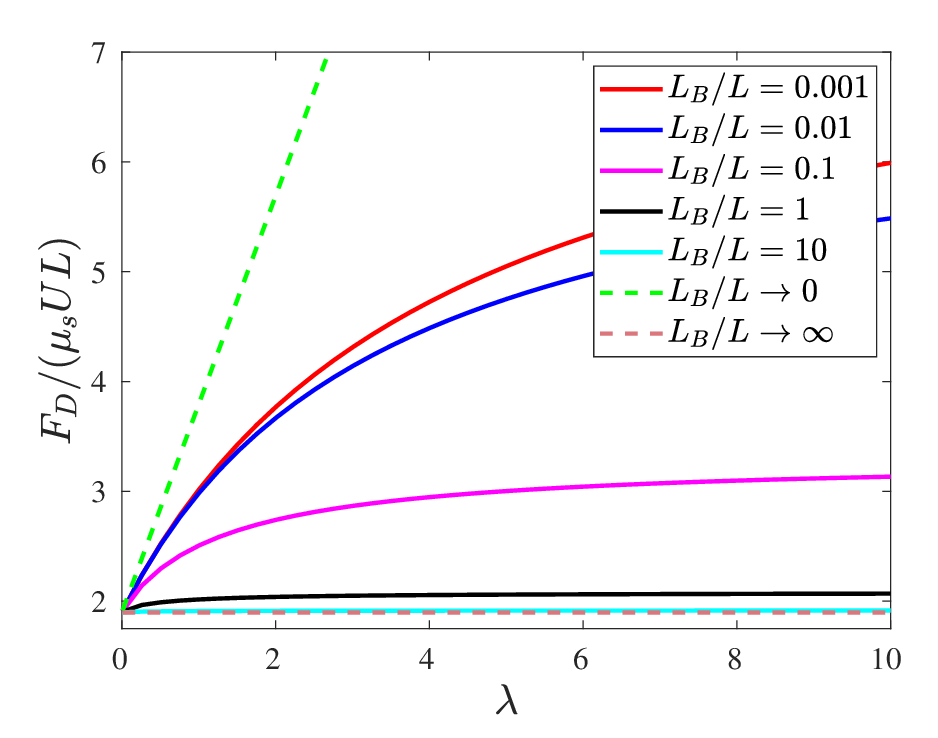}
\includegraphics[width = 0.495\textwidth]{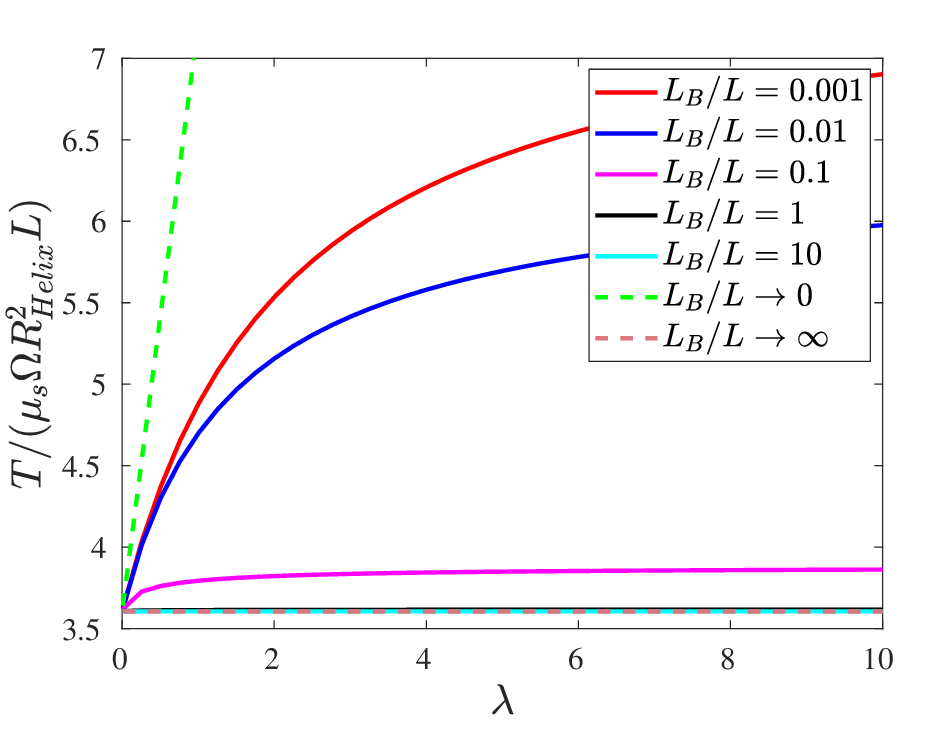}
\caption{Plots of normalised (a) thrust, (b) drag, and (c) torque for a slender helical fiber with spheroidal cross-section, as a function of $\lambda$, from the numerical solutions for axial translation ($U$) and rotation ($\Omega$) of the helix in a two-fluid medium with no polymer-fiber interaction, where the curves correspond to different $L_B/L$. Here, $R_{Helix}/L \approx 0.052$ and $a/L \approx 0.0043$.}
\label{fig:10}
\end{figure}

\begin{figure}
\centering
\includegraphics[width = 0.495\textwidth]{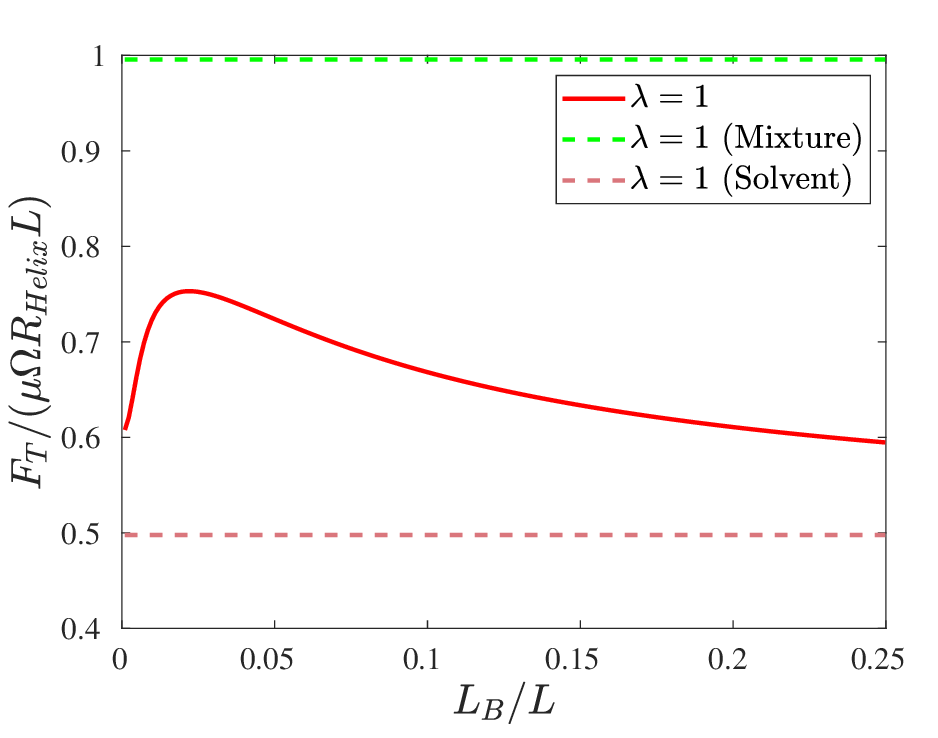}
\includegraphics[width = 0.495\textwidth]{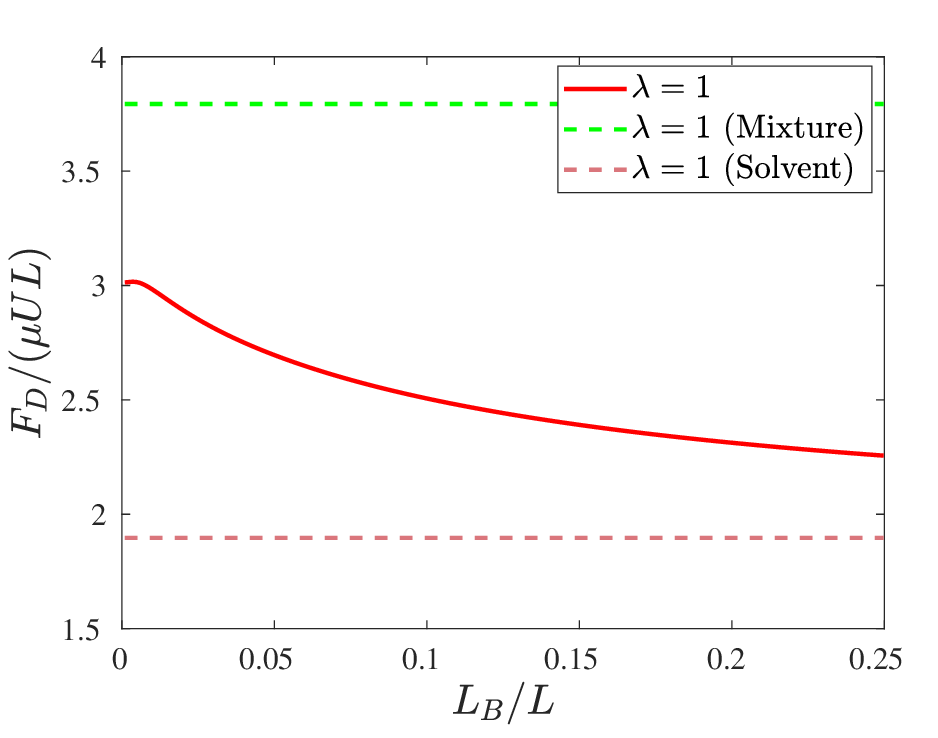}
\caption{Plots of the normalised (a) thrust and (b) drag on a helix rotating and translating in a two-fluid medium with no polymer-fiber interaction ($\lambda = 1$). The green, dashed line indicates the thrust for the same helix rotating in a mixture ($L_B/L \rightarrow 0$) and the brown, dashed line indicates the thrust in the solvent ($L_B/L \rightarrow \infty$). Here, $R_{Helix}/L \approx 0.052$ and $a/L \approx 0.0043$.}
\label{fig:10N}
\end{figure}

\begin{figure}
\centering
\includegraphics[width = 0.5\textwidth]{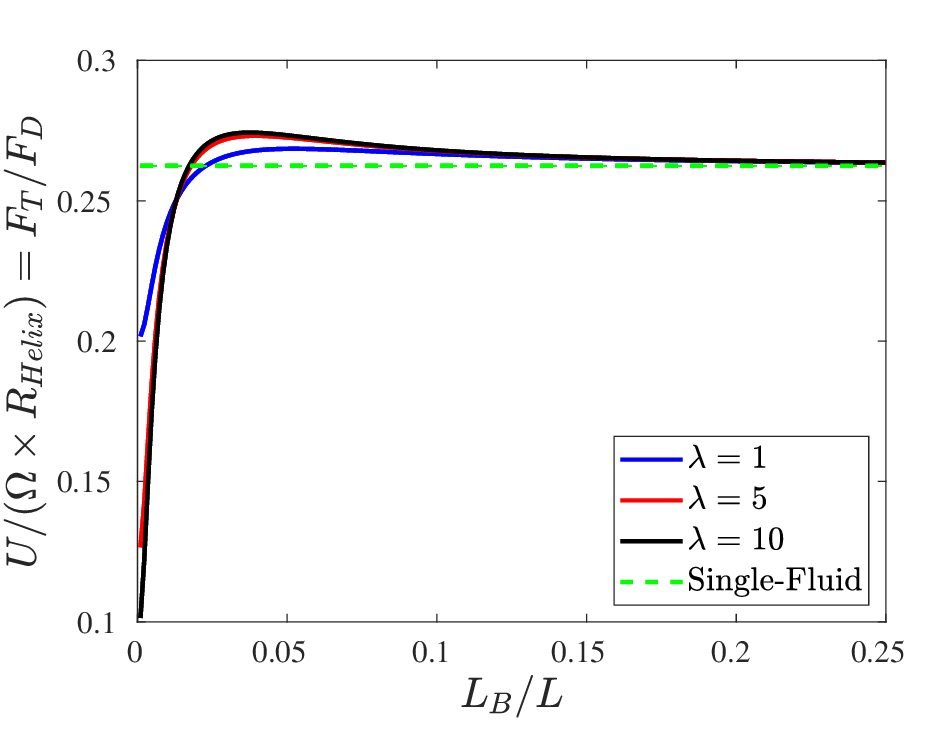}
\includegraphics[width = 0.475\textwidth]{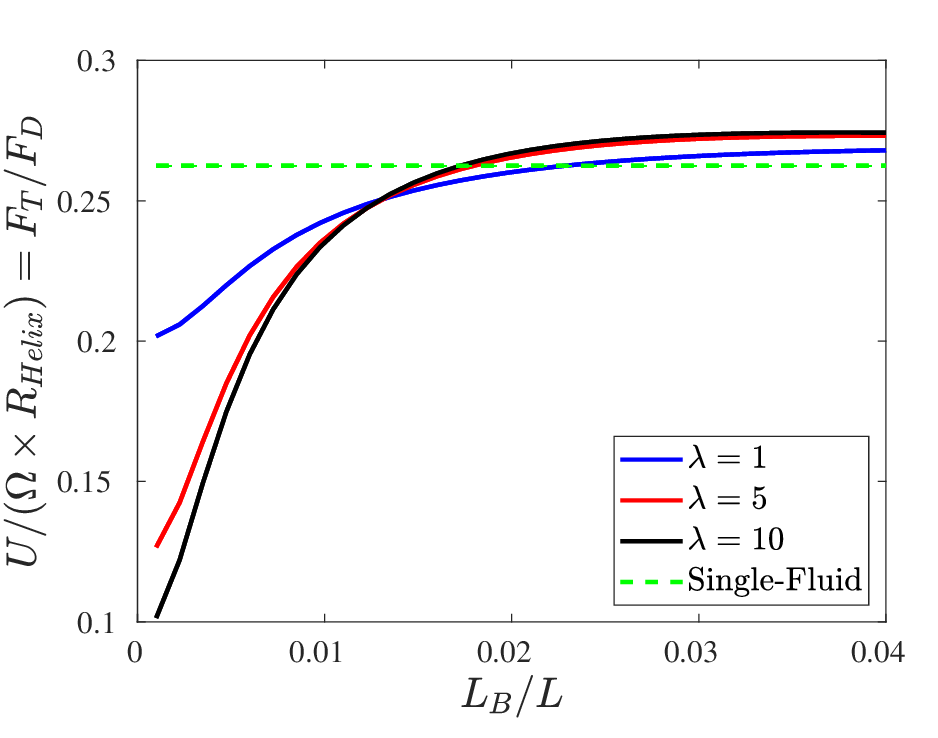}
\caption{Plots of (a) the ratio of the thrust and drag on a helix rotating and translating in a two-fluid medium with no polymer-fiber interaction. Single-fluid (green, dashed line) indicates the thrust to drag ratio for the same helix in both mixture ($L_B/L \rightarrow 0$) and solvent ($L_B/L \rightarrow \infty$). Note that this ratio is the same for mixture and solvent, as the ratio is independent of viscosity. In (b), the ratio is plotted for small $L_B/L$. Here, $R_{Helix}/L \approx 0.052$ and $a/L \approx 0.0043$.}
\label{fig:11}
\end{figure}

In Fig.\ref{fig:10}, we plot the results of SBT with no polymer-fiber interaction, where we have restricted $L_B$ to the outer region. The plots show the normalised thrust, drag and torque on the helix as a function of $\lambda$ compared with the results for a helix moving in a single fluid medium of viscosity $\mu_s + \mu_p$ (mixture) and in two independent fluids.  For the latter case, the polymer fluid is not interacting with the helix and so the helix only sees a single-fluid solvent  medium with viscosity $\mu_s$. Here, we see that the quantities vary non-monotonically with $\lambda$ for a given $L_B/L$ and importantly, the thrust also varies non-monotonically with $L_B/L$ at a given $\lambda$. This is clearly seen in Fig.\ref{fig:10N}, where we plot the thrust and drag for $\lambda = 1$ as a function of $L_B/L$. While the trends in the thrust could be easily understood from the leading order solution to the force strengths in the case of a slipping polymer, here we see that such an approach would not work. This is because, the slip boundary condition affects the solution at leading order in $\epsilon$ ($=1/\log(2 \gamma)$), but the no polymer-fiber interaction condition only affects the solution at $O(\epsilon^2)$. Thus, the change in thrust compared to the thrust on a helix in a single-fluid medium is small and is $O(\epsilon)$. This suggests the motion of the fiber in the two-fluid medium is very sensitive to the nature of the interaction between the fiber and the polymer. Fig.\ref{fig:11} is a plot of the ratio of thrust and drag as a function of $L_B/L$ for different $\lambda$ compared with the same for helix in a mixture of two fluids ($L_B/L \rightarrow 0$) and in the solvent ($L_B/L \rightarrow \infty$). Since both these limits correspond to a single-fluid medium, this ratio is the same for both limits (as it does not depend on the viscosity). From these plots we see that the ratio of thrust to drag varies non-monotonically, first increasing and then decreasing with $L_B/L$, and approaches the value in a single-fluid medium for both $L_B/L \rightarrow 0$ and $L_B/L \rightarrow \infty$. Note that for very small values of $L_B/L$ (Fig.\ref{fig:11}(b)), the ratio becomes smaller than the ratio for a mixture, because the SBT was derived assuming $L_B$ in the outer region. This non-monotonic variation in the thrust to drag ratio results from the non-monotonic variation of the drag anisotropy for a rotating helix in the two-fluid medium. While the leading order solution resulted in the same drag anisotropy as in a single-fluid medium (Eq.\ref{Eq.64}), the two-fluid effects present at higher orders in $\epsilon$ result in a slightly increased drag anisotropy relative to the single fluid case.  This anisotropy reaches a maximum at $L_B \sim R_{Helix}$ (since the flow disturbance due to rotation decays over a distance $R_{Helix}$), before decreasing again to the single fluid value as less polymers are disturbed by the rotating helix when $L_B$ is increased beyond $R_{Helix}$.

\newpage
\section{A swimming bacterium in a two-fluid medium} \label{5}
We have shown from our SBT calculations that a force-free helical fiber moves with a larger velocity because of the presence of microstructure, regardless of the type of interaction with the polymer. This effect of microstructure, modeled by our two-fluid equations, is therefore crucial to understand motility of bacteria in entangled polymer solutions. In this section, we calculate the swimming parameters of force- and torque-free bacterial motion in a two-fluid medium using Resistive Force Theory (RFT) and compare it against the experimentally observed trends for a bacterium swimming in a concentrated polymer solution. The basic idea in RFT is to calculate the resistance coefficients for motion of the flagellar bundle and the cell (head) and use them to calculate the velocity and other swimming parameters, while ensuring that the motion as a whole is force- and torque-free. In this calculation, the hydrodynamic interactions between the cell and the flagellar bundle and also between the different segments of the bundle are neglected. While this is not an accurate description, we shall see that this calculation can still capture the qualitative trends observed in experiments, where the entangled polymer solutions have a macro-rheology that is almost Newtonian\citep{Poon,Breuer}. With hydrodynamic interactions neglected, the motion of a segment of a bundle can be split into motion tangential and normal to the local centerline orientation, giving us two coefficients of resistance ($C_N$, $C_T$) proportional to the local velocity, in these two directions. And for the cell, we again have two coefficients $\alpha_C$ and $\beta_C$ for translation and rotation. Using these coefficients, one can describe the motion of a bacterium swimming with speed, $v$, cell angular speed, $\omega_{Cell}$, and flagellar angular speed, $\omega_f$ as:
\begin{equation}
    \begin{pmatrix}
    \bm{F}_{C} \\
    \bm{T}_{C}
    \end{pmatrix} = \begin{pmatrix}
                     \alpha_C & 0\\
                     0 & \beta_C
                     \end{pmatrix} \begin{pmatrix}
                                   \bm{v} \\
                                   \bm{\omega}_{Cell}
                                   \end{pmatrix} \label{Eq.86}
\end{equation}
for the head of the bacterium, and
\begin{equation}
    \begin{pmatrix}
    \bm{F}_{f} \\
    \bm{T}_{f}
    \end{pmatrix} = \begin{pmatrix}
                     A & B\\
                     B & D
                     \end{pmatrix} \begin{pmatrix}
                                   \bm{v} \\
                                   \bm{\omega}_f
                                   \end{pmatrix} \label{Eq.87}
\end{equation}
for the flagellar bundle, where $\bm{v} = [v, 0, 0]$, $\bm{\omega}_f = [\omega_f, 0 ,0]$ and $\bm{\omega}_{Cell} = [\omega_{Cell}, 0 ,0]$. The coefficients $A,B,D$ for the flagellar bundle are given by:
\begin{align}
 A &= -C_N L \sin \psi \tan \psi (1 + \zeta \cot^2 \psi) \label{Eq.88}\\
 B &= -C_N L \frac{p}{2\pi}\sin \psi \tan \psi (1 - \zeta) \label{Eq.89}\\
 D &= -C_N L (\frac{p}{2\pi})^2 \sin \psi \tan \psi (1 + \zeta \cot^2 \psi) \label{Eq.90}
\end{align}
with $L, p, \psi$ being the length, pitch and pitch angle of the flagellar bundle and $\zeta = C_N/C_T$. The expressions for the resistance coefficients can be obtained from the leading order solutions to our SBT equations (Eq.\ref{Eq.22}-\ref{Eq.23},\ref{Eq.51}-\ref{Eq.52},\ref{Eq.63}) and depend on whether $L_B$ is in the inner or outer region. For the flagellar bundle with characteristic radius $a$ these are given by,
\begin{equation}
    C_N = \begin{cases}
            \frac{4 \pi  (\lambda +1) \mu_s}{\log (2 \gamma )}& \text{for } L_B/a \gg O(1) \\
            \frac{4 \pi  (\lambda +1) \mu_s}{\log (2 \gamma )}& \text{for } L_B/a \sim O(1)
          \end{cases} \label{Eq.88}
\end{equation}
\begin{equation}
    C_T = \begin{cases}
            \frac{2 \pi \mu_s}{\log (2 \gamma )}& \text{for } L_B/a \gg O(1) \\
            \frac{2 \pi  (\lambda +1) \mu_s}{\log (2 \gamma )}& \text{for } L_B/a \sim O(1)
          \end{cases} \label{Eq.89}
\end{equation}
for the case with a slipping polymer ($\gamma$ is the aspect ratio for the bundle), where the coefficients correspond to the cases with screening length in the inner ($L_B/a \sim O(1)$) and outer ($L_B/a \gg O(1)$) regions. For the case where the polymer has no interaction with the bundle the coefficients are,
\begin{align}
 C_N = \frac{4 \pi \mu_s}{\log (2 \gamma)} \label{Eq.90}\\
 C_T = \frac{2 \pi \mu_s}{\log (2 \gamma)} \label{Eq.91}.
\end{align}
For the cell, assumed to be spherical in shape for now, good approximations to the resistance coefficients are those corresponding to a sphere translating and rotating in a mixture of solvent and polymer fluids with viscosity $\mu_s(1+\lambda)$, given by:
\begin{align}
    \alpha_C = -6 \pi \mu_s (1 + \lambda) R_{Cell} \label{Eq.92} \\
    \beta_C = -8 \pi \mu_s (1 + \lambda) R_{Cell}^3 \label{Eq.93},
\end{align}
since in the physical picture presented earlier, the two-fluid behavior only applies to the flagellar bundle whose radius is comparable to the length scale of the microstructure of polymer solution. However, one can also find the effect of microstructure at the scale of the head, as was done by \citet{Kudo}, and calculate the swimming parameters for the case of a head translating and rotating in a two-fluid medium (with a slipping polymer), for which the resistance coefficients are given by:
\begin{align}
    \alpha_C = -\frac{6 \pi  (\lambda +1) R_{\text{Cell}} \mu _s \left(R_{\text{Cell}} \sqrt{\frac{\lambda +1}{\lambda  L_B^2}}+2 \lambda +3\right)}{R_{\text{Cell}} \sqrt{\frac{\lambda +1}{\lambda  L_B^2}}+3 \lambda +3} \label{Eq.94} \\
    \beta_C = -8 \frac{8 \pi  R_{\text{Cell}}^3 \mu _s \left(\frac{3 \lambda  L_B^2}{R_{\text{Cell}}^2}+(\lambda +1) \left(\frac{3 \sqrt{\frac{\lambda }{\lambda +1}} L_B}{R_{\text{Cell}}}+1\right)\right)}{\frac{3 \lambda
   L_B^2}{R_{\text{Cell}}^2}+\frac{3 \sqrt{\frac{\lambda }{\lambda +1}} (\lambda +1) L_B}{R_{\text{Cell}}}+1}. \label{Eq.95}
\end{align}
These coefficients were calculated by solving for flow due to a translating and rotating sphere in two-fluid medium as described in Section 1. A motivation for this calculation is the fact that entangled polymer solutions are known to slip at solid bodies with length scales much larger than the entanglement length scale, as was reported by \citet{Archer}. It is not known whether polymers slip at the surface of bacterial cells, since no experiments have addressed this question, but future experiments with bacteria in entangled polymer solutions can shed light on this aspect. The force-free and torque-free conditions are given by:
\begin{align}
 &\bm{F}_c + \bm{F}_f = 0 \label{Eq.98}\\
 &2\bm{T}_c = \bm{T}_m \label{Eq.99}\\
 &2\bm{T}_f = -\bm{T}_m \label{Eq.100}
\end{align}
where $\bm{T}_m$ is the torque supplied by the motor. It has been known that the torque generated by the motor has two regimes depending on the angular velocity of the motor $\bm{\omega}_m = \bm{\omega}_f - \bm{\omega}_{Cell}$ \citep{11}. The motor torque behavior is given by:
\begin{equation}
 \bm{T}_m =
 \begin{cases}
     \bm{T}_0 &\text{ for } |\bm{\omega}_m| \leq \omega_0\\
     \bm{T}_0 \left( 1 + \frac{\omega_0 - \omega_m}{\omega_{max} - \omega_0} \right) &\text{ for } |\bm{\omega}_m| > \omega_0
 \end{cases} \label{Eq.102}
\end{equation}
Here, $|\bm{T}_0|$ is the knee-torque, $\omega_0$ is the knee-rotation rate and $\omega_{max}$ is the maximum rotation rate of the flagellar motor, which are constants for a particular bacterial species swimming in a motility buffer (a Newtonian medium with negligible nutrient content, which optimally supports bacterial motility and chemotaxis but does not support bacterial
growth) at a particular temperature. Note that these constants are not sensitive to the viscosity of the buffer. In our RFT calculations, we assume the motor torque to be the input, having the form given in Eq.\ref{Eq.102}. The above equations were solved simultaneously and the resulting values of the swimming speed, cell rotation rate and flagellar rotation rate are calculated for different scenarios. In calculating these parameters, the dimensions of the bacterium that appear in the expressions ($R_{Cell},L,p,\psi,\gamma,a$) correspond to the measured values of wild-type \textit{E. Coli} \citep{11,Poon} and are the same as given in Fig.\ref{fig:6N} (also tabulated in Table.\ref{Tab.1}) and the input torque profile (with $|\bm{T}_0|$, $\omega_0$ and $\omega_{max}$) are obtained from the experimentally measured values reported in \citep{11,Poon,Sowa}, corresponding to {\it E.Coli} swimming in motility buffer at room temperature ($T = 298 K$) that results in a constant driving potential for the motor (proton-motive force) \citep{Berry}.
\begin{table}
	\centering
	\begin{tabular}{ | c | c |}
        \hline
		{\bf Parameter} & {\bf Value}  \\
		\hline
		$a$ & $0.03 \mu m$ \\
		$L$ & $7 \mu m$ \\
		$p$ & $2 \mu m$  \\
		$\psi$ & $41^\circ$ \\
		$\gamma$ & $240$ \\
		$R_{Cell}$ & $1.5 \mu m$ \\
		$\mu_s$ & $1 mPa\,s$  \\
		$|\bm{T}_0|$ & $1250 pN\,nm$  \\
	  $\omega_0$ & $350\pi \, rad/s$  \\
		$\omega_{max}$ & $600\pi \, rad/s$  \\ 		
		\hline
	\end{tabular}
\caption{Values of the various parameters corresponding to {\it E.Coli} used in RFT calculation.}
\label{Tab.1}
\end{table}

\subsection{Mixture behavior at the bacterial cell}
\begin{figure}
\centering
\includegraphics[width = 0.33\textwidth]{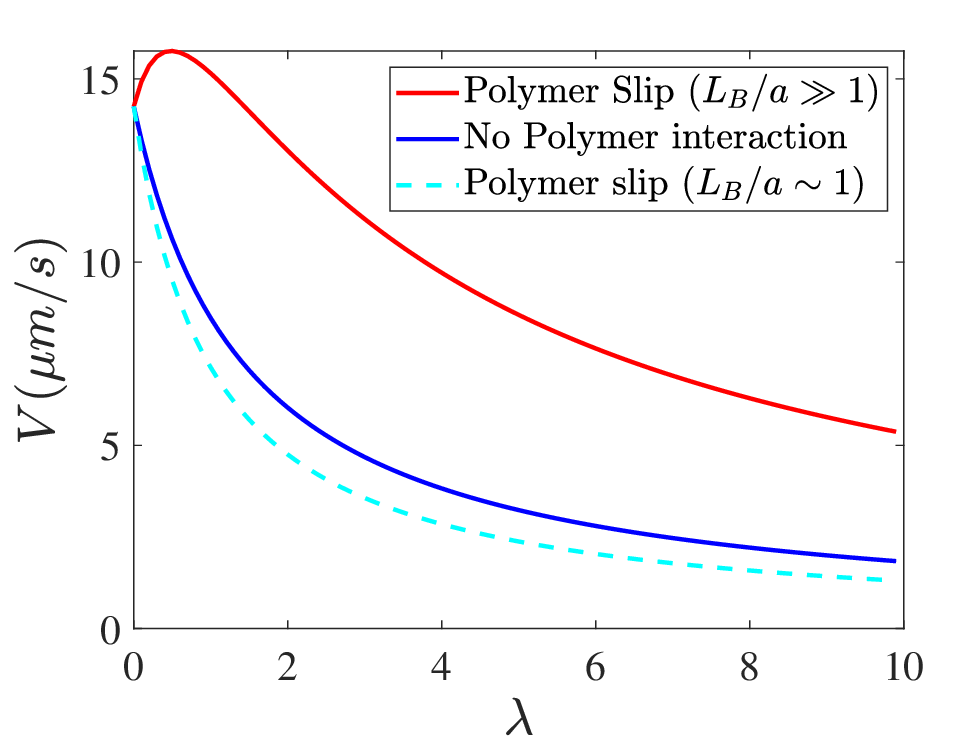}
\includegraphics[width = 0.325\textwidth]{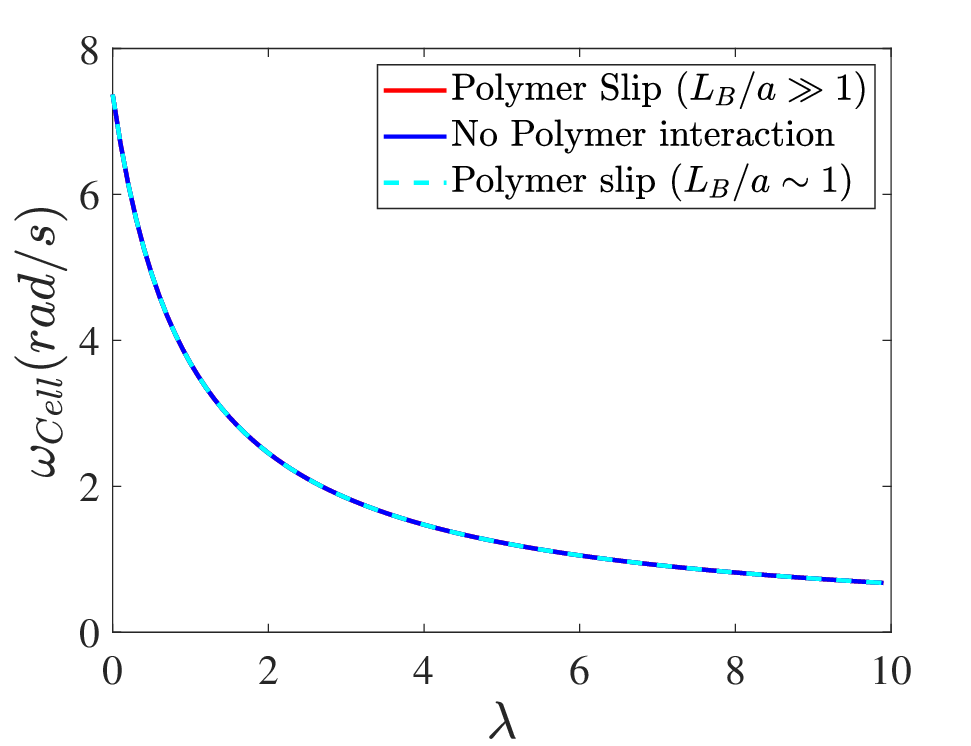}
\includegraphics[width = 0.325\textwidth]{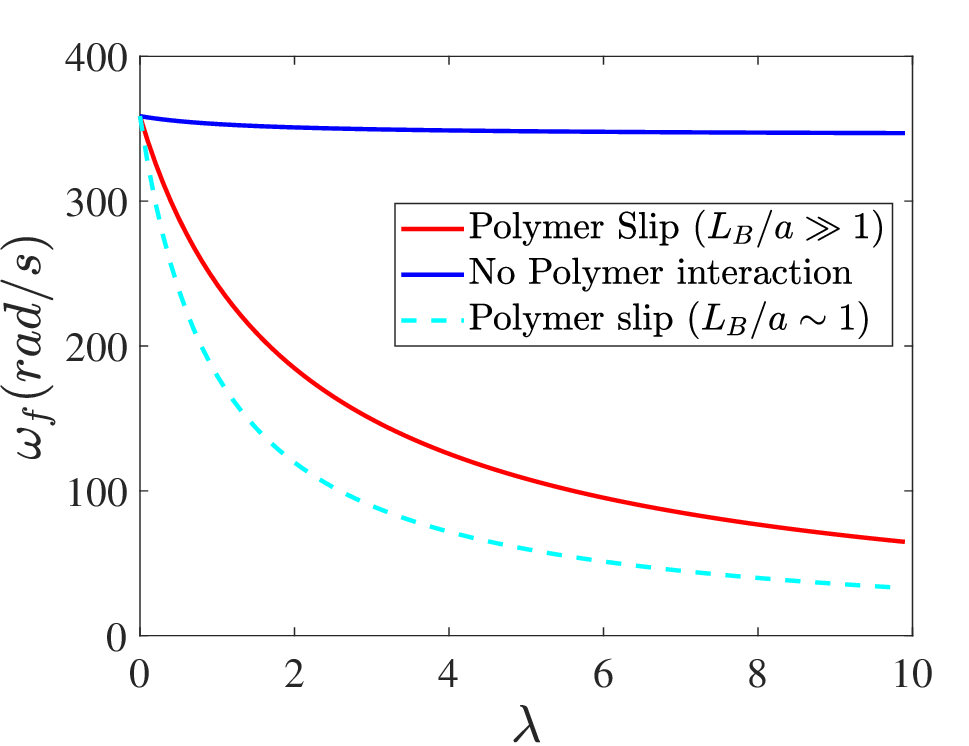}
\caption{Plots of (a) the swimming velocity, (b) the angular velocity of the cell, and (c) the angular velocity of the flagellar bundle of a bacterium in a two-fluid medium with a slipping polymer fluid ($L_B/a \sim 1$ - cyan, $L_B/a \gg 1$ - red), and with no polymer-bundle interaction (blue), where the cell sees the mixture of the two fluids. The scenario with polymer slip ($L_B/a \gg 1$) leads to a two-fold increase in swimming velocity for $\lambda > 1$.}
\label{fig:12}
\end{figure}

\begin{figure}
\centering
\includegraphics[width = 0.33\textwidth]{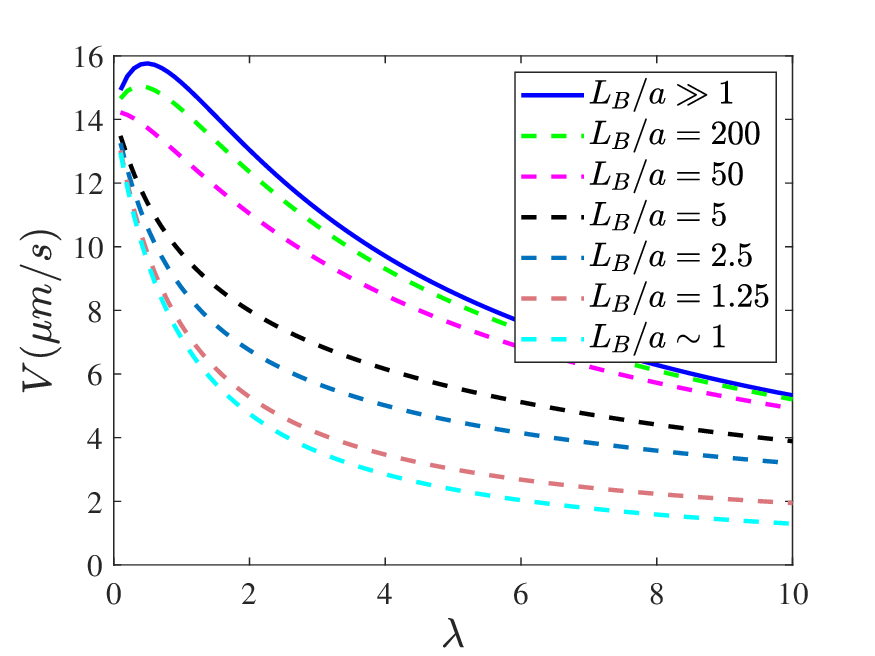}
\includegraphics[width = 0.325\textwidth]{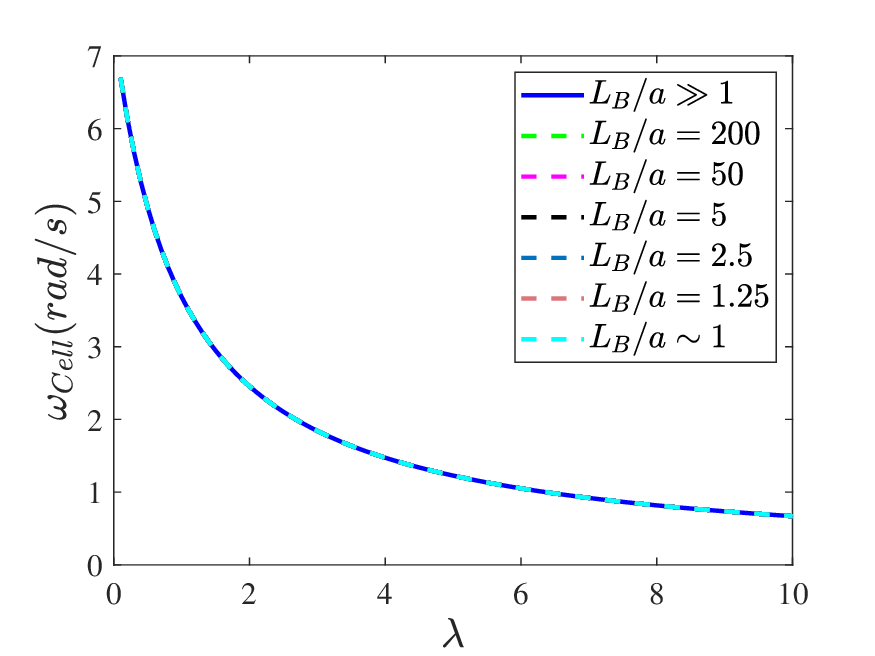}
\includegraphics[width = 0.325\textwidth]{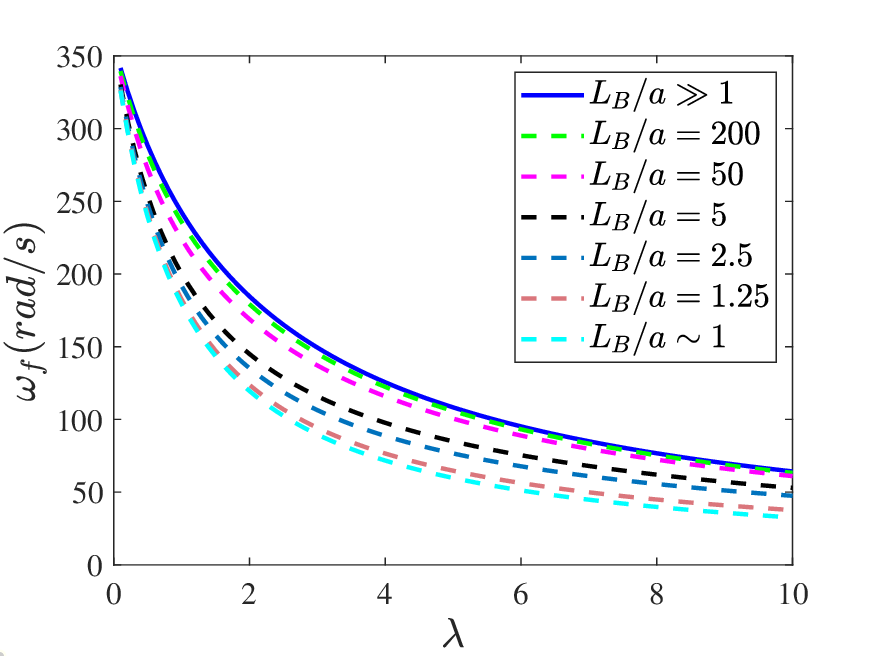}
\caption{Plots of (a) the swimming velocity, (b) the angular velocity of the cell, and (c) the angular velocity of the flagellar bundle of a bacterium in a two-fluid medium with a slipping polymer, where the cell sees the mixture of the two fluids. Here $L_B/a \gg 1$ and $L_B/a \sim 1$ correspond to the screening length being in the outer and inner region respectively, while the other curves correspond to screening length in matching region.}
\label{fig:13}
\end{figure}
First, the swimming parameters are presented for the case where the cell moves in a mixture, but the bundle sees two-fluid behavior. The swimming velocity and angular velocities of the cell and bundle are given in Fig.\ref{fig:12} for this case, where the three curves in each plot corresponds to the three physical scenarios considered for the flagellar bundle namely: (i) a slipping polymer with $L_B/a \sim O(1)$, (ii) a slipping polymer with $L_B/a \gg 1$ and (iii) a non-interacting polymer. For cases (i) and (iii), the resistance coefficients (of the flagellar bundle) to the leading order correspond to the resistance coefficients for a fiber moving in a single fluid medium, with viscosities $\mu_s(1+\lambda)$ (mixture) and $\mu_s$ (solvent) respectively, while for case (ii), the resistance coefficients involve the effect of slipping polymer at leading order. This is evident in the plots of swimming velocity and angular velocities as functions of $\lambda$ in Fig.\ref{fig:12}, where we see that scenario (ii) results in an  enhancement in swimming velocity compared to scenarios (i) and (iii), with (i) being the same as the bacterium (both head and bundle) swimming in a mixture. The observed trends can be explained in terms of the drag anisotropy on the slender fiber, which is directly proportional to $\lambda$ for scenario (ii) and is independent of $\lambda$ for scenario (i) and (iii). Even though the flagellar bundle has the same drag anisotropy in cases (i) and (iii), the fact that it moves entirely in the solvent for case (iii) results in the slight enhancement (at a given $\lambda$) compared to (i), where the bundle moves in a mixture of two fluids. The nearly constant angular velocity of flagellar bundle with $\lambda$ for scenario (iii) is a consequence of this fact. Also, in Fig.\ref{fig:13}, we have plotted the swimming parameters for the case where the cell moves through a mixture but the flagellar bundle moves through a two-fluid medium with slipping polymer, with the resistance coefficients $C_N, C_T$, now given by Eq.\ref{Eq.55G}, \ref{Eq.55H} respectively. These coefficients correspond to the coefficients valid for $L_B \in [0,\infty]$, and thus lead to swimming parameters that extend between the two limiting cases (case (i) and (ii)) considered in Fig.\ref{fig:12}.  These intermediate swimming parameters result from the evolution of the anisotropic drag as $L_B$ passes through the matching region.  Thus, we see that for the case with polymer slip, one can go from the swimming velocity corresponding to a mixture to an enhanced swimming velocity at a given $\lambda$, depending on the screening length $L_B$.

\subsection{Two-fluid behavior at the bacterial cell}
\begin{figure}
\centering
\includegraphics[width = 0.33\textwidth]{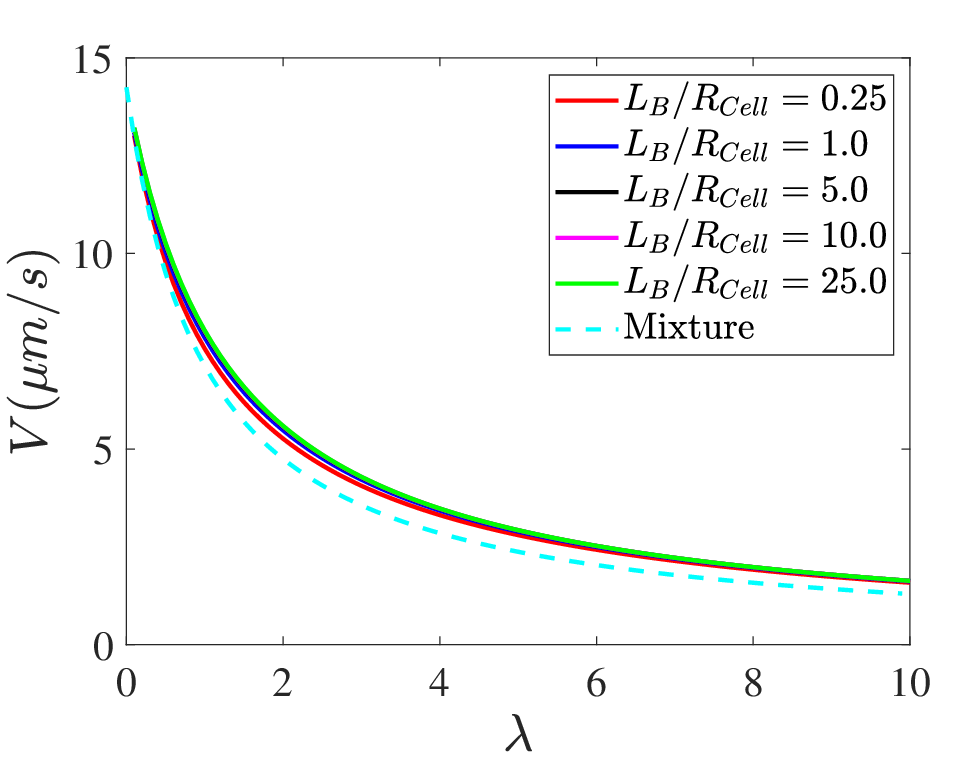}
\includegraphics[width = 0.325\textwidth]{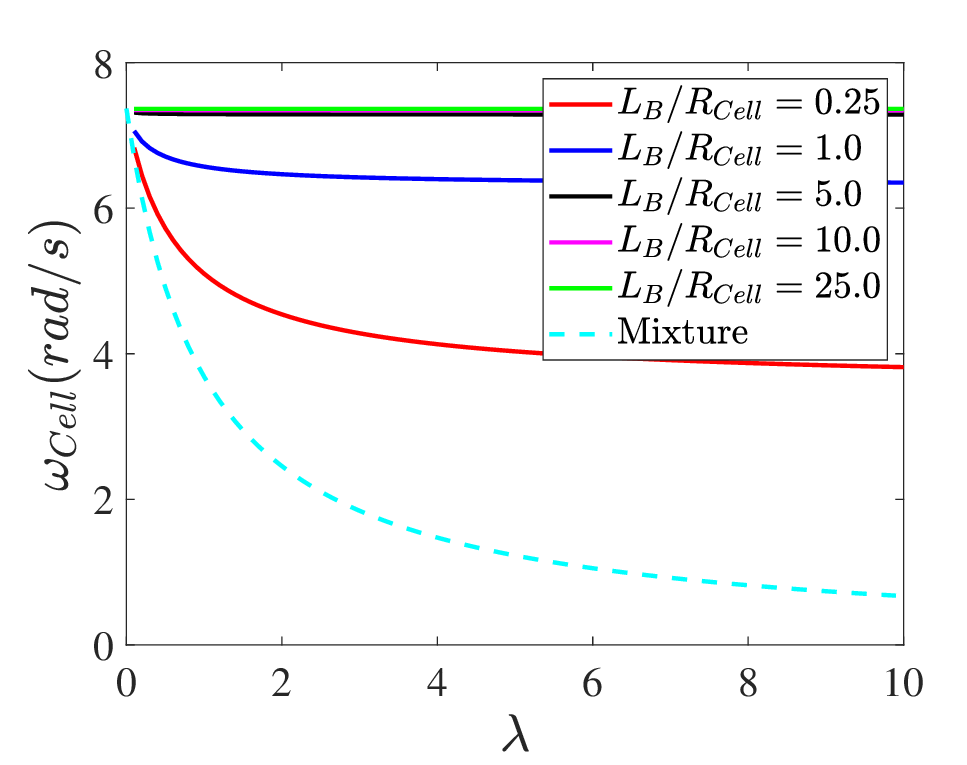}
\includegraphics[width = 0.325\textwidth]{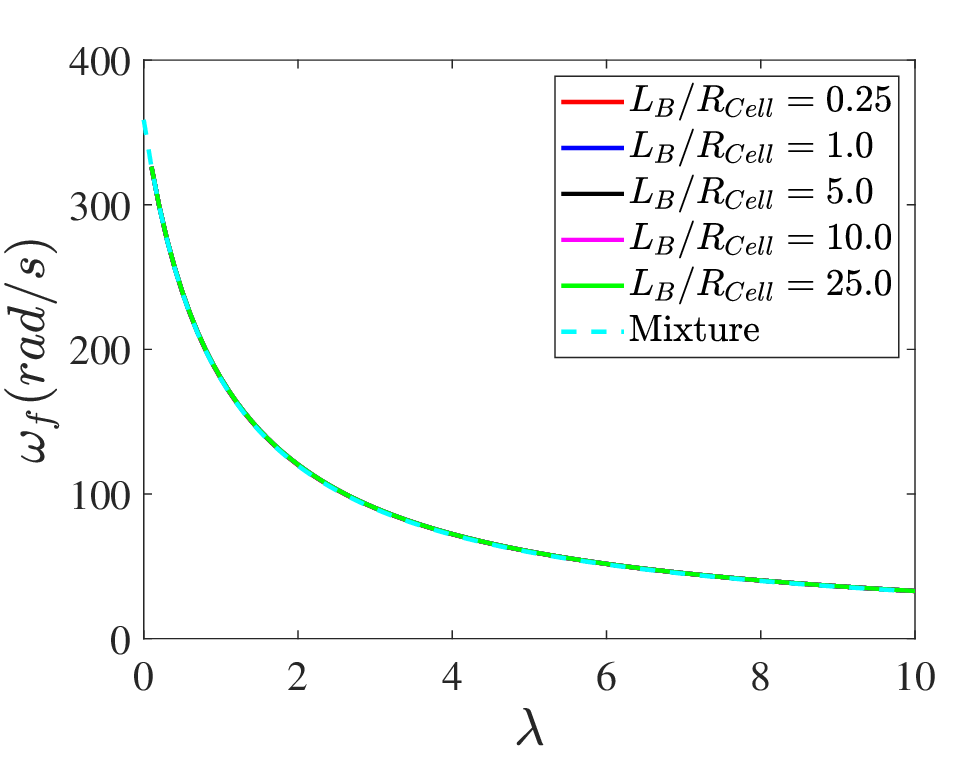}
\caption{Plots of (a) the swimming velocity and the angular velocities of (b) the bacterial head and (c) the flagellar bundle for different values of $L_B/R_{Cell}$, for a bacterium in a two-fluid medium with slipping polymer ($L_B/a \sim 1$) compared with the case where the bacterium swims in a mixture of two fluids (cyan, dashed curve).}
\label{fig:14}
\end{figure}

\begin{figure}
\centering
\includegraphics[width = 0.33\textwidth]{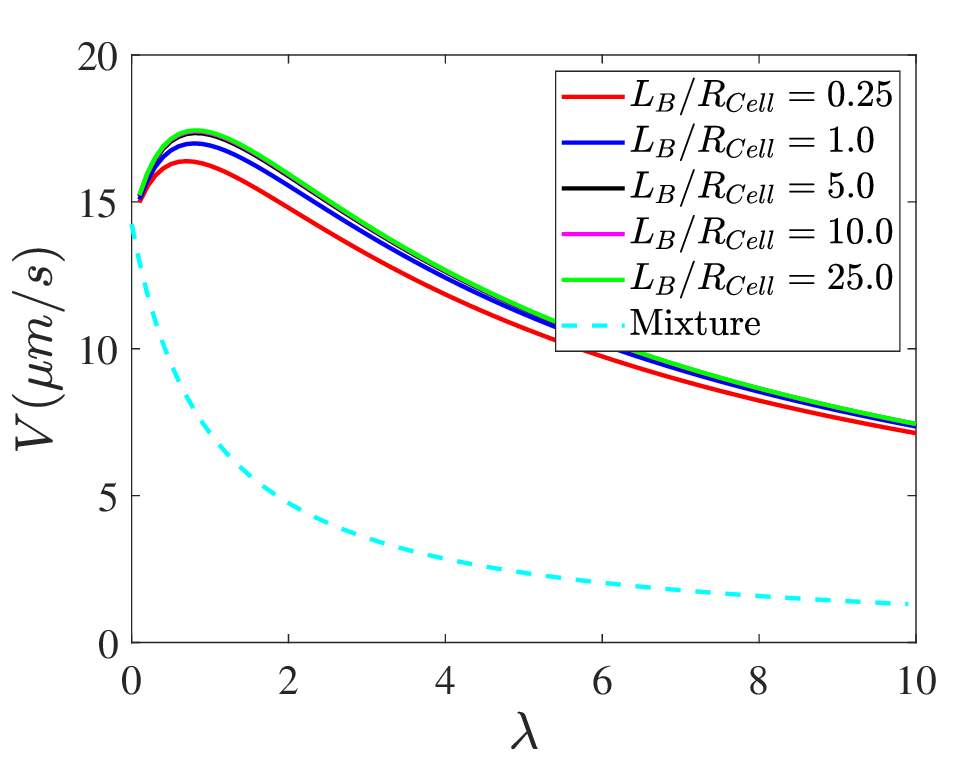}
\includegraphics[width = 0.325\textwidth]{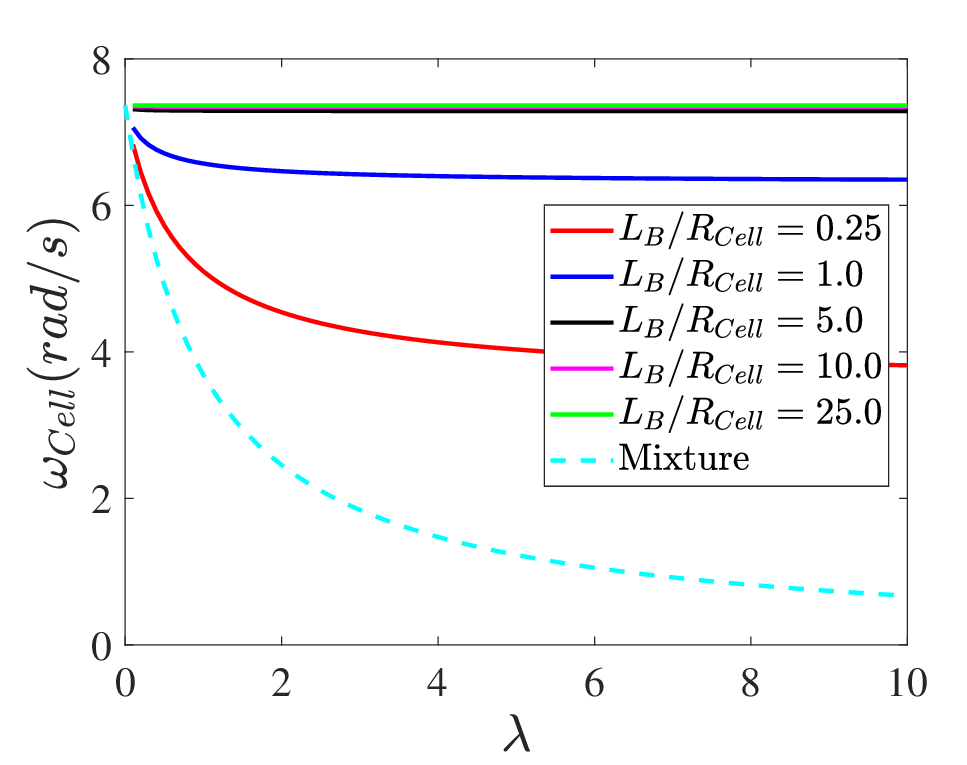}
\includegraphics[width = 0.325\textwidth]{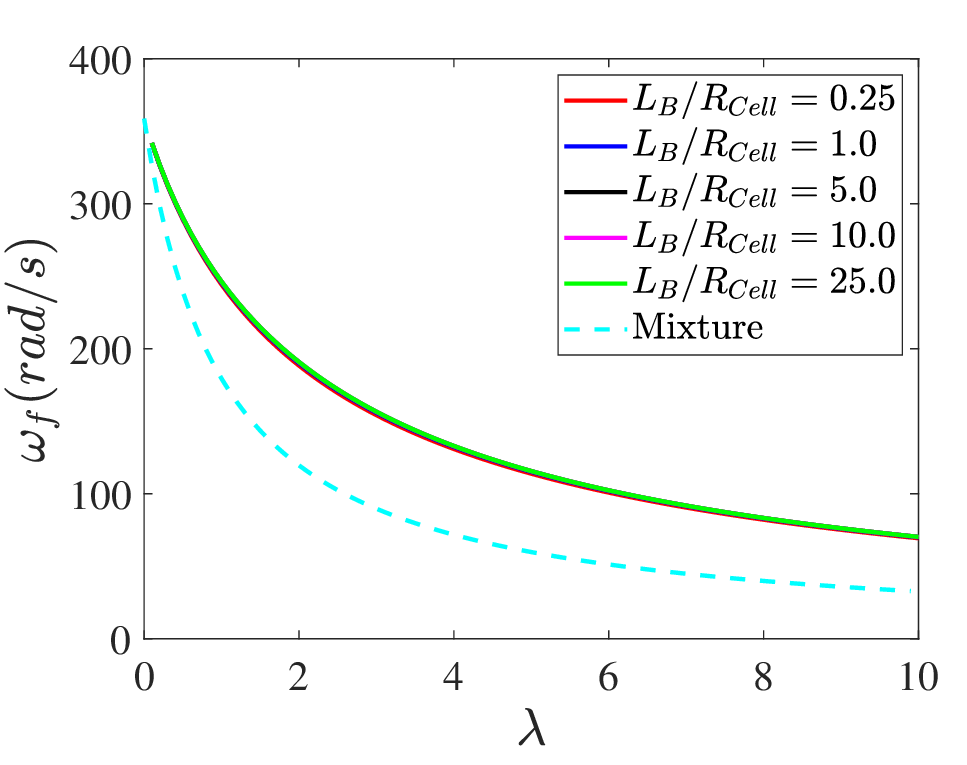}
\caption{Plots of (a) the swimming velocity and the angular velocities of (b) the bacterial head and (c) the flagellar bundle for different values of $L_B/R_{Cell}$, for a bacterium in a two-fluid medium with slipping polymer ($L_B/a \gg 1$) compared with the case where the bacterium swims in a mixture of two fluids (cyan, dashed curve).}
\label{fig:15}
\end{figure}
\begin{figure}
\centering
\includegraphics[width = 0.33\textwidth]{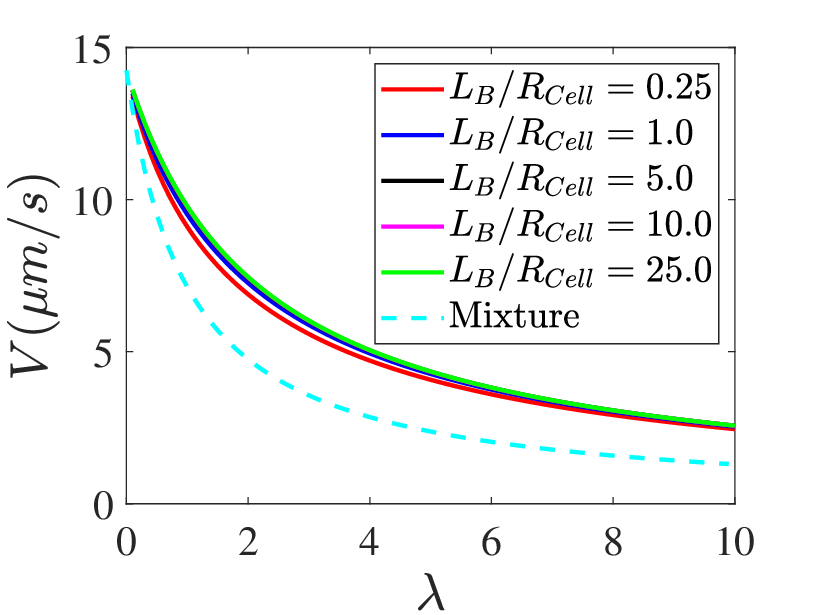}
\includegraphics[width = 0.325\textwidth]{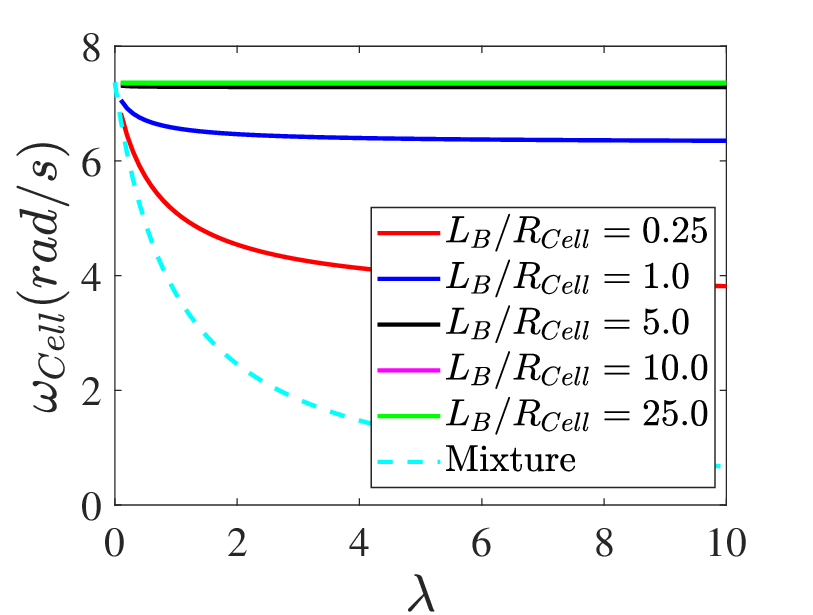}
\includegraphics[width = 0.325\textwidth]{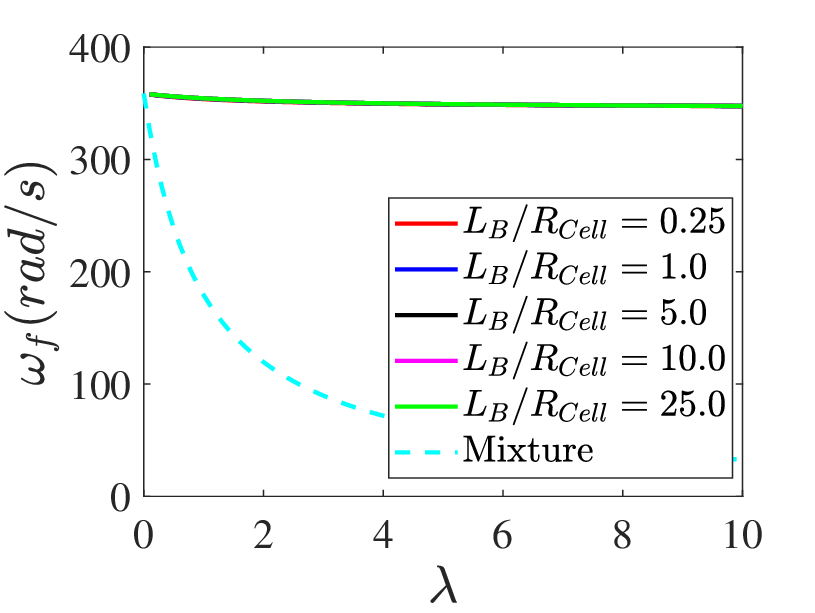}
\caption{Plots of (a) the swimming velocity and the angular velocities of (b) the bacterial head and (c) the flagellar bundle for different values of $L_B/R_{Cell}$, for a bacterium in a two-fluid medium, with non-interacting polymer (on bundle) and slipping polymer on head, compared with the case where the bacterium swims in a mixture of two fluids (cyan, dashed curve).}
\label{fig:16}
\end{figure}

We now look at the effect of two-fluid behavior (microstructure) at the scale of the head on the swimming parameters. For this calculation, we still have the following three cases for the flagellar bundle: (i) Polymer slip at the bundle with $L_B/a \sim 1$, (ii) Polymer slip at the bundle with $L_B/a \gg 1$, and (iii) no polymer-bundle interaction. Now, for each of these three cases, the resistance coefficients for the head correspond to those given in Eq.\ref{Eq.94}-\ref{Eq.95} (slipping polymer on the head). The results of the calculations are shown for case (i) in Fig.\ref{fig:14} and for case (ii) in Fig.\ref{fig:15} as functions of $\lambda$ for different values $L_B/R_{Cell}$. We see from the plots that the two-fluid behavior at the scale of head does not qualitatively change the trends for either scenario, showing that the effect of slipping polymer is more dominant at the scale of flagellar bundle. Similar trends are obtained for case (iii) shown in Fig.\ref{fig:16} for which the polymer does not interact directly with the flagella and slips on the cell. Note that in all these figures (Fig.\ref{fig:14}-\ref{fig:16}), the cyan, dashed curve corresponds to the case where both cell and flagellar bundle swim through the mixture of viscosity $\mu_s(1+\lambda)$. Therefore, we see that the two-fluid model of an entangled polymer solution predicts an enhancement in the swimming velocity of a force-free and torque-free bacterium, when the polymer solution has a microstructure with a length scale comparable to or larger than the flagellar bundle diameter. These results are consistent with the observed trends for the force-free motion of a helix in the two-fluid medium described in the previous sections.

\subsection{Comparison with earlier studies}
\begin{figure}
\centering
\includegraphics[width=0.45\textwidth]{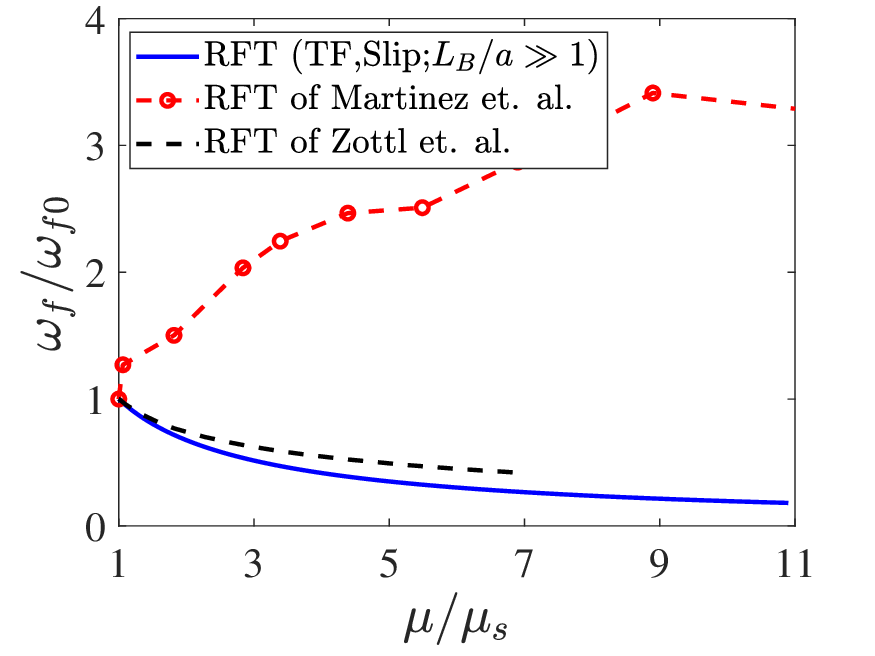}
\includegraphics[width=0.45\textwidth]{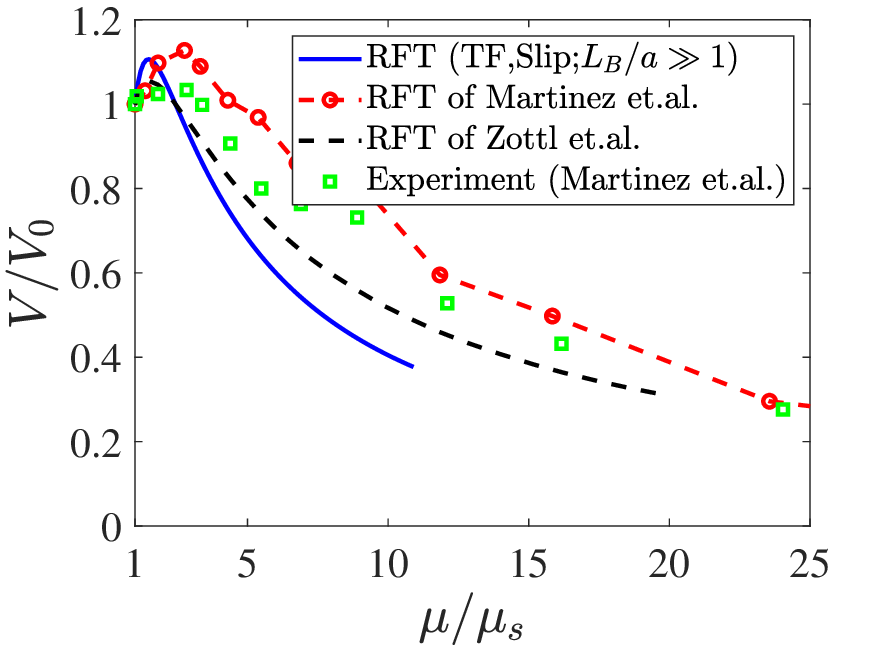}
\caption{Plots of the (a) angular velocity of the flagellar bundle and (b) the velocity of the bacterium, normalised by the respective values in a solvent of viscosity $\mu_s$, as a function of normalised viscosity $\mu/\mu_s$, calculated using the three versions of RFT. Our two-fluid RFT (labelled TF) pertains to the case with polymer slip at the bundle ($L_B/a \gg 1$), while the head \textquoteleft sees' a mixture. The velocity is compared with experimental measurements of \citet{Poon}.}
\label{fig:17}
\end{figure}

Finally, we compare these calculations with experimentally observed trends and with previous RFT calculations, which have sought to explain the motion of a bacterium swimming in a concentrated polymer solution. These include the works of \citet{Kudo,Poon,Yeomans} dealing with {\it E.Coli} motion in concentrated polymer solutions, where the authors have performed RFT calculations assuming bacterium sized pores, shear-thinning and physical depletion of polymers near the flagellar bundle respectively. With the exception of the calculation by \citet{Yeomans}, these studies predict a nonphysical trend, where one of the swimming parameters (the cell angular velocity ($\omega_{Cell}$) in \citet{Kudo} and the flagellar angular velocity ($\omega_{f}$) in \citet{Poon}) increases with medium viscosity. In the calculation of \citet{Poon}, RFT relations were used to fit experimentally observed values for the swimming velocity by using experimentally observed cell angular velocities and the authors show that the fit is satisfactory when one uses $\mu_s$ as the viscosity seen by the flagellar bundle. However, they do not measure flagellar bundle rotation rates in the experiment. Using the RFT equations of \citet{Poon} to obtain the bundle angular velocities from the measured cell angular velocities results in an increasing $\omega_f$ with viscosity. This is a non-physical trend because it implies that the motor angular velocity $\omega_M = \omega_{f}- \omega_{Cell}$ increases with viscosity. This is shown in Fig.\ref{fig:17}(a), where the normalised angular-velocities calculated by the three versions of RFT are shown as a function of normalised viscosity. The resistance coefficients and the input parameters used in the two-fluid RFT calculations for this comparison are the same as those used by \citet{Poon} and \citet{Yeomans} and are shown in Table \ref{Tab.2}. In these calculations, the bacterium has a prolate spheroidal head, with semi-major and semi-minor radii $A_{Cell}$ and $B_{Cell}$ whose resistance coefficients are given by:
\begin{align}
    \alpha_C &= -\frac{4 \pi \mu_s (1+\lambda) U A_{Cell}}{\log(\frac{2 A_{Cell}}{B_{Cell}}) - \frac{1}{2}} \\
    \beta_C &= -\frac{16 \pi}{3} \mu_s (1+\lambda) U B_{Cell}^2 A_{Cell}.
\end{align}

\begin{table}
	\centering
	\begin{tabular}{ | c | c |}
        \hline
		{\bf Parameter} & {\bf Value}  \\
		\hline
		$a$ & $0.03 \mu m$ \\
		$L$ & $7 \mu m$ \\
		$p$ & $2 \mu m$  \\
		$\psi$ & $41^\circ$ \\
		$\gamma$ & $240$ \\
		$A_{Cell}$ & $1.2 \mu m$ \\
        $B_{Cell}$ & $0.43 \mu m$ \\
		$\mu_s$ & $1 mPa\,s$  \\
		$|\bm{T}_0|$ & $1450 pN\,nm$  \\
	  $\omega_0$ & $350\pi \, rad/s$  \\
		$\omega_{max}$ & $600\pi \, rad/s$  \\ 		
		\hline
	\end{tabular}
\caption{Values of the parameters used by \citet{Poon} in RFT calculations.}
\label{Tab.2}
\end{table}
While the RFT of \citet{Yeomans} predicts a similar trend as ours, their calculation is based on an assumption of physical depletion of polymers which, given the coarse-grained model used for the polymers in their simulations, might overestimate the actual depletion near the flagellar bundle (if any). The depletion distance calculated by \citet{Yeomans} from their simulation is $\sim 0.35 R_{Helix}$, which is extracted from a coarse grain MD simulation where polymers are modeled as chains having 12 monomer beads. In Fig.\ref{fig:17}(b), we compare the plot of normalised $V$ against $\mu/\mu_s$ measured experimentally by \citet{Poon} against our two-fluid RFT and the calculations of \citet{Poon,Yeomans}. From the plot, we see that our model also qualitatively follows the experimentally observed trend. Thus the presence of porous microstructure at the length scale of the flagellar bundle, due to entanglement in polymer, also predicts a similar enhancement in swimming velocity as observed in the experiments. 

\vspace{-0.15in}
\section{Conclusions}
In this report, we have developed a two-fluid model to capture the effect of the microstructure of an entangled polymer solution and analysed the motion of a swimming \textit{E.Coli} using slender-body theory. The model predicts an enhancement in swimming velocity, which results directly as a consequence of the microstructure. The two-fluid model does not suffer from the inconsistencies in earlier theoretical models based on shear thinning and depletion near the flagellar bundle\,\citep{Kudo, Yeomans, Lauga3, Poon}. In our model, the flagellar bundle \lq sees' a different viscosity as a consequence of the microstructure of the polymer solution and exerts different continuum stresses on the polymer and solvent, which are hydrodynamically coupled. Therefore, this model better represents the underlying physical conditions in a complex fluid with a microstructure. 

A key assumption in our model lies in the nature of interaction of the polymer with the flagellar bundle. The choices made in our calculations, those of slip or no direct interaction, require validation from experiments, which will also shed light on the nature of interaction between flagellar filaments and the polymers during swimming. The choice of slip between the polymer and the bundle used in this work corresponds to a limiting case, while in reality, the polymer might satisfy a Maxwell-like slip boundary condition\,\citep{Archer}. This can also be easily incorporated into our model, provided the slip length at the flagellar bundle and the head (if slip is present) are known. Regardless, it is easily seen that even with a partial slip of the polymer, the results will remain qualitatively similar to the calculations shown here, with slip resulting in an enhancement of swimming velocity. A stricter no polymer-bundle interaction condition also predicts a slight enhancement in swimming speed, from a leading order RFT calculation, but this does not follow the experimentally observed trends. Thus slip of polymer near the flagellar bundle might be a more plausible condition in those experiments, apart from other non-Newtonian effects. These results, thus shed light on a possible mechanism of swimming speed enhancement observed in experiments, due to the microstructure of an entangled polymer solution.

In the future, it would be valuable to study the motion of a bacterium swimming in an entangled polymer solution, with both non-Newtonian and microstructure effects incorporated in a two-fluid model.   This might be accomplished by incorporating the slender-body theory presented in this work into a numerical solver for a rigid body (cell) moving through a two-fluid polymer solution. The parallel, finite-difference solver for spheroidal-particle-resolved simulations in an inertia-less, unbounded non-Newtonian fluid medium developed by \citep{Arjun} might provide the basis for such a calculation. 

\section*{Acknowledgement}
This work was supported by NSF grant number 2135617.

\appendix
\section{A translating and rotating sphere in the two-fluid medium} \label{App.A}
For a sphere translating with a velocity $\bm{U}$ in a two-fluid medium, the dimensionless governing equations for the solvent and polymer phases are given by:
\begin{align}
  \nabla^2 \bm{u}_s - \nabla p_s - \frac{1}{L_B^2} (\bm{u}_s - \bm{u}_p) = 0 \label{A1} \\
  \lambda \nabla^2 \bm{u}_p - \lambda \nabla p_p + \frac{1}{L_B^2} (\bm{u}_s - \bm{u}_p) = 0 \label{A2}
\end{align}
where, $\bm{u}_s, \bm{u}_p, p_s, p_p$ correspond to the solvent and polymer velocities and pressures respectively. Assuming solutions of the form
\begin{align}
  \bm{u}_s &= \bm{f} \cdot(\nabla \nabla - \bm{I} \nabla^2) f_s(r) \label{A3} \\
  p_s &= \bm{f}\cdot\nabla h_s(r) \label{A4} \\
  \bm{u}_p &= \bm{f}\cdot(\nabla \nabla - \bm{I} \nabla^2) f_p(r) \label{A5} \\
  p_p &= \bm{f}\cdot\nabla h_p(r) \label{A6}
\end{align}
where $\bm{f} = \bm{U}$ for this case (but it can be an arbitrary vector that depends on the boundary condition in general), the governing equations reduce to,
\begin{align}
  &\nabla^2 f_s - h_s - \frac{1}{L_B^2} (f_s - f_p) = 0 \label{A7} \\
  &\lambda \nabla^2 f_p - \lambda h_p + \frac{1}{L_B^2} (f_s - f_p) = 0 \label{A8}\\
  &\nabla^2 h_s = \lambda \nabla^2 h_p = 0 \label{A9}
\end{align}
These equations can be solved by defining a mixture flow $f_m = f_s + \lambda f_p$, $h_m = h_s + \lambda h_p$, and a difference flow $f_d = f_p - f_s$ and $h_d = h_p - h_s$ for which the equations reduce to,
\begin{align}
  &\frac{1}{r^2} \frac{d}{d r} \left( r^2 \frac{d f_m}{d r} \right) - h_m = 0 \label{A10} \\
  &\frac{1}{r^2} \frac{d}{d r} \left( r^2 \frac{d f_d}{d r} \right) - \left( \frac{1 + \lambda}{\lambda L_B^2} \right) f_d - h_d = 0 \label{A11}
\end{align}
which are the well-known Stokes' equation and the Brinkman equation respectively. Note that here and in Eq.\ref{A7} and Eq.\ref{A8}, the Laplacian only involves the radial derivative owing to spherical symmetry in the problem. The boundary conditions used are:
\begin{align}
  \bm{u}_s, \bm{u}_p \rightarrow 0 \; &\text{as} \; r \rightarrow \infty \label{A12} \\
  \bm{u}_p \cdot \bm{n} = \bm{U}\cdot\bm{n} \; &\text{at} \; r = a \label{A13} \\
  \bm{u}_s = \bm{U} \; &\text{at} \; r = a \label{A14} \\
  (\bm{I} - \bm{nn})\cdot(\bm{\sigma}_p \cdot \bm{n}) = 0 \; &\text{at} \; r = a \label{A15}
\end{align}
where the last boundary condition corresponds to zero polymeric tangential stress at the sphere's surface. Solving Eq.\ref{A9} and Eq.\ref{A10} subject to these BCs gives,
\begin{align}
  \bm{u}_s = k_1 \; \bm{f} + k_2 (\bm{f} \cdot \bm{n})\bm{n} \label{A16}\\
  \bm{u}_p = k_3 \; \bm{f} + k_4 (\bm{f} \cdot \bm{n})\bm{n} \label{A17}
\end{align}
where,
\begin{align}
  \textstyle
  k_1 = \frac{e^{-\sqrt{k} r} \left(e^{\sqrt{k} r} \left(k \left(a \sqrt{k}+3\right) \left(a^2+3 r^2\right)-6 \lambda  \left(a \sqrt{k}-k r^2+1\right)\right)+6 \lambda  e^{a \sqrt{k}} \left(k r^2+\sqrt{k} r+1\right)\right)}{4 k r^3 \left(a \sqrt{k}+3 \lambda +3\right)} \label{A18}\\
  \textstyle
  k_2 = \frac{e^{-\sqrt{k} r} \left(3 e^{\sqrt{k} r} \left( - k \left(a \sqrt{k} + 3 \right) \left(a^2-r^2\right)+2 \lambda \left(3 a \sqrt{k}+k r^2+3\right)\right)-6 \lambda  e^{a \sqrt{k}} \left(k r^2+3 \sqrt{k} r+3\right)\right)}{4 k r^3 \left(a \sqrt{k}+3 \lambda +3\right)} \label{A19}\\
  \textstyle
  k_3 = \frac{e^{-\sqrt{k} r} \left(e^{\sqrt{k} r} \left(a^3 k^{3/2}+3 a^2 k+3 a \sqrt{k} \left(k r^2+2\right)+3 k (2 \lambda +3) r^2+6\right)-6 e^{a \sqrt{k}} \left(k r^2+\sqrt{k} r+1\right)\right)}{4 k r^3 \left(a \sqrt{k}+3 \lambda +3\right)} \label{A20}\\
  \textstyle
  k_4 = \frac{3 e^{-\sqrt{k} r} \left(e^{\sqrt{k} r} \left(a^3 \left(-k^{3/2}\right)-3 a^2 k+a \sqrt{k} \left(k r^2-6\right)+k (2 \lambda +3) r^2-6\right)+2 e^{a \sqrt{k}} \left(k r^2+3 \sqrt{k} r+3\right)\right)}{4 k r^3 \left(a \sqrt{k}+3 \lambda
   +3\right)} \label{A21}
\end{align}
with $k = \frac{1 + \lambda}{\lambda L_B^2}$. Using this velocity field, the drag on the sphere $\bm{F}_{drag} = \iint (\bm{\sigma}_p + \bm{\sigma}_s ) \bm{.n} \; a^2 \sin \theta d \theta d \phi $ is given as
\begin{align}
 \frac{\bm{F}_{drag}}{\bm{F}_N} = \frac{(\lambda +1) \left(\frac{a \sqrt{\frac{\lambda +1}{\lambda }}}{L_B}+2 \lambda +3\right)}{\frac{a \sqrt{\frac{\lambda +1}{\lambda }}}{L_B}+3 \lambda +3} \label{A23}
\end{align}
where $\bm{F}_N = (6 \pi \mu_s a \bm{U})$ is the Stokes drag force on a sphere in the solvent.

The limiting forms of this expression are:
\begin{align}
 \lim_{\lambda \rightarrow 0} \frac{\bm{F}_{drag}}{\bm{F}_N} &= 1 \label{A24} \\
 \lim_{\lambda \rightarrow \infty} \frac{\bm{F}_{drag}}{\bm{F}_N} &=  \frac{2 \lambda}{3}. %+ \left( 1 + \frac{a}{9 L_B} \right) + O\left( \frac{1}{\lambda}\right).
 \label{A25}
\end{align}
The first limit corresponds to the value of drag on a sphere through a pure solvent and the second limit to the drag on a spherical bubble translating through a polymer fluid (owing to Eq.\ref{A15}). Also,
\begin{align}
 \lim_{L_B \rightarrow 0} \frac{\bm{F}_{drag}}{\bm{F}_N} &= (1 + \lambda) %- \lambda^{3/2} \sqrt{1 + \lambda} \left( \frac{L_b}{a}\right)
 \label{A26} \\
 \lim_{L_B \rightarrow \infty} \frac{\bm{F}_{drag}}{\bm{F}_N} &= 1 + \frac{2 \lambda}{3} \label{A27}
\end{align}
which correspond, respectively,  to the mixture-like behavior of the two fluids and to a sphere moving through two independent fluids exerting additive drag forces.

For the rotating sphere, the governing equations Eq.\ref{A1}-\ref{A2} are solved using the same procedure as before with the boundary conditions now being,
\begin{align}
 \bm{u}_s(r) = (\bm{\omega \times n}) \; &\text{at} \; r = a \label{A28} \\
 \bm{u}_s, \bm{u}_p \rightarrow 0 \; &\text{as} \; r \rightarrow \infty \label{A29}\\
 (\bm{I} - \bm{nn})\cdot(\bm{\sigma}_p \cdot \bm{n}) = 0 \; &\text{at} \; r = a \label{A30}
\end{align}
Here, solutions of the form
\begin{align}
 \bm{u}_s &= \bm{\omega \times n} f_s(r) \label{A31}\\
 \bm{u}_p &= \bm{\omega \times n} f_p(r) \label{A32}\\
 p_p &= p_s = 0 \label{A33}
\end{align}
are assumed using spherical symmetry. Using these boundary conditions, the solution is obtained as:
\begin{align}
 \bm{u}_s = \bm{\omega \times n} \frac{a^3 e^{-\frac{r}{\sqrt{k}}} \left(\left(a^2+3 a \sqrt{k}+3 k\right) \left(-e^{\frac{r}{\sqrt{k}}}\right)-3 \sqrt{k}
   \lambda  e^{\frac{a}{\sqrt{k}}} \left(\sqrt{k}+r\right)\right)}{r^3 \left(a^2+3 a \sqrt{k} (\lambda +1)+3 k (\lambda
   +1)\right)} \label{A34} \\
\bm{u}_p = \bm{\omega \times n} \frac{a^3 e^{-\frac{r}{\sqrt{k}}} \left(3 \sqrt{k} e^{\frac{a}{\sqrt{k}}} \left(\sqrt{k}+r\right)-\left(a^2+3 a \sqrt{k}+3 k\right) e^{\frac{r}{\sqrt{k}}}\right)}{r^3 \left(a^2+3 a \sqrt{k} (\lambda +1)+3 k (\lambda +1)\right)} \label{A35}
\end{align}
where $k = \left( \frac{1 + \lambda}{\lambda L_B^2} \right)^{-1}$. Using this velocity field, the torque on the sphere $\bm{T}_{sph} = \iint \bm{r \times}(\bm{\sigma}_p + \bm{\sigma}_s ) \; a^2 \sin \theta d \theta d \phi $ is given as:
\begin{align}
 \frac{\bm{T}_{sph}}{\bm{T}_N} = \frac{(\lambda +1) \left(\frac{a^2}{L_B^2}+\frac{3 a \sqrt{\frac{\lambda }{\lambda +1}}}{L_B}+\frac{3 \lambda }{\lambda +1}\right)}{\frac{a^2}{L_B^2}+\frac{3 a \sqrt{\frac{\lambda }{\lambda +1}} (\lambda +1)}{L_B}+3 \lambda } \label{A37}
\end{align}
where $\bm{T}_N=8 \pi \mu_s a^3 \bm{\omega}$ is the torque on a sphere in the solvent.
From the expression for torque, we see that:
\begin{align}
 \lim_{L_B \rightarrow 0} \frac{\bm{T}_{sph}}{\bm{T}_N} = 1 + \lambda \label{A38} \\
 \lim_{L_B \rightarrow \infty} \frac{\bm{T}_{sph}}{\bm{T}_N} = 1 \label{A39}
\end{align}
consistent with the expected behavior. From Fig.\ref{fig:1}(a)-(b), we see that the drag force on a translating sphere  and torque on a rotating sphere of given radius $a$ decreases with  increasing in $L_B$. Also note that the decrease in torque is more significant than the drop in drag for a given viscosity ratio $\lambda$.

\subsection{Solutions for the case with a non-interacting polymer}
To obtain solutions for the case with a non-interacting polymer fluid, we assume that the polymer fluid exists everywhere in the domain, including within the sphere. Thus, one has velocity field $\bm{u}_p^{in}$ and pressure field $p_p^{in}$ within the sphere, which satisfy Stokes equations.  We have a two-fluid medium outside the sphere which satisfy Eq.\ref{A1}-\ref{A2}. The boundary conditions for this case are given by:
\begin{align}
  \bm{u}_s, \bm{u}_p \rightarrow 0 \; &\text{as} \; r \rightarrow \infty \label{A40} \\
  \bm{u}_p\ = \bm{u}_p^{in} \; &\text{at} \; r = a \label{A41} \\
  \bm{u}_s = \bm{U} \; &\text{at} \; r = a \label{A42} \\
  \bm{\sigma}_p \cdot \bm{n} = 0 \; &\text{at} \; r = a \label{A43}
\end{align}
Assuming a growing velocity and pressure field for the polymer inside the sphere, given by:
\begin{align}
\bm{p}_p^{in} &= c_1 \bm{U} \cdot \bm{r} \label{A44} \\
\bm{u}_p^{in} &= \left( c_2  -\frac{c_3}{3}r^2 \right) \bm{U} + \left( \frac{c_1}{2} + c_3 \right) (\bm{U} \cdot \bm{r})\bm{r} \label{A45}
\end{align}
subject to $\nabla \cdot \bm{u}_p^{in} = 0$, and assuming the same form for $\bm{u}_s$, $\bm{u}_p$, $p_s$ and $p_p$ as in Eq.\ref{A3} - \ref{A6}, we can solve the above system of equations to obtain the drag force as:
\begin{align}
\frac{\bm{F}_{drag}}{\bm{F}_N} = \frac{(\lambda +1) \left(\frac{a^3 \sqrt{\frac{\lambda +1}{\lambda }}}{L_B^3}+\frac{a^2 (2 \lambda +3)}{L_B^2}+\frac{18 a \lambda  \sqrt{\frac{\lambda +1}{\lambda }}}{L_B}+18 \lambda \right)}{\frac{a^3 \sqrt{\frac{\lambda +1}{\lambda }}}{L_B^3}+\frac{3 a^2 (\lambda +1)}{L_B^2}+\frac{18 a \lambda  \sqrt{\frac{\lambda +1}{\lambda }}}{L_B}+18 \lambda  (\lambda +1)}
\end{align}
A similar procedure can be used to calculate the torque on a rotating sphere for this case, and it can be shown to result in the same expression as Eq.\ref{A37}.

\section{Numerical scheme for SBT with validation}\label{App.B}
The numerical approach to solve the integral equations for force strength (Eq.\ref{Eq.20}, \ref{Eq.44}, \ref{Eq.45}, \ref{Eq.61}) uses the helical phase $\psi = ks \cos \theta$, where $k = 2 \pi/p$ ($p$ is the pitch) to parameterize spatial locations as,
\begin{equation}
    \bm{r}(\psi) = R\left[\cos \psi, \sin \psi, \psi \cot \theta \right] \label{Eq.65}
\end{equation}
so an integral equation for force strength, such as Eq.\ref{Eq.20}, becomes:
\begin{equation}
    \begin{split}
    \bm{U}_n = &\frac{\bm{f}_n}{4 \pi (1 + \lambda)}\bm{.}\left[(\bm{I} + \bm{t}_n\bm{t}_n) \log 2\gamma  + \frac{(\bm{I} - 3 \bm{t}_n\bm{t}_n)}{2}\right] + \frac{R \Delta \psi \csc \theta}{8 \pi (1 + \lambda)} \sum_{m \neq n} \frac{\bm{I} + \bm{\hat{r}}_{nm} \bm{\hat{r}}_{nm} }{r_{nm}} \bm{.f}_m\\
    &- \sum_{m \neq n} \frac{(\bm{I} + \bm{t}_n\bm{t}_n)}{r_{nm}}\bm{.f}_n + \mathcal{O}(\Delta \psi)
    \end{split} \label{Eq.66}
\end{equation}
where $n,m = 1,2,3....pN$, $\bm{r}_{nm} = \bm{r}(\psi_n) - \bm{r}(\psi_m)$ is the position vector between spatial locations, $\bm{t}_n = \left[-sin \theta \sin \psi_n, \sin \theta \cos \psi_n, \cos \theta \right]$ is the tangential unit vector at $\bm{r}_n$, and $\Delta \psi$ is the mesh size of the helical phase. Note that the equation above assumes that the slender fiber has a spheroidal cross-section $\overline{a}(s) = 2\sqrt{s(1-s)}$. For convenience, we now move to a coordinate system which is rotated with the helical phase, such that the surface velocity $\bm{U}_n$ is invariant along the helix. We use these invariant velocity components to create a linear mapping between the velocity and force per unit length, which can be evaluated for a speciﬁed helical geometry, helical axial velocity $U$, and rotation rate $\Omega$ to give the thrust, torque, and drag. The transformation to this rotated coordinate system is achieved by means of the rotation operator $\mathcal{R}(\psi)$ defined as:
\begin{equation}
    \mathcal{R}(\psi) = \begin{pmatrix}
                        \cos \psi & -\sin \psi & 0 \\
                        \sin \psi & \cos \psi & 0 \\
                        0 & 0 & 1
                        \end{pmatrix}\label{Eq.67}
\end{equation}
The transformed velocity and force vectors are now denoted by
\begin{equation}
    \bm{U}'_n = \mathcal{R}(-\psi_n) \bm{.U}_n, \quad \bm{f}'_n = \mathcal{R}(-\psi_n) \bm{.f}_n \label{Eq.68}
\end{equation}
such that $\bm{U}'_n$ is invariant along the helical filament. For a rigid helix that rotates at rate $\Omega$ and translates at speed $U$ along its axial direction, we have
\begin{equation}
    \bm{U}'_n = \left[0, \Omega\,R, U  \right]^T \label{Eq,68}
\end{equation}
and
\begin{equation}
    \sum_{n = 1}^{pN} \bm{f}'_n R \Delta \psi \csc \theta = \left[0, \frac{T}{R}, F_z \right]^T \label{Eq.69}
\end{equation}
The integral equation (Eq.\ref{Eq.20}) now becomes,
\begin{equation}
    \begin{split}
    \bm{U}'_n = &\frac{\bm{f}'_n}{4 \pi (1 + \lambda)}\cdot\left[(\bm{I} + \bm{t}'\bm{t}') \log 2\gamma  + \frac{(\bm{I} - 3 \bm{t}'\bm{t}')}{2}\right] - \sum_{m \neq n} \frac{(\bm{I} + \bm{t}'\bm{t}')}{r_{nm}} \cdot \bm{f}'_n \\
    &+ \frac{R \Delta \psi \csc \theta}{8 \pi (1 + \lambda)} \sum_{m \neq n} \frac{\mathcal{R}(\psi_m - \psi_n) + \mathcal{R}(-\psi_n) \bm{.\hat{r}}_{nm} \bm{\hat{r}}_{nm}\bm{.}\mathcal{R}(\psi_m) }{r_{nm}} \cdot \bm{f}'_m + \mathcal{O}(\Delta \psi)
    \end{split} \label{Eq.70}
\end{equation}
where $\bm{t}' = \left[0, \sin \theta, \cos \theta \right]$ is the tangent vector invariant along the helical fiber. This results in a linear mapping between $\bm{U}'$ and $\bm{f}'$ given by
\begin{equation}
    \begin{pmatrix}
    \bm{U}'_1 \\
    \bm{U}'_2 \\
    .\\
    .\\
    .\\
    \bm{U}'_{pN} \\
    \end{pmatrix} = \mathcal{G}.\begin{pmatrix}
                                \bm{f}'_1 \\
                                \bm{f}'_2 \\
                                .\\
                                .\\
                                .\\
                                \bm{f}'_{pN} \\
                                \end{pmatrix} \label{Eq.71}
\end{equation}
For a known motion of helix $\bm{U}'_n = \bm{U}^0 = \left[0, \Omega\,R, U \right]^T$, we have:
\begin{equation}
    \begin{pmatrix}
    \bm{f}'_1 \\
    \bm{f}'_2 \\
    .\\
    .\\
    .\\
    \bm{f}'_{pN} \\
    \end{pmatrix} = \mathcal{G}^{-1}.\begin{pmatrix}
                                     \bm{U}^0 \\
                                     \bm{U}^0 \\
                                     .\\
                                     .\\
                                     .\\
                                     \bm{U}^0 \\
                                     \end{pmatrix}. \label{Eq.72}
\end{equation}
or in simple terms,
\begin{equation}
\bm{\mathcal{F}} = \mathcal{G}^{-1} \cdot \bm{\mathcal{U}} \label{Eq.72N}
\end{equation}
where $\bm{\mathcal{F}}$ is the unknown vector of force strengths having dimension $3pN$ and $\bm{\mathcal{U}}$ is the given velocity vector of dimension $3pN$ at each grid point on the fiber surface. The total axial hydrodynamic force $F_z$, and net torque $T$ are therefore given by Eq.\ref{Eq.69}, which includes the thrust and torque due to rotation and the drag due to translation of the helix.

The procedure highlighted above can be generalised to the two-fluid case, for both polymer slip and no polymer-fiber interaction scenarios. For the first case, the uniformly valid SBT equation (Eq.\ref{Eq.54A}) is used, while for the latter scenario, the SBT equation with $L_B$ restricted to the outer region (Eq.\ref{Eq.61}) is used for numerical calculation. In the first case, one can calculate the force strengths given by the three versions of SBT ($L_B \sim O(a)$ (Eq.\ref{Eq.20}), $L_B \gg O(a)$ (Eq.\ref{Eq.44},\ref{Eq.45N}) and $a \ll L_B \ll l$ (Eq.\ref{Eq.53A},\ref{Eq.53B})) individually and combine the total force strengths according to Eq.\ref{Eq.54A}.

For the SBT with $L_B \sim O(a)$, the equation for (total) force strength ($\bm{f}_s + \lambda \bm{f}_p$) is similar to the SBT equation in a single fluid case and so the equation can be discretised into the form given by Eq.\ref{Eq.72N}, with $\mathcal{G}$ now being a function of $h(\lambda,L_B)$ and $g(\lambda, L_B)$ given in Eq.\ref{Eq.20}. But for the other two versions of SBT (Eq.\ref{Eq.44},\ref{Eq.45N} and Eq.\ref{Eq.53A},\ref{Eq.53B}) one has two force strengths $\bm{f}_s$ and $\bm{f}_p$ at each point on the fiber, so that the unknown vector of force strengths in Eq.\ref{Eq.72} is now of dimension $6pN$ (twice as much as the dimension of unknown vector $\bm{\mathcal{F}}$ for the single fluid case) with:
\begin{equation}
    \bm{\mathcal{\overline{F}}} = \begin{pmatrix}
    \bm{f}^{s'}_{1} \\
    \bm{f}^{s'}_{2} \\
    .\\
    .\\
    .\\
    \bm{f}^{s'}_{pN} \\
    \bm{f}^{p'}_1 \\
    \bm{f}^{p'}_2 \\
    .\\
    .\\
    .\\
    \bm{f}^{p'}_{pN} \\
    \end{pmatrix} = \mathcal{\overline{G}}^{-1} \cdot \begin{pmatrix}
                                     \bm{U}^0 \\
                                     \bm{U}^0 \\
                                     .\\
                                     .\\
                                     .\\
                                     \bm{U}^0 \\
                                     \bm{U}^0 \cdot (\bm{I} - \bm{t}'\bm{t}')\\
                                     \bm{U}^0 \cdot (\bm{I} - \bm{t}'\bm{t}')\\
                                     .\\
                                     .\\
                                     .\\
                                     \bm{U}^0 \cdot (\bm{I} - \bm{t}'\bm{t}')\\
                                     \end{pmatrix} = \bm{\mathcal{\overline{U}}}
\end{equation}
 where $\mathcal{\overline{G}}$ now is given by Eq.\ref{Eq.44},\ref{Eq.45N} for SBT with $L_B/a \gg O(1)$ and by Eq.\ref{Eq.53A},\ref{Eq.53B} for SBT with $a \ll L_B \ll l$.

\subsection{Validation of the numerical scheme}
\begin{figure}
\includegraphics[scale = 0.375]{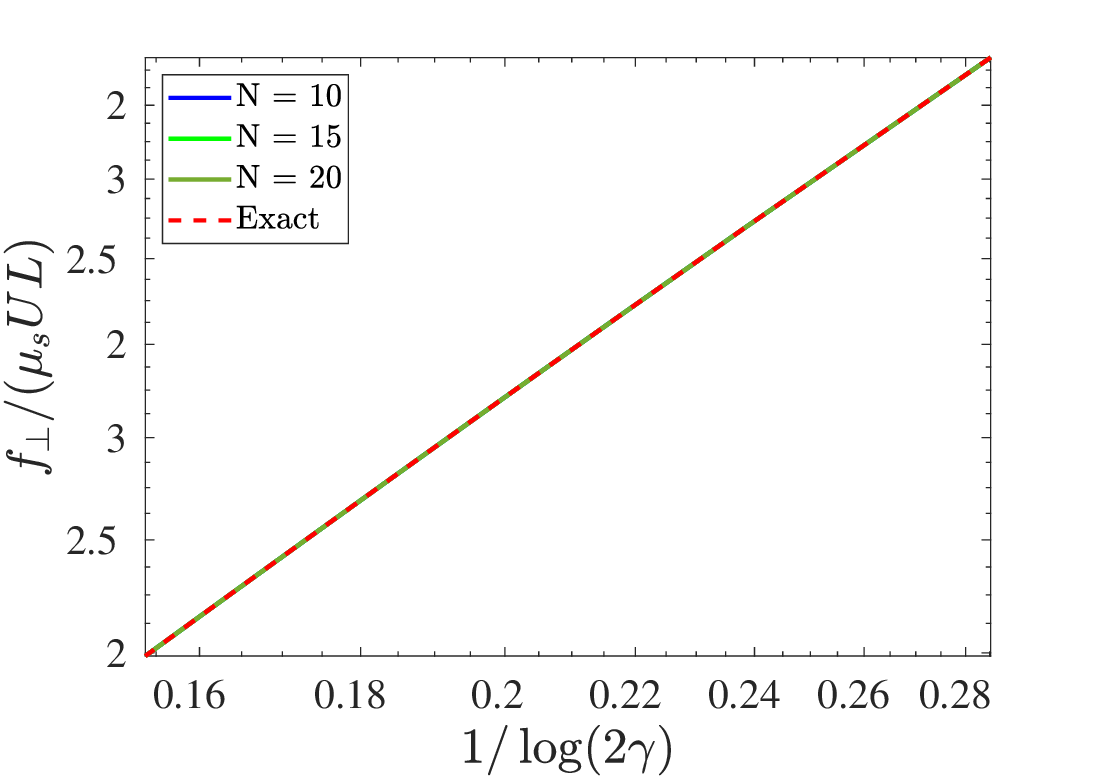}
\includegraphics[scale = 0.375]{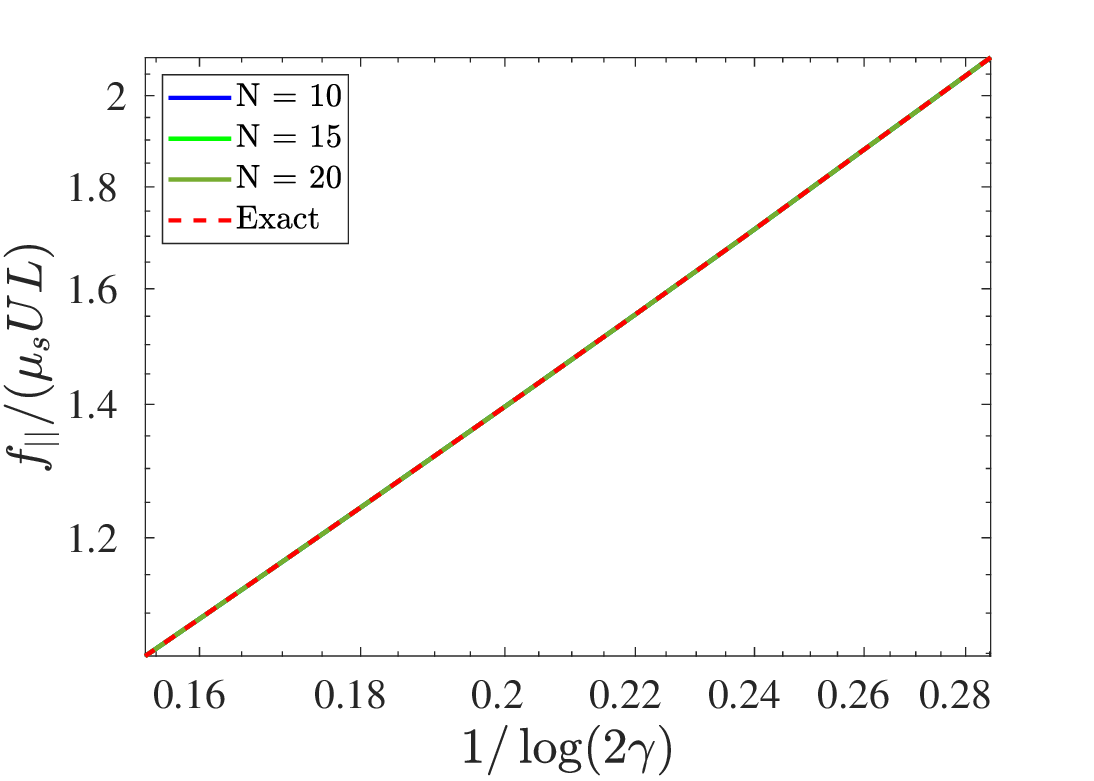}
\caption{Comparison of the (a) normal and (b) tangential components (to the spheroid axis) of the force on a straight, slender spheroid evaluated using the numerical scheme with $N = 10, 15, 20$ grid points per pitch and the exact theoretical expression given in Eq.\ref{Eq.73}.}
\label{fig:B1}
\end{figure}

\begin{figure}
\centering
\includegraphics[scale = 0.4]{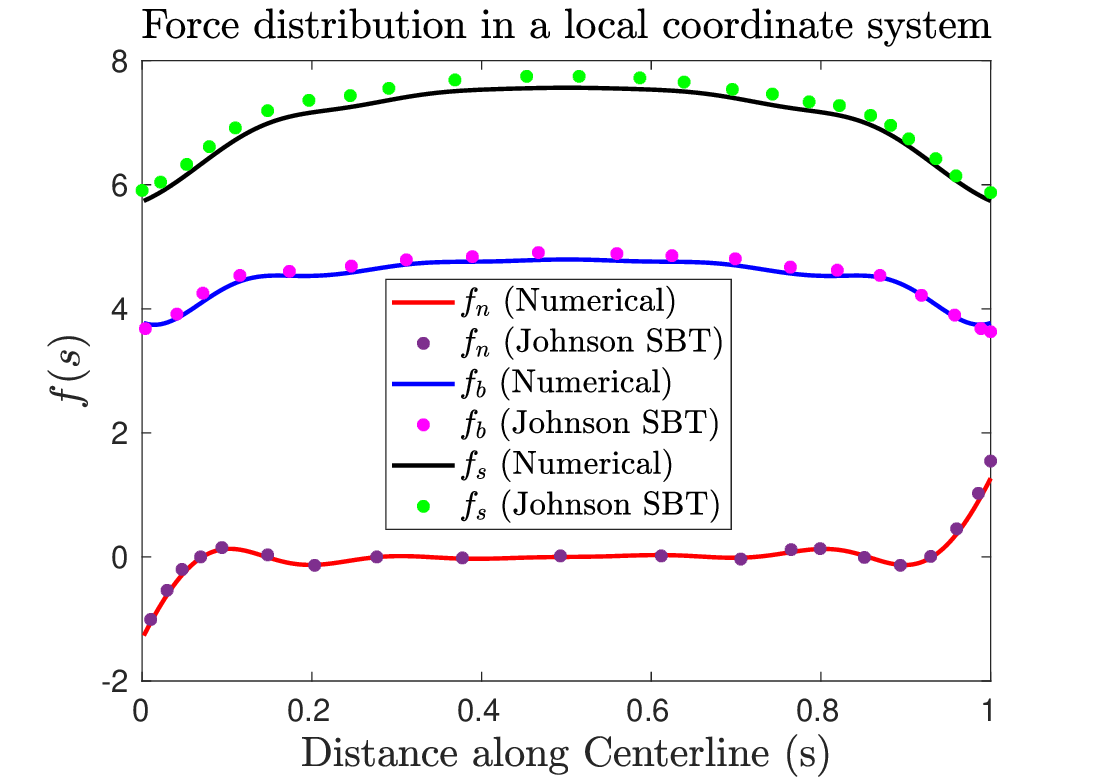}
\caption{Comparison of the components of the force strength along the centerline of a helical fiber in a local coordinate system from our numerical implementation with the results of \citet{Johnson}. The helix has a prolate spheroidal cross-section with $l/p$ = 5, $R_{Helix}/p = 0.25$ and $a/p = 0.01$ and undergoes force-free motion.}
\label{fig:B2}
\end{figure}

We first validate the numerical scheme for a straight slender spheroid  with a cross-section given by $\overline{a}(s) = 2\sqrt{s(1-s)}$ in a single Newtonian fluid medium. From the SBT of \citet{Batchelor, Cox} it is known that for uniform flow past a slender, smooth spheroid, the non-dimensional force per unit length from SBT (accurate to all orders in $O(1/ \log 2\gamma)$) is given by:
\begin{equation}
\bm{U} = \frac{1}{4 \pi}\bm{f} \cdot (\bm{I} + \bm{e_z e_z}) \log 2\gamma + \frac{1}{8 \pi} \bm{f} \cdot(\bm{I} - 3\bm{e_z e_z}) \label{Eq.73}
\end{equation}
where $\bm{e_z}$ now denotes the unit vector along the axis of the spheroid. This is compared with the result from the numerical calculation of total force on a straight spheroid in Fig.\ref{fig:B1}. The plot suggests that the numerical result agrees extremely well with the theoretical prediction for $N \geq 10$, which we also find to hold true for helical fibers (shown later in Fig.\ref{fig:B2}, where the numerical calculation is compared with the theoretical prediction of \citet{Johnson}). Thus, in our subsequent calculations, we use $N = 30$ grid points per pitch for the helical fiber. Note that for a fiber with a cylindrical cross-section, the numerical implementation is more involved due to end effects, which introduce errors of $O(\gamma^{-1} \log (1/\gamma))$, owing to a force singularity at the end \citep{Mackaplow94}.  Fibers with spheroidal cross-section remain numerically well behaved for large $\gamma$. Note that for the case of straight fibers, the numerical scheme was implemented in a local cylindrical coordinate system instead of the helical one described earlier.

Next, we validate the numerical scheme by calculating the total forces and torque on a helical fiber held in a uniform flow of a single Newtonian fluid and compare the results with those given by \citet{Johnson}. In his work, the author considers a helical fiber (with spheroidal cross-section) of length $l = 5 p$, (where $p$ is the pitch of the helix), radius $R_{Helix} = 0.25 p$ and cross-section radius at the mid-point $a = 0.01 p$, that rotates due to an external torque and translates  with a velocity such that the helix is force-free. We have considered the same case and have calculated the force strength along the fiber centerline. \citet{Johnson} calculates the force strength in a local coordinate system along the centerline, where the coordinate directions are tangent, normal and binormal to the centerline. We can calculate the same by using a simple coordinate transformation from our local helical coordinate system. The results from our calculation are plotted against the predictions of \citet{Johnson} in Fig.\ref{fig:B2}, and we see that our numerical implementation works well for fibers with curved centerlines as well.

\section{Slender-body theory for slipping polymer with $L_B$ in the matching region}\label{App.C}
To derive the slender-body equation with screening length in the matching region, we divide the domain into inner, outer and Brinkman regions as shown in Fig.\ref{fig:2A}. In the inner region, two independent fluids undergo two-dimensional flow relative to an infinite cylinder.  In the outer region, the solvent and polymer move with the same three-dimensional mixture velocity due to the distribution of singularities along the centerline. In the Brinkman region, the flow is the two-dimensional flow driven by a point singularity in the two-fluid medium. The dimensionless inner, outer and Brinkman solutions are given below for the solvent and polymer fluids, where the same scales used for the inner and outer region in Section \ref{3} are used here and for the Brinkman region the length is non-dimensionalised by $L_B$, while the other scales are kept the same.
\subsection{Inner region}
For the solvent, the inner solution in the limit of $\rho/a \gg  1$ is given by:
\begin{equation}
    \bm{u}_s^{in} = \bm{U} - \frac{\bm{f}_s\cdot(\bm{I} + \bm{e_z e_z})}{4 \pi} \log \rho + \frac{\bm{f}_s}{4 \pi}\cdot\left[ \bm{nn} - \frac{\bm{I} - \bm{e_ze_z}}{2}\right] + O(\frac{1}{\rho^2})\label{Eq.C1}
\end{equation}
and for the polymer, one has
\begin{equation}
    \bm{u}_p^{in} = \bm{U}\cdot(\bm{I} - (1-c)\bm{e_ze_z}) - \frac{\bm{f}_p\cdot(\bm{I} - \bm{e_z e_z})}{4 \pi} \log \rho + \frac{\bm{f}_p}{4 \pi}\cdot\left[ \bm{nn} - (\bm{I} - \bm{e_ze_z}) \right] + O(\frac{1}{\rho^2}) \label{Eq.C2}
\end{equation}
\subsection{Outer region}
In the outer region, the solvent and polymer flow with the same velocity (the velocity of mixture) due to the distribution of singularities along the fiber centerline. The inner limit of the outer solution ($\rho/l \ll 1$) for the solvent and polymer are therefore:
\begin{equation}
    \begin{split}
    &\bm{u}_s^{out}(\bm{r}) = \bm{u}_p^{out}(\bm{r}) = \bm{U}_{\infty}(\bm{r}) + \frac{\bm{f}\cdot(\bm{I + e_z e_z})}{4 \pi (1+\lambda)} \left(\log \left(\frac{2(\sqrt{s(1-s)})}{\rho} \right) \right) - \frac{\bm{f}\cdot\bm{e_z e_z}}{4 \pi (1+\lambda)} + \\
    &\frac{\bm{f.nn}}{4 \pi (1+\lambda)} +
    \frac{1}{8\pi (1+\lambda)} \int \left[\left( \frac{\bm{I}}{|\bm{r_c}(s) - \bm{r_c}(s')|} + \frac{(\bm{r_c}(s) - \bm{r_c}(s'))(\bm{r_c}(s) - \bm{r_c}(s'))}{|\bm{r_c}(s) - \bm{r_c}(s')|^3} \right)\cdot\bm{f}(\bm{r_c}(s')) \right.\\
    &\left.- \left(\frac{(\bm{I} + \bm{e_z e_z})}{|s - s'|} \right)\cdot\bm{f}(\bm{r_c}(s)) \right] ds'
    \end{split} \label{Eq.C3}
\end{equation}
where $\bm{f} = \bm{f}_s + \lambda \bm{f}_p$.

\subsection{Brinkman region}
In the Brinkman region ($a \ll L_B \ll l$), the solvent and polymer are coupled by the two-fluid equations, and the flow is the two-dimensional flow driven by a point singularity. Thus, the dimensionless solvent and polymer velocity fields are:
\begin{equation}
\begin{split}
    \bm{u}_s^{Br} &= \bm{f}_s \cdot \bm{G}_{SS} + \lambda \bm{f}_p \cdot \bm{G}_{PS} +c_1\\
    &=\bm{f}_s \cdot \bm{G}_{St} + \frac{\lambda}{1+\lambda} (\bm{f}_s - \bm{f}_p) \cdot [\bm{G}_{Br} - \bm{G}_{St}] +c_1
\end{split}\label{Eq.C4}
\end{equation}
\begin{equation}
\begin{split}
    \bm{u}_p^{Br} &= \bm{f}_s \cdot \bm{G}_{SP} + \lambda \bm{f}_p \cdot \bm{G}_{PP} + c_2\\
    &=\bm{f}_p \cdot \bm{G}_{St} + \frac{1}{1+\lambda} (\bm{f}_p - \bm{f}_s) \cdot [\bm{G}_{Br} - \bm{G}_{St}] +c_2
\end{split}\label{Eq.C5}
\end{equation}
where $c_1$ and $c_2$ are arbitrary constants which will be determined by matching, and
\begin{equation}
    \bm{G}_{St} = \frac{1}{4 \pi} \left( (\bm{I} + \bm{e_z e_z}) \log\left( 1/\rho \right) + \bm{nn}\right) \label{Eq.C6}
\end{equation}
and
\begin{equation}
    \bm{G}_{Br} = \frac{1}{4 \pi} \left( (\bm{I} - \bm{e_z e_z})   A_1 \left( \rho/L_B \right) + \bm{nn} A_2 \left( \rho/L_B \right) + 2\,K_0 (\rho/L_B) \bm{e_z e_z} \right) \label{Eq.C7}
\end{equation}
where,
\begin{align}
A_1 (\chi) &= 2 \left(K_0(\chi) + \frac{K_1(\chi)}{\chi} - \frac{1}{\chi^2} \right) \label{Eq.C8} \\
A_2 (\chi) &= 2 \left(-K_0(\chi) - \frac{2 K_1(\chi)}{\chi} + \frac{2}{\chi^2} \right) \label{Eq.C9}
\end{align}
with $K_\nu$ being the modified Bessel function of order $\nu$. Here, the length ($\rho$) is non-dimensionalised with $L_B$. The solution in the Brinkman region should be matched to both the inner and outer solution in order to obtain the SBT equation for $\bm{f}_s$ and $\bm{f}_p$. Thus in this procedure, one has two matching conditions, instead of one. In the first matching, the Brinkman solution in the limit of $\rho /L_B \ll 1$ is matched with the outer limit ($\rho /a \gg 1$) of the inner solution. This yields,
\begin{equation}
\begin{split}
    \bm{U} &= c_1 + \frac{\bm{f}_s \cdot (\bm{I} + \bm{e_z e_z}) }{4 \pi} \log \left( \frac{L_B}{a} \right) +\frac{\bm{f}_s \cdot (\bm{I} - \bm{e_z e_z})}{8 \pi} \\
    &+ \frac{\lambda}{4 \pi (1+\lambda)} (\bm{f}_s - \bm{f}_p) \cdot \left[ (\bm{I} + \bm{e_z e_z}) \left[ -\Gamma + \log 2 \right] - \frac{(\bm{I} - \bm{e_z e_z})}{2} \right],
\end{split} \label{Eq.C10}
\end{equation}
and
\begin{equation}
\begin{split}
    \bm{U}\cdot(\bm{I} - (1-c) \bm{e_z e_z}) &= c_2 + \frac{\bm{f}_p}{4 \pi} \log \left( \frac{L_B}{a} \right) + \frac{\bm{f}_p}{4 \pi} \cdot (\bm{I} - \bm{e_z e_z}) + \\
    &+ \frac{(\bm{f}_p - \bm{f}_s)}{4 \pi (1+\lambda)} \cdot \left[ (\bm{I} + \bm{e_z e_z}) \left[ -\Gamma + \log 2 \right] - \frac{(\bm{I} - \bm{e_z e_z})}{2} \right].
\end{split} \label{Eq.C11}
\end{equation}
Here the term with $\log (L_B/a)$ arises because the lengths are non-dimensionalised with $a$ and $L_B$ in the inner and Brinkman region respectively. For the second matching, the Brinkman solution in the limit of $\rho/L_B \gg 1$ should be matched with the inner limit ($\rho/l \ll 1$) of the outer solution. The resulting equations in this case are:
\begin{equation}
\begin{split}
    c_1 &= \bm{U}_{\infty} + \frac{\bm{f}_s + \lambda \bm{f}_p}{4\pi (1+\lambda)} \left[ \log (2\gamma) + \log \left(\frac{\sqrt{s(1-s)}}{\overline{a}(s)} \right) \right] +
    \frac{\bm{f}_s + \lambda \bm{f}_p}{4\pi (1+\lambda)} \log \left(\frac{a}{L_B} \right) - \frac{(\bm{f}_s + \lambda \bm{f}_p) \cdot \bm{e_z e_z}}{4 \pi (1+\lambda)} \\
    &+ \frac{1}{8\pi (1 + \lambda)} \int \left[\left( \frac{\bm{I}}{|\bm{r_c}(s) - \bm{r_c}(s')|} + \frac{(\bm{r_c}(s) - \bm{r_c}(s'))(\bm{r_c}(s) - \bm{r_c}(s'))}{|\bm{r_c}(s) - \bm{r_c}(s')|^3} \right)\cdot(\bm{f}_s + \lambda \bm{f}_p)(\bm{r_c}(s')) \right.\\
    &\left.- \left(\frac{(\bm{I} + \bm{e_z e_z})}{|s - s'|} \right)\cdot(\bm{f}_s + \lambda \bm{f}_p)(\bm{r_c}(s)) \right] ds'
\end{split} \label{Eq.C12}
\end{equation}
and
\begin{equation}
\begin{split}
    c_2 &= \bm{U}_{\infty} + \frac{\bm{f}_s + \lambda \bm{f}_p}{4\pi (1+\lambda)} \left[ \log (2\gamma) + \log \left(\frac{\sqrt{s(1-s)}}{\overline{a}(s)} \right) \right] +
    \frac{\bm{f}_s + \lambda \bm{f}_p}{4\pi (1+\lambda)} \log \left(\frac{a}{L_B} \right) - \frac{(\bm{f}_s + \lambda \bm{f}_p) \cdot \bm{e_z e_z}}{4 \pi (1+\lambda)} \\
    &+ \frac{1}{8\pi (1 + \lambda)} \int \left[\left( \frac{\bm{I}}{|\bm{r_c}(s) - \bm{r_c}(s')|} + \frac{(\bm{r_c}(s) - \bm{r_c}(s'))(\bm{r_c}(s) - \bm{r_c}(s'))}{|\bm{r_c}(s) - \bm{r_c}(s')|^3} \right)\cdot(\bm{f}_s + \lambda \bm{f}_p)(\bm{r_c}(s')) \right.\\
    &\left.- \left(\frac{(\bm{I} + \bm{e_z e_z})}{|s - s'|} \right)\cdot(\bm{f}_s + \lambda \bm{f}_p)(\bm{r_c}(s)) \right] ds'
\end{split} \label{Eq.C13}
\end{equation}
Substituting Eq.\ref{Eq.C12},\ref{Eq.C13} in Eq.\ref{Eq.C10} and \ref{Eq.C11} gives us Eq.\ref{Eq.53A},\ref{Eq.53Bo} respectively.

%------------------------------------------------****-----------------------------------------

\bibliographystyle{jfm}

\begin{thebibliography}{99}

\expandafter\ifx\csname natexlab\endcsname\relax
\def\natexlab#1{#1}\fi
\expandafter\ifx\csname selectlanguage\endcsname\relax
\def\selectlanguage#1{\relax}\fi

% \bibitem[Armitage \& Schmitt(1997)]{Armitage97}
% {\sc Armitage, J.P. and Schmitt, R.} 1997 {Bacterial chemotaxis: Rhodobacter sphaeroides and Sinorhizobium meliloti—variations on a theme?.}, {\it Microbiol.}, {\bf 143}, 3671-3682.

\bibitem[Batchelor(1970)]{Batchelor}
{\sc Batchelor, G.K.} 1970 {Slender-body theory for particles of arbitrary cross-section in Stokes flow.}, {\it J. Fluid Mech.}, {\bf 44}(3), 419-440.

\bibitem[Berg \&  Anderson(1973)]{10a}
{\sc Berg, H.C. and Anderson, R.A.} 1973 {Bacteria swim by rotating their flagellar filaments.}, {\it Nature}, {\bf 245}, 380–382.

\bibitem[Berg \& Turner(1979)]{Berg}
{\sc Berg, H.C. and Turner, L.} 1979 {Movement of microorganisms in viscous environments.}, {\it Nature}, {\bf 278}, 349-351.

\bibitem[Berg(2003)]{11}
{\sc Berg, H.C.} 2003 {The rotary motor of bacterial flagella.}, {\it Ann. Rev. Biochem.}, {\bf 72}, 19-54.

\bibitem[Borker \& Koch(2019)]{Borker}
{\sc Borker, N.S. and Koch, D.L.} 2019 {Slender body theory for particles with non-circular cross-sections with application to particle dynamics in shear flows.}, {\it J. Fluid Mech.}, {\bf 877}, 1098-1133.

% \bibitem[Bray(2000)]{Bray2000}
% {\sc Bray, D.} 2000 {\it Cell Movements}, {Garland, NY}.

\bibitem[Brinkman(1947))]{Brinkman}
{\sc Brinkman, H.C.} 1947 {A calculation of the viscous force exerted by a flowing fluid on.a dense swarm of  particles.}, {\it App. Sci. Res.}. {\bf A1}, 27-34.

% \bibitem[Canale-Parloa(1984)]{Canale-Parola84}
% {\sc Canale-Parloa, E.} 1984 {The spirochetes.}, {\it Bergey’s Manual of
% Systematic Bacteriology} {ed. N.R. Krieg and J.G. Holt, Williams and Wilkins, Baltimore, MD}, 38-70.

% \bibitem[Celli et al.(2009)]{Celli09}
% {\sc Celli, J.P., Turner, B.S., Afdhal, N.H., Keates, S., Ghiran, I., Kelly, C.P., Ewoldt, R.H., McKinley, G.H., So, P., Erramilli, S. and Bansil, R.} 2009 {Helicobacter pylori moves through mucus by reducing mucin viscoelasticity}, {\it Proc. Nat. Acad. Sci.}, {\bf 106}(34), 14321-14326.

\bibitem[Chwang \& Wu(1971)]{Chwang}
{\sc Chwang, A.T. and Wu, T.Y.} 1971 {A note on the helical locomotion of micro-organisms.}, {\it Proc. Roy. Soc. Lond. B}, {\bf 178}, 327-346.

\bibitem[Chen et.al.(2020)]{Pak}
{\sc Chen, Y., Lordi, N., Taylor, M. and Pak, O.S.} 2021 {Helical locomotion in a porous medium.}, {\it Phys. Rev. E}, {\bf 102}, 043111.

\bibitem[Cone(2009)]{Cone}
{\sc Cone, R.} 2009 {Barrier properties of mucus.}, {\it Adv. Drug Del. Rev.}, {\bf 61}, 75-85.

\bibitem[Cox(1970)]{Cox}
{\sc Cox, R.G.} 1970 {The motion of long slender bodies in a viscous fluid - Part 1. General theory.}, {\it J. Fluid Mech.}, {\bf 44}(4), 791-810.

\bibitem[Das \& Lauga(2018)]{Das}
{\sc Das, D. and Lauga, E.} 2018 {Computing the motor torque of Escherichia coli}, {\it Soft Matter}, {\bf 14}, 5955-5967.

\bibitem[Doi(1990)]{Doi}
{\sc Doi, M.}, {Effects of viscoelasticity on polymer diffusion} in {\it Dynamics and patterns in complex fluids}, {A. Onuki and K. Kawasaki (eds.)}, Springer, 1990.

\bibitem[Du et.al(2012)]{Du}
{\sc Du, J., Keener, J. P., Guy, R. D. and Fogelson, A. L.} 2012 {Low Reynolds number swimming in viscous two-phase fluids}, {\it Phys. Rev. E}, {\bf 85}, 036304.

% \bibitem[Dasgupta et al.(2013)]{Dasgupta}
% {\sc Dasgupta, M., Liu, B., Fu, H.C., Berhanu, M., Breuer, K.S., Powers, T.R. and Kudrolli, A.}, {Speed of a swimming sheet in Newtonian and viscoelastic fluids}, {\it Phys. Rev. E}, {\bf 87}, 013015.

% \bibitem[Eastham, Mohammadigoushki \& Shoele(2022)]{Hadi22}
% {\sc Eastham, P.S., Mohammadigoushki, H. and Shoele, K.} 2022 {Squirmer locomotion in a yield stress fluid.}, {\it J. Fluid Mech.}, {\bf 948}, A54.

\bibitem[Espionsa-Garcia et al.(2013)]{Zenit}
{\sc Espinosa-Garcia, J., Lauga, E. and Zenir, R.} 2013 {Fluid elasticity increases the locomotion of flexible swimmers.}, {\it Phys. Fluids}, {\bf 25}, 031701.

%\bibitem[Fan, Dhont \& Tuiner(2007)]{Tuinier}
%{\sc Fan, T.H., Dhont, J.K.G. and Tunier, R.} 2007 {Motion of a sphere through a polymer solution.}, {\it Phys. Rev. E.}, {\bf 75}, 011803.

% \bibitem[Figueroa-Morales et al.(2019)]{Aranson19}
% {\sc Figueroa-Morales, N., Dominguez-Rubio, L, Ott, T.L. and Aranson, I.S.} 2019 {Mechanical shear controls bacterial penetration in mucus.}, {\it Sci. Rep.}, {\bf 9}, 9713.

\bibitem[Fu et al.(2007)]{Fu1}
{\sc Fu, H. C., Powers, T. R. and Wolgemuth, C. W.} 2007 {Theory of swimming filaments in viscoelastic media.}, {\it Phys. Rev. Lett.}, {\bf 99}(25), 258101.

\bibitem[Fu et.al(2009)]{Fu2}
{\sc Fu, H. C., Wolgemuth, C. W. and Powers, T. R.} 2009 {Swimming speeds of filaments in nonlinearly viscoelastic fluids.}, {\it Phys. Fluids}, {\bf 21}(3), 033102.

\bibitem[Fu et.al(2010)]{PowersTF}
{\sc Fu, H. C., Shenoy, V. B. and Powers, T. R.} 2010
{Low Reynolds number swimming in gels}, {\it Euro. Phys. Lett.}, {\bf 91}, 24002.

\bibitem[Ghosh \& Ghosh(2021)]{9}
{\sc Ghosh, A. and Ghosh, A.} 2021 {Mapping viscoelastic properties using helical magnetic nanopropellers}, {\it Trans. Ind. Nat. Acad. Eng.}, {\bf 6}, 429-438.

\bibitem[Gray \& Hancock(1955)]{Gray}
{\sc Gray, J. and Hancock, G.J.} 1955 {The propulsion of sea-urchin spermatozoa.}, {\it J. Exp. Biol.}, {\bf 32}(4), 802-814.

\bibitem[Hinch \& Acrivos(1980)]{Hinch}
{\sc Hinch, E.J. and Acrivos, A.} 1980 {Long slender drops in a simple shear flow.}, {\it J. Fluid Mech.}, {\bf 98}(2), 305-328.

\bibitem[Howells(1974)]{Howells}
{\sc Howells, I.D.} 1974 {Drag due to the motion of a Newtonian fluid through a sparse random array of small fixed rigid objects.}, {\it J. Fluid Mech.}, {\bf 64}(3), 449-475.

\bibitem[Howells(1998)]{Howells98}
{\sc Howells, I.D.} 1998 {Drag on fixed beds of fibres in slow flow.}, {\it J. Fluid Mech.}, {\bf 355}, 163-192.

\bibitem[Huang et al.(2019)]{8}
{\sc Huang, H.W., Uslu, F.E., Katsamba, P. Lauga, E. Sakar, M.S. and Nelson, B.J.} 2019 {Adaptive locomotion of artificial microswimmers.}, {\it Sci. Adv.}, {\bf 5}, 1532.

\bibitem[Jarrel \& McBride(2008)]{2}
{\sc Jarrell, K.F. and McBride, M.J.} 2008 {The surprisingly diverse ways that prokaryotes move.}, {\it Nature Rev. Microbiol.}, {\bf 6}, 466476.

\bibitem[Jean et al.(1996)]{1}
{\sc Jean, A.B., Tanya, R., Lin, J.C.T. and MacDonald, J.M.} 1996 {Bacterial
Food-borne Disease: Medical Costs and Productivity Losses.}, {\it Food and Consumer Economics
Division, Economic Research Service, U.S. Department of Agriculture. Agricultural Economic
Report No. 741.}

\bibitem[Johnson(1980)]{Johnson}
{\sc Johnson, R.E.} 1980 {An improved slender-body theory for Stokes flow.}, {\it J. Fluid Mech.}, {\bf 99}(2), 411-431.

% \bibitem[Kamal, Gravelle \& Botto(2020)]{Kamal20}
% {\sc Kamal, C., Gravelle, S. and Botto, L.} 2020 {Hydrodynamic slip can align thin nanoplatelets in shear ﬂow}, {\it Nat. Communications}, {\bf 11}, 2425.

% \bibitem[Kamal, Gravelle \& Botto(2021)]{Kamal21}
% {\sc Kamal, C., Gravelle, S. and Botto, L.} 2021 {Alignment of a flexible plate-like particle in shear flow: Effect of surface slip and edges}, {\it Phys. Rev. Fluids}, {\bf 6}(8), 084102.

\bibitem[Kamdar et al.(2022)]{Cheng}
{\sc Kamdar, S., Shin, S., Leishangthem, P., Francis, L.F., Xu, X. and Cheng, X.} 2022 {The colloidal nature of complex fluids enhances bacterial motility}, {\it Nature}, {\bf 603}, 819-822.

\bibitem[Kearns(2010)]{3}
{\sc Kearns, D.B.} 2010 {A field guide to bacterial swarming motility.}, {\it Nature Rev. Microbiol.}, {\bf 8}, 634664.

\bibitem[Keller \& Rubinow(1976)]{Keller}
{\sc Keller, J. and Rubinow, S.} 1976 {Slender body theory for slow viscous flow.}, {\it J. Fluid Mech.}, {\bf 75}(4), 705-714.

\bibitem[Kirch et al.(2012)]{Lehr}
{\sc Kirch, J., Schneider, A., Abou, B., Hopf, A., Schaefer, U.F., Schneider, M., Schall, C., Wagner, C. and Lehr, C.M.} 2012 {Optical tweezers reveal relationship between microstructure and nanoparticle
penetration of pulmonary mucus.}, {\it Proc. Nat. Acad. Sci.}, {\bf 109}(45), 18355-18360.

% \bibitem[Koyasu \& Shirakihara(1984)]{Koyasu84}
% {\sc Koyashu, S. and Shirakihara, Y.} 1984 {Caulobacter crescentus
% flagellar filament has a right-handed helical form.}, {\it J. Mol. Biol.}, {\bf 173}, 125-130.

% \bibitem[Lai et al.(2009)]{Lai09}
% {\sc Lai, S.K., Wang, Y.Y, Wirtz, D. and Hanes, J.} 2009 {Micro- and macro-rheology of mucus.}, {\it Adv. Drug Del. Rev..}, {\bf 61}, 86-100.

\bibitem[Lai et al.(2011)]{Lai}
{\sc Lai, S., Wang, Y., Hida, K., Cone, R. and Hanes, J.} 2011 {Nanoparticles reveal that human
cervicovaginal mucus is riddled with pores larger than viruses.}, {\it Proc., Nat. Acad., Sci.}, {\bf 108}(34), 14371-14375.

\bibitem[Lauga(2007)]{Lauga}
{\sc Lauga, E.} 2007 {Propulsion in a viscoelastic fluid}, {\it Phys. Fluids}, {\bf 19}(8), 083104.

\bibitem[Lauga \& Powers(2009)]{13}
{\sc Lauga, E. and Powers, T.R.} 2009 {The hydrodynamics of swimming microorganisms.}, {\it Rep. Prog. Phys.}, {\bf 72}, 096601.

\bibitem[Leshansky(2009)]{Leshansky}
{\sc Leshansky, A.L.} 2009 {Enhanced low-Reynolds-number propulsion in heterogeneous viscous environments}, {\it Phys. Rev. E}, {\bf 80}, 051911.

% \bibitem[Li et al.(2021)]{Gaojin}
% {\sc Li, G., Lauga, E. and Ardekani, A.M.} 2021 {Microswimming in viscoelastic fluids.}, {\it J. Non-Newt. Fluid Mech.}, {\bf 297}, 104655.

\bibitem[Liu et al.(2011)]{Liu}
{\sc Liu, B., Powers, T.R. and Breuer, K.S.} 2011 {Force-free swimming of a model helical
flagellum in viscoelastic fluids.}, {\it Proc. Nat. Acad. Sci.}, {\bf 108}(49), 19516-19520.

\bibitem[Ludwig (1930)]{Ludwig30}
{\sc Ludwig, W.} 1930 {Zur theorie der Flimmerbewegung (Dynamik, Nutzeffekt, Energiebilanz).}, {\it J. Comparative Physiol. A.}, {\bf 13}, 397-504.

\bibitem[Mackaplow et al.(1994)]{Mackaplow94}
{\sc Mackaplow, M.B., Shaqfeh, E.S.G. and Schiek, R.L.} 1994 {A Numerical Study of Heat and Mass Transport in Fibre Suspensions}, {\it Proc. Roy. Soc. Lond. A}, {\bf 447}, 77-110.

\bibitem[Magariyama \& Kudo(2002)]{Kudo}
{\sc Magariyama, Y. and Kudo, S.} 2002 {A mathematical explanantion of increase in bacterial swimming speed with viscosity in linear-polymer solutions.}, {\it Biophys. J.}, {\bf 83}, 733-739.

\bibitem[Man \& Lauga(2015)]{Lauga3}
{\sc Man, Y. and Lauga, E.} 2015 {Phase-separation models for swimming enhancement in complex fluids.}, {Phys. Rev. E}, {\bf 92}, 023004.

\bibitem[Martinez et al.(2014)]{Poon}
{\sc Martinez, V.A., Schwarz-Link, J., Reufer, A.M., Wilson, L.G., Morozov, A.N. and Poon, W.C.K.}
2014 {Flagellated bacterial motility in polymer solutions.}, {\it Proc. Nat. Acad. Sci.}, {\bf 111}(50), 17771-17776.

\bibitem[McShane et al.(2021)]{5}
{\sc McShane, A., Bath, J., Jaramilo, A.M., Ridley, C., Walsh, A.A., Evans, C.M., Thornton, D.J. and Ribbeck, K.} 2021 {Mucus.}, {\it Curr. Biol.}, {\bf 31}, R931-R947.

\bibitem[Mhetar \& Archer(1998)]{Archer}
{\sc Mhetar, V. and Archer, L.A.} 1998 {Slip in entangled polymer solutions.}, {\it Macromol.}, {\bf 31}, 6639-6649.

% \bibitem[Moschopoulos et al.(2023)]{Tsam23}
% {\sc Moschopoulos, P., Kouni, E., Psaraki, K., Dimakopoulos, Y. and Tsamopoulos, J.} 2023 {Dynamics of elastoviscoplastic filament stretching.}, {\it Soft Matt.}, {\bf 19}, 4717-4736.

% \bibitem[Nazari, Shoele \& Mohammadigoushki(2023)]{Hadi23}
% {\sc Nazari, F., Shoele, K. and Mohammadigoushki, H.} 2023 {Helical Locomotion in Yield Stress Fluids.}, {\it Phys. Rev. Lett.}, {\bf 130}, 114002.

\bibitem[Ottemann \& Miller(1997)]{4}
{\sc Ottemann, K.M. and Miller, J.F.}, {Roles for motility in bacterial-host interactions.}, {\it Mol. Microbiol.}, {\bf 24}, 11091117.

\bibitem[Patteson et al.(2015)]{Patteson15}
{\sc Patteson, A.E., Gopinath, A., Goulian, M. and Araatia, P.E.} 2015 {Running and tumbling with E. coli in polymeric solutions.}, {\it Sci. Rep.}, {\bf 5}, 15761.

%\bibitem[Patteson et al.(2016)]{Patteson}
%{\sc Patteson, A.E., Gopinath, A. and Araatia, P.E.} 2016 {Active colloids in complex fluids.}, {\it Curr. Opim. Collo. Interf. Sci.}, {\bf 21}, 86-96.

\bibitem[Purcell(1977)]{10}
{\sc Purcell, E.M.} 1977 {Life at low Reynolds number.}, {\it Am. J. Phys.}, {\bf 45}, 3-11.

\bibitem[Qu \& Breuer(2020)]{Breuer}
{\sc Qu, Z. and Breuer, K.S.} 2020 {Effects of shear-thinning viscosity and viscoelastic stresses on flagellated bacteria motility.}, {\it Phys. Rev. Fluids.}, {\bf 5}, 073103.

\bibitem[Qu et al.(2018)]{Qu18}
{\sc Qu, Z., Temel, F.Z., Henderikx, R. and Breuer, K.S.} 2018 {Changes in the flagellar bundling time account for variations in swimming behavior of flagellated bacteria in viscous media,.}, {\it Proc. Natl. Acad. Sci. USA}, {\bf 115}(8), 1707-1712.

% \bibitem[Quicke et al.(1992)]{Quicke92}
% {\sc Quicke, D.L.J., Ingram, S.N., Baillie, H.S. and Gaitens, P.V.} 1992 {Sperm structure and ultrastructure in Hymenoptera (insecta).}, {\it Zoologica Scr.}, {\bf 21}, 318-402.

% \bibitem[Quraishi, Jones \& Mason(1998)]{Quraishi98}
% {\sc Quraishi, M.S., Jones, N.S., and Mason, J.} 1998 {The rheology of nasal mucus: a review.}, {\it Clin. Otol. Allied Sci.}, {\bf 23}, 403-413.

\bibitem[Riley \& Lauga(2014)]{Lauga2}
{\sc Riley, E.E. and Lauga, E.} 2014 {Enhanced active swimming in viscoelastic fluids.}, {Europhys. Lett.}, {\bf 108}, 34003.

\bibitem[Rodenborn et al.(2013)]{Swinney}
{\sc Rodenborn, B., Chen, C. H., Swinney, H. L., Liu, B. and Zhang, H. P.} 2013 {Propulsion of microorganisms by a helical ﬂagellum.}, {\it Proc. Nat. Acad. Sci.}, {\bf 110}(5), E338-E347.

\bibitem[Sharma \& Koch(2023)]{Arjun}
{\sc Sharma, A. and Koch, D.L.} 2023 {Finite difference method in prolate spheroidal coordinates for linear flow past a spheroid in viscoelastic fluids with zero to moderate inertia}, {\it J. Comp. Phys.} (Accepted).

% \bibitem[Sharma et al.(2023)]{Sabarish23}
% {\sc Sharma, A. Narayanan, S.V., Hormozi, S. and Koch, D.L.} 2023 {Manuscript under preparation}.

\bibitem[Sowa \& Berry (2008)]{Sowa}
{\sc Sowa, Y. and Berry, R.M.} 2008 {Bacterial flagellar motor.}, {\it Quart. Rev. Biophys.}, {\bf 41}(2), 103-132.

\bibitem[Spagnolie et al.(2013)]{Spagnolie}
{\sc Spagnolie, S.E., Liu, B. and Powers, T.R.} 2013 {Locomotion of helical bodies in viscoelastic fluids: enhanced swimming at large helical amplitudes.}, {\it Phys. Rev. Lett.}, {\bf 111}(6), 068101.

\bibitem[Spagnolie \& Underhill(2023)]{SpagnolieARFM}
{\sc Spagnolie, S.E. and Underhill, P.T.} 2023 {Swimming in Complex fluids}, {\it Ann. Rev. Condens. Matter Phys.}, {\bf 14}, 381-415.

% \bibitem[Suarez \& Ho(2003)]{Suarez03}
% {\sc Suarez, S.S. and Ho, H.C.} 2003 {Hyperactivated motility in sperm.}, {\it Reprod. Dom. Anim.}, {\bf 38}, 119-124.

% \bibitem[Suarez \& Pacey(2006)]{Suarez06}
% {\sc Suarez, S.S. and Pacey, A.A.} 2006 {Sperm transport in the female reproductive tract.}, {\it Human Reprod. Update.}, {\bf 12}, 23-37.

\bibitem[Subramanian \& Nott(2011)]{12}
{\sc Subramanian, G. and Nott, P.R.} 2011 {The fluid dynamics of swimming microorganisms and cells.}, {\it J. Ind. Ins. Sci.}, {\bf 91}(3), 383-413.

\bibitem[Teran et al.(2010)]{Teran}
{\sc Teran,J., Fauci, L. and Shelley, M.} 2010 {Viscoelastic fluid response can increase the
speed and efficiency of a free swimmer.}, {\it Phys. Rev. Lett.}, {\bf 104}(3), 1-4.

\bibitem[Thomases \& Guy(2014)]{Guy}
{\sc Thomases, B. and Guy, R.D.} 2014 {Mechanisms of elastic enhancement and eindrance for finite-length undulatory swimmers in viscoelastic Fluids.}, {\it Phys. Rev. Lett.}, {\bf 113}, 098102.

% \bibitem[Tung et al.(2017)]{Wu}
% {\sc Tung, C.K., Lin, C., Harvey, B., Fiore, A.G., Ardon, F., Wu, M. and Suarez, S.S.} 2017 {Fluid viscoelasticity promotes collective swimming of sperm.}, {\it Sci. Rep.}, {\bf 7}(1), 1-9.

\bibitem[Turner et al.(2000)]{Bergfig}
{\sc Turner, L., Ryu, W.S. and Berg, H.C.} 2000 {Real-Time Imaging of Fluorescent Flagellar Filaments.}, {\it J. Bacteriol.}, {\bf 182}, 10.

\bibitem[Wada(2010)]{Wada}
{\sc Wada, H.} 2010 {Pumping visoelastic two-fluid media}, arXiv:1004.1254.

\bibitem[Werlang et al.(2019)]{6}
{\sc Werlang, C., Carcarmo-Oyarce, G. and Ribbeck, K.} 2019 {Engineering mucus to study and influence the microbiome.}, {\it Nature Rev. Mat.}, {\bf 4}, 135-145.

% \bibitem[Werner \& Simmons(2008)]{Werner08}
% {\sc Werner, M. and Simmons, L.W.} 2008 {Insect sperm motility.}, {\it Biol. Rev.}, {\bf 83}, 191-208.

\bibitem[Xiea et al.(2020)]{7}
{\sc Xiea, S., Xiab, T., Lia, S., Moa, C., Chena, M. and Lia, X.} 2020 {Bacteria-propelled microrockets to promote the tumor accumulation and intracellular drug uptake}, {\it Chem. Eng. J.}, {\bf 392}, 123786.

\bibitem[Xing et al.(2006)]{Berry}
{\sc Xing, J., Bai, F., Berry, R. and Oster, G.} 2005 {Torque-speed relationship of bacterial flagellar motor}, {\it Proc. Nat. Acad. Sci.}, {\bf 103}(5), 1260-1265.

\bibitem[Zottl \& Yeomans(2019)]{Yeomans}
{\sc Zottl, A. and Yeomans, J.M.} 2019 {Enhanced bacterial swimming speeds in macromolecular polymer solutions.}, {\it Nat. Phys.}, {\bf 15}(6), 554-558.

\end{thebibliography}

\end{document}